\documentclass[11pt]{elsarticle}
 \usepackage[margin=2cm]{geometry}
 \journal{Physics Reports}

\usepackage{amsmath}
\usepackage{amssymb,slashed}
\usepackage{dcolumn}	
\usepackage{bm}			
\usepackage{bbm} 		
\usepackage{multirow}
\usepackage{blkarray}
\usepackage{feynmp}
\usepackage[normalem]{ulem}
\usepackage[usenames,dvipsnames]{xcolor}
\definecolor{hgreen}{rgb}{0,.3,0}
\definecolor{hred}{rgb}{.3,0,0}
\definecolor{hblue}{rgb}{0,0,.3}
\definecolor{LightGray}{gray}{0.95}
\usepackage[colorlinks=true,linkcolor=hblue,citecolor=hgreen,filecolor=hblue,
urlcolor=hred]{hyperref}
\makeatletter
\def\endfmffile{%
	\fmfcmd{\p@rcent\space the end.^^J%
		end.^^J%
		endinput;}%
	\if@fmfio
	\immediate\closeout\@outfmf
	\fi
	\ifnum\pdfshellescape>\z@
	\immediate\write18{mpost \thefmffile}%
	\fi}
\makeatother

\usepackage{hyperref,graphicx}           
\usepackage{pbox}
\usepackage{caption}
\usepackage{subcaption}
\usepackage{xcolor}
\usepackage[utf8]{inputenc}

\newcommand{\beq}{\begin{equation}}
\newcommand{\eeq}{\end{equation}}
\newcommand{\ev}[1]{\langle #1 \rangle}

\begin{document}

\title{Review of Neutral Naturalness}

\author[a]{Brian Batell}\ead{batell@pitt.edu}
\affiliation[a]{
organization={Pittsburgh Particle Physics, Astrophysics, and Cosmology Center, Department of Physics and Astronomy, University of Pittsburgh}, 
city={Pittsburgh},
postcode={15260},
state={Pennsylvania},
country={USA}
}
\author[a]{Matthew Low}\ead{matthew.w.low@pitt.edu}
\author[b]{Ethan T. Neil}\ead{ethan.neil@colorado.edu}
\affiliation[b]{
organization={Department of Physics, University of Colorado}, 
city={Boulder},
postcode={80309},
state={Colorado},
country={USA}
}
\author[c]{Christopher B. Verhaaren\corref{cor1}}\ead{verhaaren@physics.byu.edu}
\affiliation[c]{
organization={Department of Physics and Astronomy,\\ Brigham Young University}, 
city={Provo},
postcode={84602},
state={Utah},
country={USA}
}
\cortext[cor1]{Corresponding author}

\date{\today}
\begin{abstract}
The hierarchy between the mass parameter of the Higgs boson and larger mass scales becomes ever more puzzling as experiments explore higher energies. Neutral naturalness is the umbrella term for symmetry-based explanations for these hierarchies whose quark symmetry partners are not charged under the SU(3)$_c$ color gauge group of the Standard Model. Though the first manifestations of this idea predate the physics runs of the Large Hadron Collider, since the Higgs discovery this paradigm has grown and developed to include a wide variety of concrete realizations with connections to intriguing collider signals. Determining the phenomenology of such models often requires the characterization\textemdash typically relying on lattice calculations\textemdash of a new confining gauge symmetry. This presents additional motivation to further develop our understanding of nonperturbative field theory as well as to pursue specific lattice studies. The wide range of suggested hidden sectors also produces a variety of dark matter candidates, intersections with astrophysics and cosmology, and ties to neutrinos and flavor. In this review, we orient the reader within both this growing collection of specific models and the physical phenomena they produce. We also survey the often less familiar dynamics of hidden-sector glueballs and quirks. In addition to providing a guide to past efforts, we reveal interesting directions for further study.
\end{abstract}



\maketitle
\tableofcontents

\section{Introduction\label{s.intro}}
Despite its phenomenal experimental successes, the Standard Model (SM) can only be considered as a low-energy, effective field theory (EFT) description of microphysics. As typically accepted, the SM does not specify how the neutrinos obtain mass nor explain the observed asymmetry between matter and antimatter in the Universe. The cosmological dark matter cannot be composed of SM fields and any explanation of the quark and lepton flavor structure or the absence of measurable CP violation in Quantum Chromodynamics (QCD) must include physics beyond the SM. In addition to these known unknowns, other particles and interactions may be essential to Nature's structure, even if we have not yet recognized what role they play.

Extensions of the SM are likely tied to a high-energy (and thus short-distance) scale above which new particles and interactions are manifest. Although other high-energy scales may appear, the one scale above which new phenomena must arise is the Planck scale $M_\text{Pl}$, which is of order $10^{19}$ GeV. Above this energy scale, the detailed physics associated with quantum fluctuations of gravity must be revealed. For most properties of the SM, consistent with the idea of scale separation within EFT, any possible effects of Planck-scale physics on low-energy experiments at some energy $E$ is suppressed by powers of $E / M_\text{Pl}$, and are thus negligibly small.  

However, an exception occurs for the Higgs field, which is unique within the SM: it is the only elementary scalar field and its mass term ($\mu^2$) is the only dimensionful quantity in the SM Lagrangian. The generic EFT expectation is that this mass (and hence the mass of the Higgs boson) is set by the largest scale it interacts with; in a hypothetical unification of the SM with quantum gravity, this would be the Planck scale. The apparent catastrophic failure of this generic expectation, due to the fact that the Higgs mass of order $100$ GeV is much smaller than $M_\text{Pl}$, presents an immediate puzzle known as the \textit{electroweak hierarchy problem}. 

The logic of this argument can be seen from the one-loop corrections to the Higgs potential. The largest effect, shown in Fig.~\ref{fig:HiggsLoop}, is due to the top quark because of its large coupling to the Higgs (though loops due to other fermions, gauge bosons, and self-interactions also contribute). The top-loop contribution indicates that the physical Higgs mass parameter $\mu^2\sim (90\text{ GeV})^2$ is the combination of some tree-level mass $\mu_0^2$ and the one-loop term
\beq
\mu^2=\mu_0^2+\frac{3\lambda_t^2}{8\pi^2}\Lambda_\text{UV}^2~,
\eeq
where $\Lambda_\text{UV}$ stands in for a physical mass scale above which the SM is significantly modified, sometimes called the cutoff. The top-quark Yukawa coupling $\lambda_t \sim 1$, so it is evident from this equation that if the cutoff $\Lambda_\text{UV} \gg \mu$, then the value of the tree-level mass $\mu_0$ must be \emph{fine-tuned}, canceling precisely against the second term.  This makes the Higgs mass parameter $\mu^2$ highly sensitive to the values of the other parameters. In principle all the parameters of the SM play some role in this sensitivity, but as the top Yukawa is by far the largest coupling to the Higgs any meaningful resolution to the hierarchy problem must address the top sector first.

One way to characterize how the physical mass depends on various parameters is the measure associated with Barbieri and Giudice~\cite{Ellis:1986yg,Barbieri:1987fn} 
\beq
\Delta_x=\left|\frac{x}{\mu^2}\frac{\partial\mu^2}{\partial x}\right|~.\label{eq:BGtuning}
\eeq
The larger this value, the more sensitive $\mu^2$ is to variations in the parameter $x$, or in other words, the more finely $x$ must be tuned to obtain the given value of $\mu^2$. The authors take the value of 10 to be a rough boundary between not very tuned ($\Delta\lesssim10$) and tuned $(\Delta\gtrsim10)$. Of course, this boundary is not a logical necessity. The notion of tuning is somewhat qualitative, and how much tuning one is willing to accept varies from individual to individual.\footnote{There are also many other measures of tuning that have been considered~\cite{Anderson:1994dz,Fowlie:2024nhs} adding another level of subjectivity to these discussions.}

Applying this measure to the SM we find
\beq
\Delta_{\Lambda_\text{UV}^2}=\frac{3\lambda_t^2}{8\pi^2}\frac{\Lambda_\text{UV}^2}{\mu^2}~.
\eeq
For $\Lambda_\text{UV}\sim M_\text{Pl}$ one finds $\Delta\sim10^{32}$, which is quite clearly much larger than 10 and likely to exceed the tuning tolerance of many, if not most, physicists. Therefore, within the SM the hierarchy between the Planck scale and the weak scale is mysterious and sometimes labeled as a puzzle or a problem. Indeed, concrete calculations of the Higgs mass within SM extensions have borne out this EFT expectation~\cite{Gildener:1976ai,Weinberg:1978ym,Susskind:1978ms}, producing Higgs masses that are many orders of magnitude beyond the weak scale unless the model parameters are tuned very precisely.\footnote{Many excellent discussions of this hierarchy problem, including more historical context and further references, can be found in~\cite{Giudice:2008bi,Craig:2022eqo,Peskin:2025lsg}.} 

\begin{figure}[ht!]
\centering
\begin{tabular}{cc}
\begin{tabular}{c}
    \begin{fmffile}{Higgsloop}
\begin{fmfgraph*}(120,80)
\fmfpen{1.0}
\fmfstraight
\fmfleft{p1,i1,p2}\fmfright{p3,o1,p4}
\fmfv{l= $H^\dag$}{i1}\fmfv{l=$H$}{o1}
\fmf{dashes,tension=1}{i1,v1}
\fmf{fermion,left=1,tension=0.4,label.side=left,label=$t$,label.dist=5}{v1,v2,v1}
\fmf{dashes,tension=1}{v2,o1}
\fmfv{decor.shape=circle,decor.filled=full,decor.size=1.5thick,l=$\lambda_t$,l.a=120,l.d=7}{v1}
\fmfv{decor.shape=circle,decor.filled=full,decor.size=1.5thick,l=$\lambda_t$,l.a=60,l.d=7}{v2}
\end{fmfgraph*}
\end{fmffile}
\end{tabular}\hspace{0.45cm}
\begin{tabular}{c}
$\sim\displaystyle\frac{3\lambda_t^2}{8\pi^2}\Lambda_\text{UV}^2$\vspace{3mm}
\end{tabular}
\end{tabular}
\vspace{-0.5cm}
    \caption{Schematic contribution to Higgs mass from top quark loop. The scale $\Lambda_\text{UV}$ stands in for interactions due to high mass particles.}
    \label{fig:HiggsLoop}
\end{figure}

An explanation of the hierarchy between the Higgs mass and other scales may emerge from a deeper symmetry, beyond those of the SM itself.  For example, if the Higgs is a pseudo-Nambu-Goldstone boson (pNGB) of an approximate global symmetry, then its mass is controlled by the symmetry breaking parameter. Alternatively, if nature were to manifest supersymmetry (SUSY), then the chiral symmetry that protects fermion masses could be transferred to bosons like the Higgs. In either case, what effectively occurs is that new ``partner" particles are required by the symmetry and these eliminate the quadratic dependence on $\Lambda_\text{UV}$. Figure~\ref{fig:susyloops} schematically illustrates this idea for the case of SUSY, and again focusing on the top quark alone as the dominant source of quadratic sensitivity. In this figure we see that the contribution from the top quark loop is combined with the effects of its scalar partners, the stops. Each contributes a term that goes like $\Lambda_\text{UV}^2$, but the symmetry structure of SUSY ensures that they enter with opposite sign and exactly the same magnitude, leaving only logarithmic sensitivity to high scales.

\begin{figure}[ht!]
\centering
\begin{tabular}{cccc}
    \begin{tabular}{c}
\begin{fmffile}{Higgsloop}
\begin{fmfgraph*}(120,80)
\fmfpen{1.0}
\fmfstraight
\fmfleft{p1,i1,p2}\fmfright{p3,o1,p4}
\fmfv{l= $H^\dag$}{i1}\fmfv{l=$H$}{o1}
\fmf{dashes,tension=1}{i1,v1}
\fmf{fermion,left=1,tension=0.4,label.side=left,label=$t$,label.dist=5}{v1,v2,v1}
\fmf{dashes,tension=1}{v2,o1}
\fmfv{decor.shape=circle,decor.filled=full,decor.size=1.5thick,l=$\lambda_t$,l.a=120,l.d=7}{v1}
\fmfv{decor.shape=circle,decor.filled=full,decor.size=1.5thick,l=$\lambda_t$,l.a=60,l.d=7}{v2}
\end{fmfgraph*}
\end{fmffile}
\end{tabular}
\begin{tabular}{c}
\hspace{5mm}$+$ \hspace{5mm}
\end{tabular} 
\begin{tabular}{c}
\begin{fmffile}{susystop}
\begin{fmfgraph*}(110,80)
\fmfpen{1.0}
\fmfleft{p1,i1,i2,p3} 
\fmfright{p2,o1,o2,p4}
\fmf{phantom}{p3,v2,p4}
\fmf{dashes}{i1,v1,o1}
\fmfv{l=$H^\dag$}{i1}\fmfv{l=$H$}{o1}
\fmffreeze
\fmf{double,right,tension=0.1,l.side=right}{v2,v1}
\fmf{double,right,tension=0.1,l.side=right}{v1,v2}
\fmfv{decor.shape=circle,decor.filled=full,decor.size=1.5thick, l=$\lambda_t^2$,l.a=-90,l.d=5}{v1}
\fmfv{l=$\widetilde{t}$,l.d=15,l.a=0}{i2} 
\end{fmfgraph*}
\end{fmffile}
\end{tabular}
\hspace{0.1cm}
\begin{tabular}{c}
$\sim\displaystyle\frac{3\lambda_t^2}{8\pi^2}m^2_{\tilde t}\ln\Lambda_\text{UV}^2$\vspace{1mm}
\end{tabular}
\end{tabular}
\vspace{-0.5cm}
\caption{\label{fig:susyloops} Cancellation of quadratic sensitivity to $\Lambda_\text{UV}$ due to the top quark by its scalar partner stops $\widetilde{t}$ in SUSY. }
\end{figure}

Schematically, we have in the general case that the top quark field is understood to be part of a larger multiplet
\beq
t\Rightarrow \left.\left(\begin{array}{c}
t \\
T
\end{array}\right)\right\downarrow\hspace{-1.0mm}\text{Symmetry}~.
\eeq
The new $T$ particles that, along with the top quark, fill out a representation of this symmetry are referred to as top partners. The characteristics of these top partners depends on the symmetry involved.

While this is an elegant mechanism, the fact that we have not observed an exact partner particle to the top quark (or other SM particles) in nature suggests that the symmetry must be broken, splitting the masses apart with $m_T > m_t$, where $m_T$ is the mass of the top partner. However, the symmetry's influence remains even when broken, so that the quadratic sensitivity only extends to the difference between the Higgs mass parameter and top-partner mass scales.  Explicitly, the general outcome is that
\beq
\Delta_{\Lambda_\text{UV}^2}\sim\frac{3\lambda_t^2}{8\pi^2}\frac{m_T^2}{\mu^2}~.
\eeq
This shows that sensitivity to very high energy physics can be much weaker. However, there still appears to be significant parameter tuning if the top-partner masses are significantly larger than the Higgs mass parameter. 

\begin{figure}[ht!]
    \centering
    \raisebox{3mm}{\includegraphics[width=0.56\textwidth]{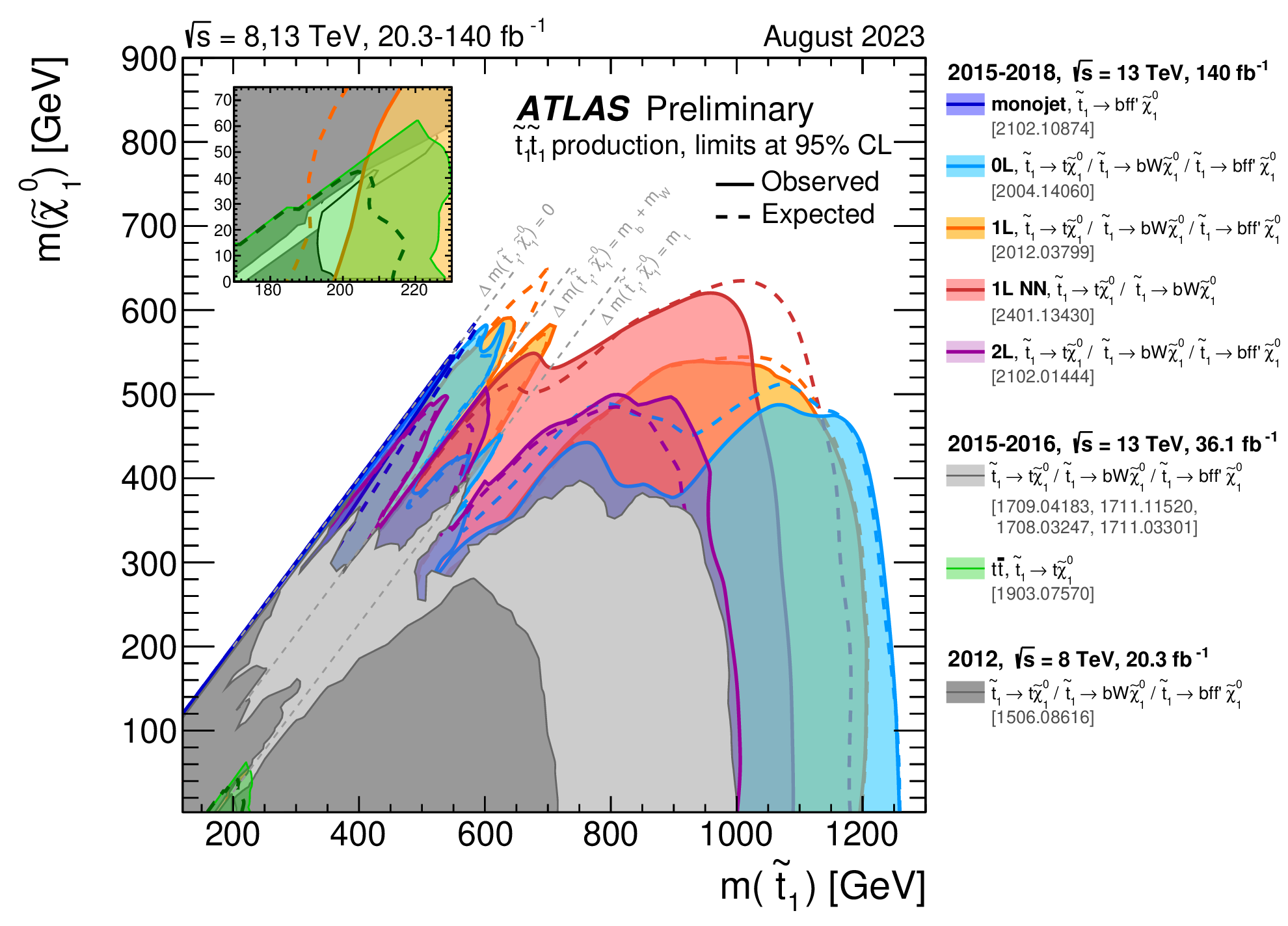}}~~
    \includegraphics[width=0.44\linewidth]{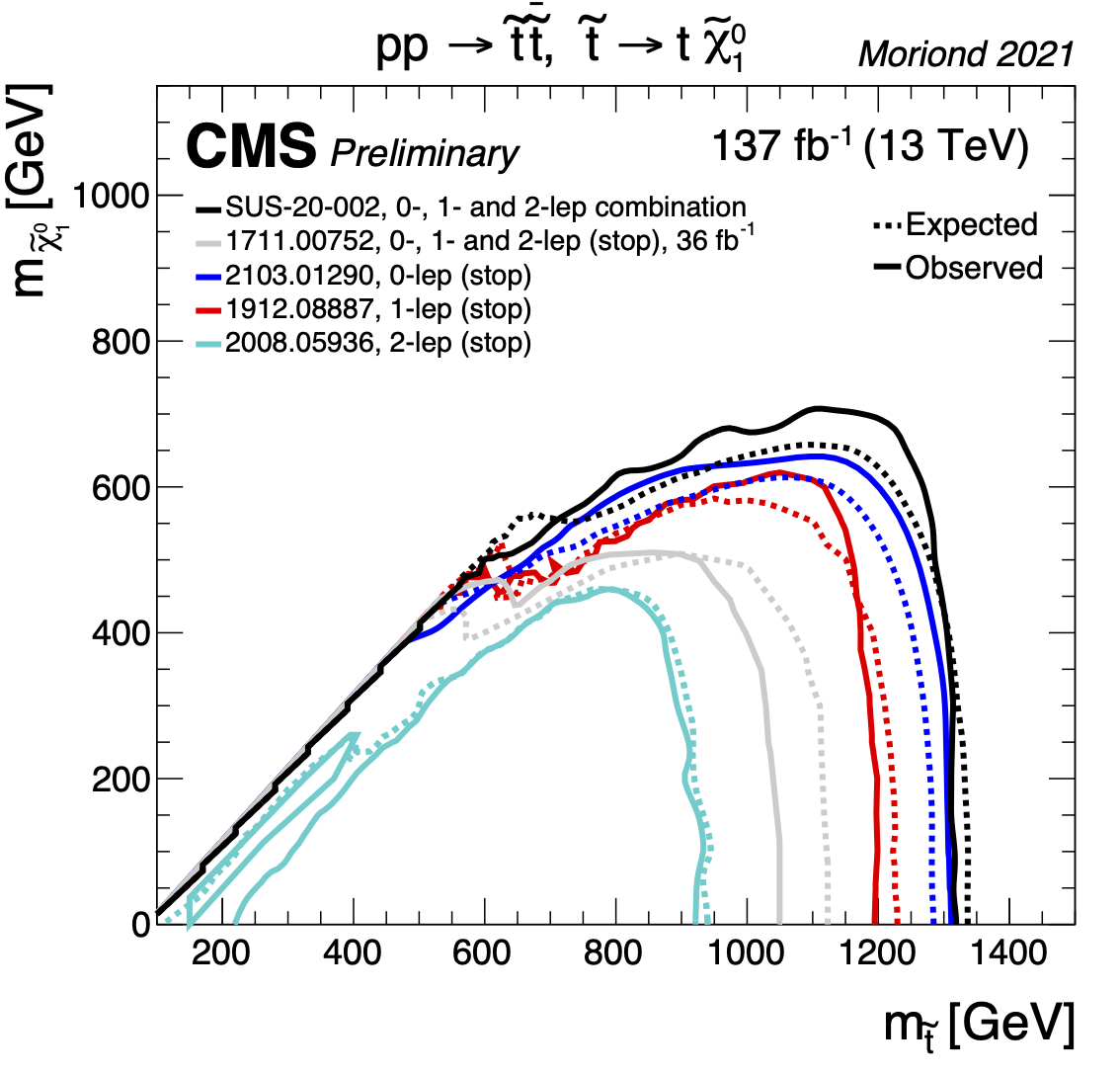}\\
    \raisebox{-3mm}{\includegraphics[width=0.52\textwidth]{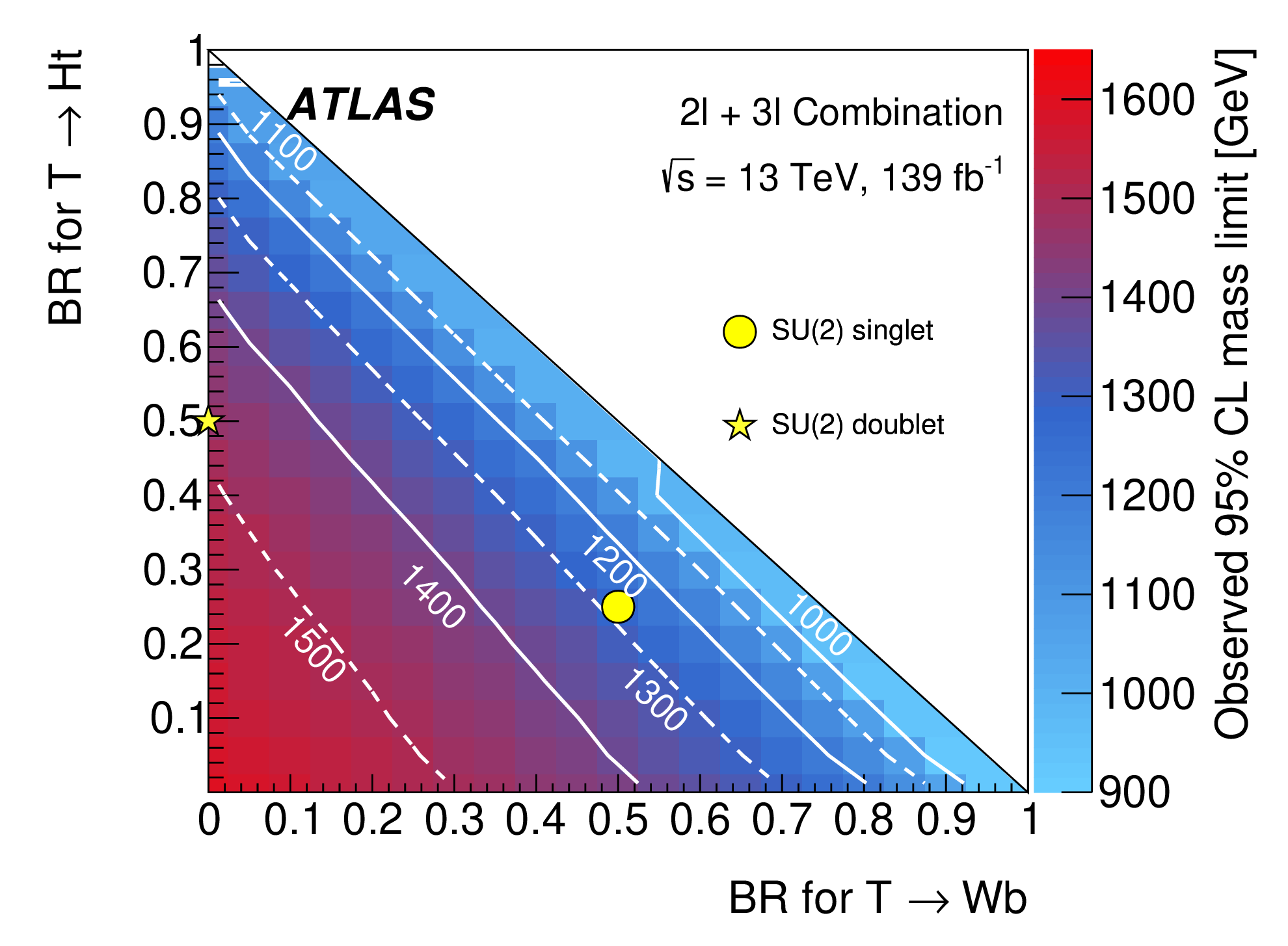}}~~
    \includegraphics[width=0.48\linewidth]{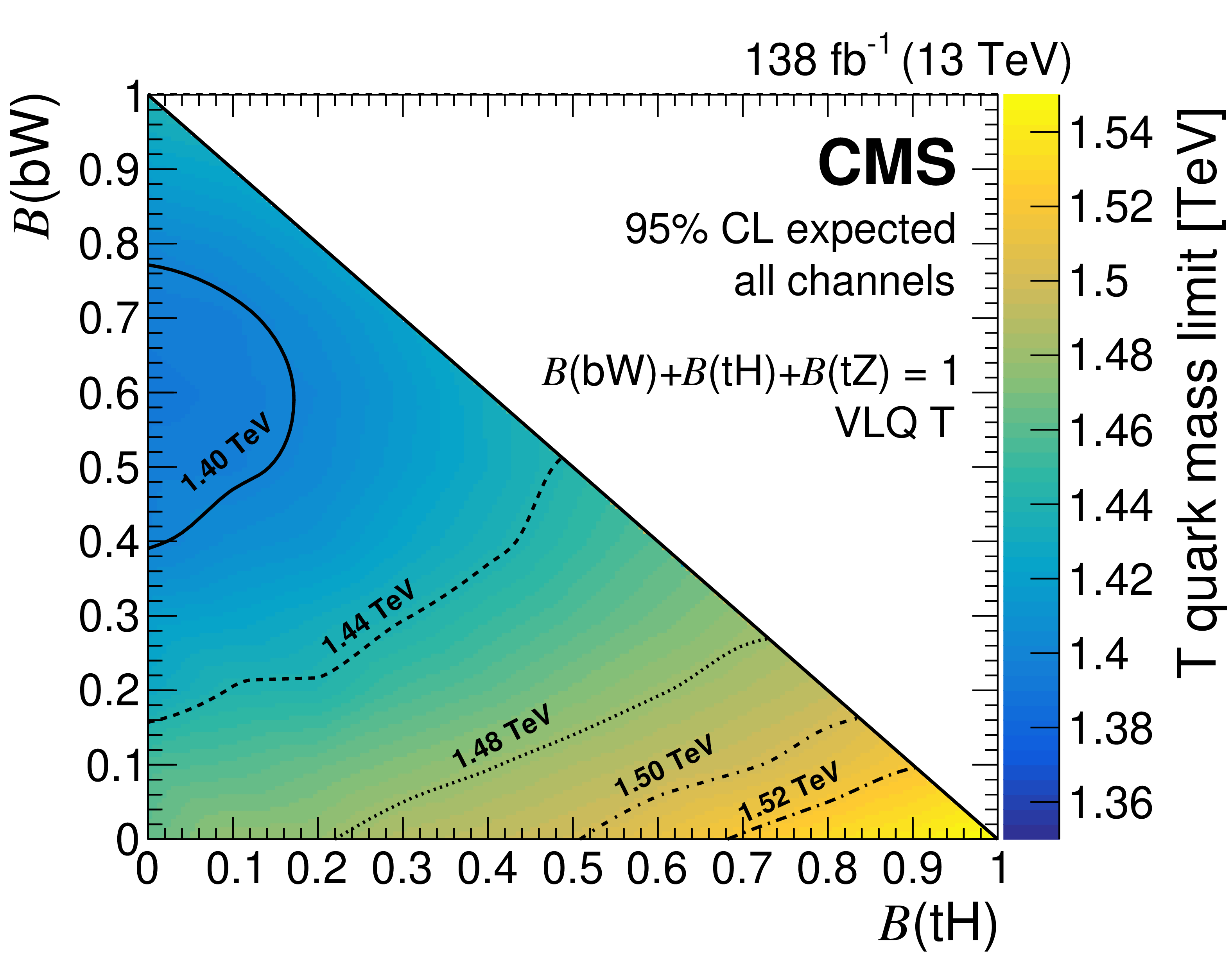}
    \caption{Top Row: LHC searches for top squarks at ATLAS~\cite{ATLAS:2024exu} and CMS~\cite{CMSstopSum}. Bottom Row: LHC searches for fermionic top partners at ATLAS~\cite{ATLAS:2024fdw} and CMS~\cite{CMS:2024bni}. Current LHC limits on stops extend up to about 1.3 TeV, depending on the lightest supersymmetric particle (LSP) mass, while bounds on fermionic top partners reach up to around 1.6 TeV, depending on their branching ratios.}
    \label{fig:LHCtopPart}
\end{figure}

Since the top quark carries color charge\textemdash charge under QCD\textemdash in typical cases the top partner $T$ should be color-charged as well (this is certainly true of the supersymmetric stops $\widetilde{t}$ in the example above.)  The Large Hadron Collider (LHC) has provided a wealth of data with which the ATLAS and CMS collaborations have sought various types of top partners, as shown in Fig.~\ref{fig:LHCtopPart}. As a hadron machine, the LHC is well equipped to discover the new colored particles predicted by many symmetry-based solutions to the hierarchy problem. The LHC sensitivity to such partners varies from model to model, but in general color-charged top partners have been excluded up to masses in the 1\textendash2 TeV range.  While the gap between the top-partner bounds and the Higgs mass is small compared to its separation from the Planck scale, it is still large enough that these models must be tuned somewhat to produce the correct Higgs mass~\cite{Barbieri:2000gf}. This is one facet of the so-called little hierarchy problem. 

\begin{figure}[ht]
\centering
\begin{tabular}{cccc}
    \begin{tabular}{c}
\begin{fmffile}{FHiggsloop}
\begin{fmfgraph*}(120,80)
\fmfpen{1.0}
\fmfstraight
\fmfleft{p1,i1,p2}\fmfright{p3,o1,p4}
\fmfv{l= $H^\dag$}{i1}\fmfv{l=$H$}{o1}
\fmf{dashes,tension=1}{i1,v1}
\fmf{fermion,left=1,tension=0.4,label.side=left,label=$t$,label.dist=5,foreground=(0.01,,0.75 ,,0.24)}{v1,v2,v1}
\fmf{dashes,tension=1}{v2,o1}
\fmfv{decor.shape=circle,decor.filled=full,decor.size=1.5thick,l=$\lambda_t$,l.a=120,l.d=7}{v1}
\fmfv{decor.shape=circle,decor.filled=full,decor.size=1.5thick,l=$\lambda_t$,l.a=60,l.d=7}{v2}
\end{fmfgraph*}
\end{fmffile}
\end{tabular}
\begin{tabular}{c}
\hspace{5mm}$+$ \hspace{5mm}
\end{tabular} 
\begin{tabular}{c}
\begin{fmffile}{Fsusystop}
\begin{fmfgraph*}(110,80)
\fmfpen{1.0}
\fmfleft{p1,i1,i2,p3} 
\fmfright{p2,o1,o2,p4}
\fmf{phantom}{p3,v2,p4}
\fmf{dashes}{i1,v1,o1}
\fmfv{l=$H^\dag$}{i1}\fmfv{l=$H$}{o1}
\fmffreeze
\fmf{double,right,tension=0.1,l.side=right,foreground=blue}{v2,v1}
\fmf{double,right,tension=0.1,l.side=right,foreground=blue}{v1,v2}
\fmfv{decor.shape=circle,decor.filled=full,decor.size=1.5thick, l=$\lambda_t^2$,l.a=-90,l.d=5}{v1}
\fmfv{l=$\widetilde{t}$,l.d=15,l.a=0}{i2} 
\end{fmfgraph*}
\end{fmffile}
\end{tabular}
\hspace{0.1cm}
\begin{tabular}{c}
$\sim\displaystyle\frac{3\lambda_t^2}{8\pi^2}m^2_{\tilde t}\ln\Lambda_\text{UV}^2$\vspace{1mm}
\end{tabular}
\end{tabular}
\vspace{-0.5cm}
\caption{\label{fig:Fsusyloops} Cancellation of quadratic sensitivity to $\Lambda_\text{UV}$ in folded SUSY. The particle loops are shaded with different colors to illustrate that the SU(3) gauge groups under which the stops are charged need not be the SU(3$)_c$ of the SM quarks.}
\end{figure}

 Beginning in 2005~\cite{Chacko:2005pe}, symmetries were found that explain these little hierarchies up to a few TeV \emph{without} colored top partners. The crucial observation, illustrated in Fig.~\ref{fig:Fsusyloops}, is that color often plays no direct role in the elimination of quadratic sensitivity to $\Lambda_\text{UV}$.  Thus, there are ways that symmetries can reduce sensitivity to UV physics without top partners charged under SM color. Because the symmetry partners of the quarks, especially the top, are color neutral, this paradigm is referred to as {\it neutral naturalness}.\footnote{This term was coined in Ref.~\cite{Craig:2014aea}.} Direct production of colorless top partners at a hadron collider is typically challenging, so the limits on such particles are weaker. Or, said another way, the explanation of why symmetry partners of SM quarks have not yet been discovered may be that they are color neutral.

Because the color-neutral top-partner masses can be lower, the tuning in the frameworks that describe them is reduced. In other words, such models are more natural, but the top partners may be difficult to produce experimentally. However, the color neutral structure can only explain a little hierarchy, typically requiring a UV completion with new colored states at the few TeV scale. Consequently, direct experimental probes of the naturalness of the Higgs sector may require even higher-energy colliders than the LHC~\cite{Curtin:2015bka} and remain a motivating target for the next generation of machines and detectors~\cite{Alekhin:2015byh,Fujii:2017vwa,Curtin:2018mvb,CEPCStudyGroup:2018ghi,CLIC:2018fvx,FCC:2018byv,AlAli:2021let,Black:2022cth,Adhikary:2024nlv,MammenAbraham:2024gun,PBC:2025sny,FCC:2025lpp,Abidi:2025dfw}. 

All known realizations of neutral naturalness feature a discrete symmetry that relates SM fields to colorless counterparts. Though these partner fields are neutral under SM color, the quark partners are typically charged under an SU(3) gauge group that is related to SM color by the discrete symmetry. Consequently, neutral naturalness motivates certain classes of rich hidden sectors, with a variety of particles and forces. What is more, the rich structure of these hidden sectors is related (at least in part) to the SM by the discrete symmetry. This leads to correlations between precision SM measurements and dark sector structure.

The Higgs always links the SM to these new sectors. The colorless partners must couple to the Higgs with SM field-like strength in order to protect the Higgs mass. This ensures the Higgs sector acts as a portal between the SM and the new sector. However, the manifestations of this connection can vary from one concrete model to another, as do other phenomenological consequences. The purpose of our review is to map this rich diversity so that more researchers can more easily investigate and extend them.

This review is organized by topic so that readers may easily navigate to the aspects of neutral naturalness they are most interested in. This leads to some redundancy, as many works are relevant to multiple sections. We begin, in Sec.~\ref{s.models}, with an overview of the various specific realizations of neutral naturalness. The twin Higgs scenario is emphasized because it provides a simple, concrete realization of many of the qualities that arise in other models. In addition, there is a growing body of modifications that have been made to its basic structure and the several UV completions that have been suggested. We then broadly classify models as being related to SUSY or to pNGB Higgs constructions. 

The following section,~\ref{s.collider}, considers the various collider signals that have been associated with these neutral naturalness scenarios. In this section the organization is by experimental signature rather than by model. Particular mention is made of hidden sector glueballs as their characteristics are unfamiliar to many. 

Similarly, the less familiar concept of quirks~\cite{Kang:2008ea}, which arise in many neutral naturalness models, is introduced in some detail in Sec.~\ref{ss.quirks}. The dynamics of these particles are reviewed because they are quite distinct from the SM particles that guide much collider intuition. We then, in Sec.~\ref{s.lattice}, discuss results from lattice gauge theory that have been important to understanding the phenomenology of these models, including providing essential input to our understanding of quirk and glueball dynamics. As yet unknown quantities that can improve the predictive power of phenomenological studies are also mentioned. 

The rich dark sectors of neutral naturalness are connected to the cosmological dark matter in Sec.~\ref{s.DM}. Here again the twin Higgs scenario has been most thoroughly investigated, though the possibilities related to other frameworks are intriguing. Other connections to astrophysics and cosmology, including gravitational waves, baryogenesis, and the Hubble tension, are discussed in Sec.~\ref{s.AstCosmo}. Intersections with neutrinos and flavor physics are outlined in Secs.~\ref{s.Neutrinos} and \ref{s.Flavor}, respectively. These relationships are less explored, but offer interesting correlations between precision SM measurements and hidden-sector observables. We conclude, in Sec.~\ref{s.Sum}, with some summarizing thoughts and discussion of interesting directions for future research.

\section{Models\label{s.models}}
The growing literature of concrete realizations of the neutral naturalness paradigm is already expansive, prohibiting a detailed explanation of each one. Therefore, we begin with a discussion of commonalities among many, if not all, models. We then make thorough investigation of the mirror twin Higgs framework, which provides a concrete realization of many of these common traits. The following sections introduce the wider variety of neutral naturalness realizations. However, we do not explain each of these in the same detail. 

\subsection{Recurring Themes\label{ss.themes}}
The broad ideas of neutral naturalness have been realized in several specific ways. Many of them center on the twin Higgs framework. However, other familiar symmetry-based solutions to the hierarchy problem, such as supersymmetry and pNGB Higgs models, also have many neutral-natural realizations.

A few characteristics are common to most concrete manifestations of neutral naturalness.  As discussed above, the schematic goal is to realize a symmetry relating the top quark $t$ to one or more top partners $T$ of the form
\beq
t\Rightarrow \left.\left(\begin{array}{c}
t \\
T
\end{array}\right)\right\downarrow\hspace{-1.0mm}\text{Symmetry}
\eeq
in such a way that $T$ does \emph{not} carry SM color charge.  In practical terms, this means that the symmetry should not commute with the SU$(3)_{A}$ color gauge group of the SM.  One simple way to realize this is to posit that there is a ``twin" color group SU$(3)_{B}$ under which $T$ is charged, together with a discrete $\mathbb{Z}_2$ symmetry that exchanges the two color groups:
\beq
\overset{G \supset \text{SU}(3)_A\times \text{SU}(3)_B\hspace{5mm}}
{\left.{\overrightarrow{\hspace{-3.5mm}
\left(\begin{array}{cc}
t & \cdots  \\
\cdots & T
\end{array}\right)\hspace{-3.5mm}}}\hspace{3.5mm}\right\downarrow}\hspace{-2.5mm}\text{Symmetry}~.
\eeq
This simple structure can be embedded within larger continuous symmetries, with additional particles; examples of this appear in some of the explicit realizations below. 

\begin{figure}[ht]
\centering
\includegraphics[width=\textwidth]{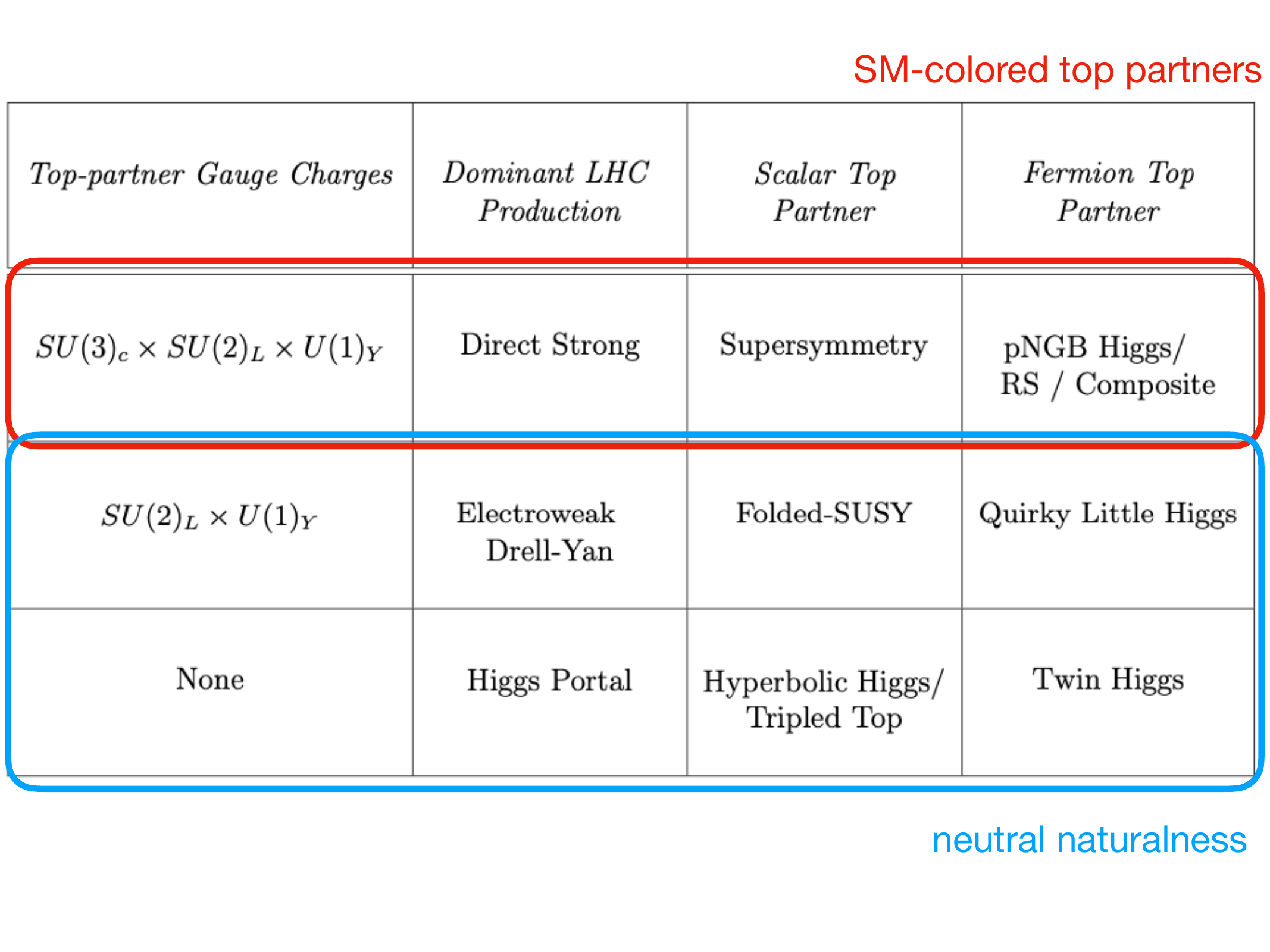}
\vspace{-1cm}
    \caption{Inspired by~\cite{Curtin:2015fna}, an illustration of the types of symmetry based solutions to the electroweak hierarchy problem, how the top partners in each might be produced at a hadron collider and some early examples of each model type.  In the first row, the top partners carry SM color charge and are thus copiously produced at the LHC in direct strong interactions.}
    \label{tab:NN_model_space}
\end{figure}

This schematic treatment is overly simplified in several ways from the more general case, although the presence of multiple sectors related by some discrete symmetry is a common starting point for neutral naturalness models. In many cases, but not all, the number of sectors is two and the symmetry is $\mathbb{Z}_2$. The sectors may differ from each other in significant ways, but aspects of the third generation, more specifically the top quark sector, are most strictly governed by the discrete symmetry. This ensures that at least some particles in the BSM sectors have top quark sized coupling to the Higgs. Therefore, the Higgs is an essential portal between the sectors; there may be others depending on the model.  An overview of the various types of neutral naturalness models with different types of top partners is shown in Fig.~\ref{tab:NN_model_space}. 

The top partners, and often other fields, in the hidden sector are typically charged under a QCD-like gauge group which is related to SM QCD by the discrete symmetry or a larger symmetry structure. However, the spectrum of particles can be quite different in the hidden sector. Considerations from cosmology, like the bounds on the effective number of light neutrinos $\Delta N_\text{eff}$~\cite{Planck:2018vyg}, and collider searches, such as the LEP bound on electrically charged particles~\cite{Egana-Ugrinovic:2018roi}, often motivate scenarios with fewer light states. This can produce a hidden confinement scale that is significantly higher than the SM equivalent and a different spectrum of bound states.

Most of the models discussed below are only effective theories up to the scale of a few TeV. However, several UV completions have been considered for twin Higgs models, as discussed in Sec.~\ref{ss.UVcomp}. Specific composite Higgs completions of pNGB Higgs constructions have also been developed, see Sec.~\ref{ss.pNGB}. In other cases definite UV completions have not been specified. However, new colored states (and a variety of other signals) are generically expected at the few TeV scale, which offers an intriguing target for the next generation of high energy colliders. 

A primary result of this section is that neutral naturalness models motivate rich hidden sectors and, as discussed in Sec.~\ref{s.DM}, provide many possible dark matter candidates. The discrete symmetry typically relates the dark matter to known SM fields and couplings. Consequently, the dark matter signals can be connected to the naturalness of the Higgs. 

Beyond dark matter, these hidden sectors present a structure at least as rich and varied as the SM. While the hidden sectors are neutral under SM color, they may share other gauge interactions. Therefore, searches for new electroweak states are highly motivated by these constructions. The experimental searches relevant to these models are discussed in Sec.~\ref{s.collider}, and are organized by signature.

\subsection{The Twin Higgs and its Variants\label{ss.twinHiggsVar}}
Neutral naturalness began\textemdash although the name did not yet exist\textemdash with the mirror twin Higgs~\cite{Chacko:2005pe} model, which also exemplifies many of the characteristics that are common to most constructions.\footnote{There are twin Higgs models that do not exhibit neutral naturalness; the top partners carry color~\cite{Chacko:2005un,Goh:2007dh,Batra:2008jy,Yu:2016cdr}. The same is true of other models that are similar in construction to many neutral naturalness models~\cite{Kats:2017ojr}.} To the SM with its $\left[\text{SU}(3)_c\times \text{SU}(2)_L\times \text{U}(1)_Y\right]_A$ gauge structure is added a twin sector with identical structure, that is, a gauge structure of $\left[\text{SU}(3)_c\times \text{SU}(2)_L\times \text{U}(1)_Y\right]_B$. The twin character of this sector is manifest in a discrete $\mathbb{Z}_2$ symmetry that relates the couplings of the two sectors. In addition to this mirror set-up, the Higgs sectors of the two copies are assumed to satisfy an approximate global symmetry. For instance, one can arrange the visible and twin sector Higgs doublets, $H_A$ and $H_B$ (respectively), in terms of a four-plet of a global SU(4)
\beq
\mathcal{H}=\left(\begin{array}{c}
H_A\\
H_B
\end{array} \right)~.
\eeq
If this approximate symmetry is spontaneously broken down to SU(3) at some scale $f$ then seven pNGBs are produced.\footnote{We have used the ${\rm SU}(4)\to{\rm SU}(3)$ breaking pattern for simplicity. However, it is preferable in strongly-coupled UV completions (see Sec.~\ref{ss.pNGB}) to use ${\rm SO}(8)\to{\rm SO}(7)$ (which also produces 7 pNGBs) as this preserves the custodial SU(2) symmetry.} Because of the SU(2$)_L\times$U(1$)_Y$ gauging in each sector, six of the pNGBs are eaten by electroweak gauge fields. This leaves one physical pNGB that is associated with the observed Higgs boson. As a pNGB, the Higgs mass is controlled by the approximate global symmetry of the potential. 

The SM fermions and their twin counterparts do not fill out a complete representation of SU(4), so one might worry that Yukawa interactions $\lambda_f\overline{f}f H$ would break the symmetry and produce large one-loop corrections to the scalar potential.  However, the discrete $\mathbb{Z}_2$ symmetry that relates the SM fermions to the twin fermions ensures that the quadratically divergent contributions to the Higgs mass from the SM and twin fermions cancel. In particular, the $\mathbb{Z}_2$ exchanges the gauged SU(2)$_{LA}\times$SU(2)$_{LB}$ factors within the global SU(4) and the SU(3)$_{cA}\times$SU(3)$_{cB}$ factors at the same time. In effect, the SM and twin fermions interact with the scalar sector as part of incomplete  SU(4) multiplets. The simplest embedding for the SU(2)$_L$ doublet quarks  takes the form
\beq
\overset{\text{SU}(3)_A\times \text{SU}(3)_B\hspace{5mm}}
{\left.{\overrightarrow{\hspace{-3.5mm}
\left(\begin{array}{cc}
q_{LA} &  \\
 & q_{LB}
\end{array}\right)\hspace{-3.5mm}}}\hspace{3.5mm}\right\downarrow}\hspace{-1mm}\text{SU(4)}~,\label{eq:MTHtops}
\eeq
while the SU(2)$_L$ singlet quarks $t_{RA}$, $t_{RB}$ form a singlet under SU(4). Here the discrete symmetry exchanges $q_{LA}\leftrightarrow q_{LB}$ and $t_{RA}\leftrightarrow t_{RB}$ as it exchanges $H_A\leftrightarrow H_B$ within the Higgs multiplet $\mathcal{H}$. So, while Eq.~\eqref{eq:MTHtops} illustrates that the quark fields are  not part of complete SU(4) multiplets, the discrete symmetry ensures that their largest contributions to the Higgs mass parameter are SU(4) symmetric.\footnote{It is possible to embed the visible and twin fermions in complete multiplets of the global SU(4) which ensures that the fermion contributions to the Higgs potential, including to the quartic, are finite at one-loop~\cite{Chacko:2005pe}.} 

\begin{figure}[ht]
\centering
\begin{tabular}{cc}
\begin{tabular}{c}
    \begin{fmffile}{twinHiggsloopA}
\begin{fmfgraph*}(120,80)
\fmfpen{1.0}
\fmfstraight
\fmfleft{p1,i1,p2}\fmfright{p3,o1,p4}
\fmfv{l= $H_A^\dag$}{i1}\fmfv{l=$H_A$}{o1}
\fmf{dashes,tension=1}{i1,v1}
\fmf{fermion,left=1,tension=0.4,label.side=left,label=$t_{RA}$,label.dist=5,foreground=(0.01,,0.75 ,,0.24)}{v1,v2}
\fmf{fermion,left=1,tension=0.4,label.side=left,label=$q_{LA}$,label.dist=5,foreground=(0.01,,0.75 ,,0.24)}{v2,v1}
\fmf{dashes,tension=1}{v2,o1}
\fmfv{decor.shape=circle,decor.filled=full,decor.size=1.5thick,l=$\lambda_{t_A}$,l.a=120,l.d=7}{v1}
\fmfv{decor.shape=circle,decor.filled=full,decor.size=1.5thick,l=$\lambda_{t_A}$,l.a=60,l.d=7}{v2}
\end{fmfgraph*}
\end{fmffile}
\end{tabular}\hspace{1.2cm}
\begin{tabular}{c}
    \begin{fmffile}{twinHiggsloopB}
\begin{fmfgraph*}(120,80)
\fmfpen{1.0}
\fmfstraight
\fmfleft{p1,i1,p2}\fmfright{p3,o1,p4}
\fmfv{l= $H_B^\dag$}{i1}\fmfv{l=$H_B$}{o1}
\fmf{dashes,tension=1}{i1,v1}
\fmf{fermion,left=1,tension=0.4,label.side=left,label=$t_{RB}$,label.dist=5,foreground=blue}{v1,v2}
\fmf{fermion,left=1,tension=0.4,label.side=left,label=$q_{LB}$,label.dist=5,foreground=blue}{v2,v1}
\fmf{dashes,tension=1}{v2,o1}
\fmfv{decor.shape=circle,decor.filled=full,decor.size=1.5thick,l=$\lambda_{t_B}$,l.a=120,l.d=7}{v1}
\fmfv{decor.shape=circle,decor.filled=full,decor.size=1.5thick,l=$\lambda_{t_B}$,l.a=60,l.d=7}{v2}
\end{fmfgraph*}
\end{fmffile}
\end{tabular}
\end{tabular}
\vspace{-0.5cm}
    \caption{Schematic contribution to Higgs mass from top quark loops in the visible ($A$) and twin ($B$) sectors. The different colored quark lines indicate that they are charged under distinct SU(3) gauge groups.}
    \label{fig:twinHiggsLoops}
\end{figure}

The consequences of this symmetry structure may be understood schematically as follows. The dominant one-loop contributions to the $\left| H_A\right|^2$ and $\left| H_B\right|^2$ operators\textemdash shown in Fig.~\ref{fig:twinHiggsLoops}\textemdash are from the visible-sector and twin-sector top quarks, $t_A$ and $t_B$, respectively. These contributions are 
\beq
\frac{3\lambda_{t_A}^2}{8\pi^2}\Lambda^2_\text{UV}\left|H_A\right|^2~, \ \ \ \text{and} \ \ \ \ \frac{3\lambda_{t_B}^2}{8\pi^2}\Lambda^2_\text{UV}\left|H_B\right|^2~,
\eeq
where $\Lambda_\text{UV}$ is the UV cutoff of the effective theory, typically taken to be a few TeV. The $\mathbb{Z}_2$ symmetry between the sectors ensures that $\lambda_{t_A}=\lambda_{t_B}\equiv\lambda_t$. This means that the sum of these two contributions is
\beq
\frac{3\lambda_t^2}{8\pi^2}\Lambda_\text{UV}^2\left(\left|H_A\right|^2+\left|H_B\right|^2 \right)=\frac{3\lambda_t^2}{8\pi^2}\Lambda_\text{UV}^2\left|\mathcal{H}\right|^2~.\label{eq:HiggsQuad}
\eeq
This result is SU(4) symmetric and, therefore, cannot contribute to the mass of any pNGB of the broken SU(4) symmetry. It is important to note that this one-loop effect does not prevent the NGBs from getting a mass. The SU(2$)_L$ gauge couplings and the Yukawa coupling to fermions break the global SU(4) and generate a $\delta(|H_A|^4+|H_B|^4)$ potential term. This gives the Higgs a $m_h\sim f\sqrt{\delta}$ mass where $\delta\sim\ln\Lambda_\text{UV}/f$ depends logarithmically on the cutoff, as shown below.

Note that in fully $\mathbb{Z}_2$ symmetric theory all contributions\textemdash from any loop order\textemdash to the Higgs potential that are quadratic in $\Lambda_\text{UV}$ must have a similar form to Eq.~\eqref{eq:HiggsQuad}. This ensures that they do not contribute to the mass of the pNGB Higgs boson. However, an exactly mirror twin sector is excluded by experimental measurements of the Higgs couplings, as further discussed below, requiring the $\mathbb{Z}_2$ symmetry be softly broken. This leads, for instance, the Yukawa couplings in the two sectors to run differently, so two-loop and higher contributions to the Higgs mass need not be symmetric under the approximate SU(4) global symmetry.

In short, in phenomenologically viable models the one-loop contributions to the Higgs mass do take the form of an SU(4) symmetric contribution, but this does not persist to higher loops. This is one reason why the mirror twin Higgs framework does not address the full gauge hierarchy problem and requires a UV completion at the few TeV scale. Several types of completions have been constructed and are discussed in Sec.~\ref{ss.UVcomp}.

In general, we can write the tree-level scalar potential as~\cite{Barbieri:2005ri}
\begin{align}
V_0=&-M^2\left|\mathcal{H}\right|^2+\lambda\left| \mathcal{H}\right|^4\nonumber\\
&+m^2\left(\left|H_A\right|^2-\left|H_B\right|^2 \right)+\delta\left(\left|H_A\right|^4+\left|H_B\right|^4 \right)~,\label{e.Hpot}
\end{align}
with the first line preserving the global SU(4), while the second line does not. We also note that the potential is $\mathbb{Z}_2$ symmetric, except for the $m^2$ term, which constitutes a soft breaking of the symmetry. In writing this potential the SU(4) symmetric, scalar mass term is given without further explanation. The origin of this scale, as in other pNGB Higgs models, depends on the UV completion. For the Higgs to be a pNGB the parameters of the second line must be small compared to those of the first. To understand this set-up in the most general way, we use a nonlinear parametrization of the pNGB Higgs fields
\beq
H_A=\bm{h}\frac{f}{\sqrt{\bm{h}^\dag\bm{h}}}\sin\left( \frac{\sqrt{\bm{h}^\dag\bm{h}}}{f}\right)~, \ \ \ \ H_B=\left(\begin{array}{c}
    0  \\
    \displaystyle f\cos\left( \frac{\sqrt{\bm{h}^\dag\bm{h}}}{f}\right)
\end{array} \right)~.
\eeq
Here $\bm{h}$ is a complex doublet field with VEV given by
\beq
\langle\bm{h}\rangle=\frac{1}{\sqrt{2}}\left(
\begin{array}{c}
0\\
v
\end{array}\right)~.
\eeq
The physical Higgs boson field $h$ is the fluctuations about this VEV in unitary gauge
\beq
\bm{h}=\frac{1}{\sqrt{2}}\left(
\begin{array}{c}
0\\
v+h
\end{array}\right)~,
\eeq
similar to the SM Higgs. Note that the VEVs of $H_A$ and $H_B$ are given by
\beq
\langle H_A\rangle=\left( \begin{array}{c}
0\\
f\sin\frac{v}{f\sqrt{2}}
\end{array}\right) ~, \ \ \ \ \langle H_B\rangle=\left( \begin{array}{c}
0\\
f\cos\frac{v}{f\sqrt{2}}
\end{array}\right) ~,
\eeq
which motivates the notation $\vartheta\equiv v/(f\sqrt{2})$. This angle describes how much of the SU(4) breaking VEV $f$ ends up in each sector's Higgs field, with $\vartheta=\pi/4$ indicating that the two VEVs are equal. 

Using these definitions the tree-level scalar potential is
\begin{align}
V_0=&-\frac{\delta f^4}{2}\sin^2\left( \frac{2\sqrt{\bm{h}^\dag\bm{h}}}{f}\right)-m^2f^2\cos\left( \frac{2\sqrt{\bm{h}^\dag\bm{h}}}{f}\right)~,\label{eq:THtreeV}
\end{align}
where we have dropped constant terms. We see immediately that the potential for $\bm{h}$ does not depend on the SU(4) symmetric parameters $M^2$ and $\lambda$. The Yukawa terms of the $A$ and $B$ sector that connect the Higgs fields to the quarks lead to the top quark couplings
\beq
\lambda_t\frac{f}{\sqrt{\bm{h}^\dag\bm{h}}}\sin\left( \frac{\sqrt{\bm{h}^\dag\bm{h}}}{f}\right)\overline{q}_{LA}\widetilde{\bm{h}}t_{RA}+\lambda_tf\cos\left( \frac{\sqrt{\bm{h}^\dag\bm{h}}}{f}\right)\overline{t}_{LB}t_{RB}+\text{ H.c.}~,\label{eq:THtop}
\eeq
where $\widetilde{\bm{h}}^i=\epsilon^{ij}\bm{h}^\dag_j$, similar to the usual Higgs coupling to up-type quarks. By expanding in the field $\bm{h}$, we find the couplings  with which we can calculate the one-loop contributions to the $\bm{h}$ propagator, as shown in Fig.~\ref{fig:twntoploop}. One finds that the quadratic divergences cancel exactly, as expected from the above schematic analysis of one-loop effects. 

\begin{figure}[ht]
  \vspace{5mm}
  \centering
\begin{tabular}{ccc}
    \begin{tabular}{c}
\begin{fmffile}{twnhiggs1}
\begin{fmfgraph*}(120,70)
\fmfpen{1.0}
\fmfleft{p1,i1,p2} \fmfright{p3,o1,p4}
\fmfv{l=$\bm{h}$}{i1}\fmfv{l=$\bm{h}$}{o1}
\fmf{dashes}{i1,v1}\fmf{dashes}{v2,o1} 
\fmf{fermion,right=1,tension=0.3,l.side=left,foreground=(0.01,,0.75 ,,0.24)}{v1,v2,v1}
\fmfv{decor.shape=circle,decor.filled=full,decor.size=1.5thick,l=$\lambda_t$,l.a=115,l.d=2}{v1} \fmfv{decor.shape=circle,decor.filled=full,decor.size=1.5thick,l=$\lambda_t$,l.a=65,l.d=2}{v2}
\fmfv{l=$q_{LA}$,l.a=0,l.d=40}{p1}
\fmfv{l=$t_{RA}$,l.a=0,l.d=40}{p2}
\end{fmfgraph*}
\end{fmffile}
\end{tabular}
\begin{tabular}{c}
\hspace{5mm}$+$ \hspace{5mm}
\end{tabular} 
\begin{tabular}{c}
\begin{fmffile}{twnhiggs2}
\begin{fmfgraph*}(110,80)
\fmfpen{1.0}
\fmfleft{p1,i1,i2,p3} 
\fmfright{p2,o1,o2,p4}
\fmf{phantom}{p3,v2,p4}
\fmf{dashes}{i1,v1,o1}
\fmfv{l=$\bm{h}$}{i1}\fmfv{l=$\bm{h}$}{o1}
\fmffreeze
\fmf{fermion,right,tension=0.1,l.side=right,foreground=blue}{v2,v1}
\fmf{fermion,right,tension=0.1,l.side=right,foreground=blue}{v1,v2}
\fmfv{decor.shape=circle,decor.filled=full,decor.size=1.5thick, l=$-\lambda_t/(2f)$,l.a=-90,l.d=5}{v1}
\fmfv{decor.shape=cross,l=$\lambda_t f$,l.a=90}{v2}
\fmfv{l=$t_{LB}$,l.d=7,l.a=0}{i2}
\fmfv{l=$t_{RB}$,l.d=6,l.a=180}{o2}
\end{fmfgraph*}
\end{fmffile}
\end{tabular}
\end{tabular}
\caption{\label{fig:twntoploop} Cancellation of quadratic divergences in the mirror twin Higgs model. The different colors emphasize that this pNGB Higgs type of cancellation occurs even when the top quark and its partner are charged under different SU(3)s.}
 \end{figure}

It is also instructive to take the couplings in~\eqref{eq:THtop} and expand $\bm{h}$ about its VEV. This leads to
\begin{align}
&\lambda_tf\overline{t}_{LA}t_{RA}\sin\frac{v+h}{f\sqrt{2}}+\lambda_tf\overline{t}_{LB}t_{RB}\cos\frac{v+h}{f\sqrt{2}}+\text{ H.c.}=\nonumber\\
&\lambda_tf\overline{t}_{LA}t_{RA}\left[\sin\vartheta+\frac{h}{f\sqrt{2}}\cos\vartheta+\ldots \right]+\lambda_tf\overline{t}_{LB}t_{RB}\left[\cos\vartheta- \frac{h}{f\sqrt{2}}\sin\vartheta+\ldots\right]+\text{ H.c.}~,
\end{align}
and similar expansions also apply to all of the fermion fields. This result illustrates a few general points. First, the ratio of the twin sector fermion mass to the visible sector fermion mass is $\cot\vartheta$. This gets larger as $\vartheta$ gets smaller. Second, the coupling of the Higgs boson $h$ to the visible sector fields is $\cos\vartheta$ times the SM prediction. In the small $\vartheta$ limit it approaches the SM value. Finally, the twin sector fermions coupling to the Higgs boson is $\sin\vartheta$ times the SM value. In this case the twin sector fields decouple from the Higgs boson as $\vartheta\to0$.

These results demonstrate how $\vartheta$ determines much of the Higgs sector. Precise measurements of the Higgs couplings to SM fields have the potential to indicate the mass of the twin sector fermions. Or turned around, the bounds on Higgs coupling deviations indicates a hierarchy between twin fermions and their visible counterparts. Such a hierarchy in the top sector leads to larger one-loop contributions to the Higgs mass, and hence a greater need for tuning of parameters to produce the observed Higgs mass.

To explicitly track how $\vartheta$ relates to the other parameters of the theory we inspect the scalar potential. Using the result of Coleman and Weinberg~\cite{Coleman:1973jx}
\beq
V_\text{CW}=-\frac{N_c}{8\pi^2}\Lambda_\text{UV}^2\text{Tr}
\mathcal{M}^2-\frac{N_c}{16\pi^2}\text{Tr}\left[\mathcal{M}^4\left(\ln\frac{\mathcal{M}^2}{\Lambda_\text{UV}^2}-\frac12 \right) \right]~,
\eeq
we include both tree-level\textemdash$V_0$ from \eqref{eq:THtreeV}\textemdash and one-loop contributions to the Higgs potential. We first note that the fermion mass matrix in the top quark sector provides the largest contribution to the Coleman-Weinberg (CW) potential. The top-sector mass matrix is given by  
\beq
\left(\overline{t}_{LA},\,\overline{t}_{LB} \right)\mathcal{M}\left(\begin{array}{c}
    t_{RA} \\
    t_{RB}
\end{array} \right)=\left(\overline{t}_{LA},\,\overline{t}_{LB} \right)\left(\begin{array}{cc}
   \lambda_t H_A   & 0 \\
    0 & \lambda_t H_B   
\end{array} \right)\left(\begin{array}{c}
    t_{RA} \\
    t_{RB}
\end{array} \right)~.
\eeq
The $\Lambda_\text{UV}^2$ term of the CW potential is independent of $h$ because
\beq
\text{Tr}\mathcal{M}^\dag\mathcal{M} =\lambda_t^2f^2\text{Tr}\left(\begin{array}{cc}
   \sin^2\left( \frac{\sqrt{\bm{h}^\dag\bm{h}}}{f}\right)  & 0 \\
   0  & \cos^2\left( \frac{\sqrt{\bm{h}^\dag\bm{h}}}{f}\right)
\end{array} \right)=\lambda_t^2f^2=m_{tA}^2+m_{tB}^2~.
\eeq
This provides another view on how the twin Higgs framework (and pNGB Higgs models more generally) ensure that the Higgs is not quadratically sensitive to high scales. 

Dropping this constant term, the potential, including the remaining term in the CW potential, can be written as
\begin{align}
V=&V_0+V_\text{CW}=-\frac{\delta f^4}{2}\sin^2\frac{2x}{f}-m^2f^2\cos\frac{2x}{f}\\
&-\frac{3\lambda_t^4f^4}{16\pi^2}\left[\sin^4\left( \frac{x}{f}\right)\left(\ln\frac{\lambda_t^2f^2\sin^2\left( \frac{x}{f}\right)}{\Lambda_\text{UV}^2} -\frac12\right)+\cos^4\left( \frac{x}{f}\right)\left(\ln\frac{\lambda_t^2f^2\cos^2\left( \frac{x}{f}\right)}{\Lambda_\text{UV}^2} -\frac12\right) \right]~,\nonumber
\end{align}
where $x=\sqrt{\bm{h}^\dag\bm{h}}$. We then use this potential to determine relations between parameters. For instance, the condition that there be a vacuum is
\beq
V'(v/\sqrt{2})=0~,
\eeq
which leads to
\beq
m^2=\delta f^2\cos2\vartheta-\frac{3\lambda_t^4f^2}{16\pi^2}\left[\cos2\vartheta\ln\frac{\lambda_t^2f^2\sin2\vartheta}{2\Lambda_\text{UV}^2}+\ln\cot\vartheta \right]~.\label{eq:THvacCond}
\eeq
In order to compare to the SM Higgs potential
\beq
V_H=-\mu_\text{SM}^2H^\dag H+\lambda_\text{SM}\left(H^\dag H \right)^2~,
\eeq
we define the Higgs mass parameter $\mu^2$ by
\beq
\mu^2=-\frac14 V''(0)-\frac14\lim_{x\to0}\frac{V'(x)}{x}~.\label{e.muDef}
\eeq
This leads to
\beq
\mu^2=2\left(\delta f^2-m^2 \right)+\frac{3\lambda_t^4f^2}{8\pi^2}\ln\frac{\Lambda_\text{UV}^2}{\lambda_t^2f^2}~.\label{eq:mu2TH}
\eeq
While this quantity is not quadratically sensitive to $\Lambda_\text{UV}$, it does exhibit a logarithmic sensitivity to high scale physics. 

Finally, the Higgs boson mass is given by
\beq
m_h^2=\frac12V''(v/\sqrt{2})~.\label{e.mhDef}
\eeq
Note that the Higgs mass being positive is equivalent to ensuring the potential extremum is a minimum. After we enforce the vacuum condition in Eq.~\eqref{eq:THvacCond} we find
\beq
m_h^2=2f^2\sin^22\vartheta\left[\delta-\frac{3\lambda_t^4}{16\pi^2}\left(1+\ln\frac{\lambda_t^2f^2\sin2\vartheta}{2\Lambda_\text{UV}^2} \right)\right] ~.\label{eq:THhiggsMass}
\eeq
This illustrates that the effective quartic coupling in the one-loop potential
\beq
\delta_\text{eff}=\delta-\frac{3\lambda_t^4}{16\pi^2}\left(1+\ln\frac{\lambda_t^2f^2\sin2\vartheta}{2\Lambda_\text{UV}^2} \right)~,
\eeq
has a mild, but nonzero, dependence on high scales through fermion loops. We can also use Eq.~\eqref{eq:THhiggsMass} solve for the tree-level quartic as
\beq
\delta=\frac{m_h^2}{2f^2\sin^2(2\vartheta)}+\frac{3\lambda_t^4}{16\pi^2}\left(1+\ln \frac{f^2\lambda_t^2\sin(2\vartheta)}{2\Lambda_\text{UV}^2} \right)~,\label{eq:THdelH}
\eeq
Similarly, we can write
\beq
m^2=\frac{m_h^2\cos2\vartheta}{2\sin^22\vartheta}+\frac{3\lambda_t^4f^2}{16\pi^2}\left(\cos2\vartheta+\ln\tan\vartheta \right)~,\label{eq:THm2}
\eeq
which is completely independent of $\Lambda_\text{UV}$.

Taken in sum, we see that the small tree-level couplings are stable at one-loop due to the discrete twin symmetry. The contributions to the symmetry breaking potential terms are at most logarithmically sensitive to $\Lambda_\text{UV}$. This scenario can be extended in a straightforward way to include not just a single twin sector, but many copies of the SM~\cite{Foot:2006ru}.

As mentioned above, when the $\mathbb{Z}_2$ symmetry is preserved both $H_A$ and $H_B$ obtain equal VEVs. This implies that
\beq
\sin\vartheta=\cos\vartheta~,
\eeq
or $\vartheta=\pi/4$. Equation~\eqref{eq:THm2} shows that this can only occur when soft symmetry breaking parameter $m^2$ vanishes. In this same limit Eq.~\eqref{eq:THdelH} show that $\delta$ is nonzero. For this value of $\vartheta$ the masses of the visible and twin sector fermions are equal and each couple to the Higgs with equal strength: half the SM value. In fact, though we have not shown it, the tree-level Higgs boson couplings to gauge bosons are reduced by the same amount. This means that $\mathbb{Z}_2$ symmetric version of the twin Higgs is in contradiction with LHC measurements of Higgs couplings~\cite{Burdman:2014zta,Ortona:2023ypp}.

However, when $m^2\neq0$ the twin symmetry is softly broken, allowing $\langle H_B\rangle>\langle H_A\rangle$. The resulting Higgs couplings are closer to the SM predictions while preserving the one-loop protection of the Higgs mass. However, choosing $m^2\neq0$ to give the correct Higgs VEV does constitute a tuning of the model.

We can calculate the tuning of $\mu^2$ according to the Barbieri-Giudice formula given in Eq.~\eqref{eq:BGtuning}. This is done by using Eq.~\eqref{eq:mu2TH} for $\mu^2$, that is the definition before any relations among the parameters from the vacuum constrain or Higgs mass are applied. Using this formula we take derivatives with respect to various parameters. The maximum of all of these is the taken to be the tuning. To leading order in $f/v$ this tuning is
\begin{align}
\Delta_{m^2}=&\frac{f^2}{v^2}-\frac{3\lambda_t^4f^2}{4\pi^2m_h^2}\ln\frac{f}{v}\ldots \\
=&\frac{m_{tB}^2}{2m_{tA}^2}-\frac{3\lambda_t^2(m_{tB}^2+m_{tA}^2)}{4\pi^2m_h^2}\ln\frac{m_{tB}}{m_{tA}\sqrt{2}}+\ldots~.
\end{align}
We have rewritten the result in terms of the top quark masses in the second line. 

This form of the tuning shows two contributions. The first is from the tree-level potential and goes like $f^2/v^2$. This tuning is common to many pNGB Higgs models. The second is from top quark loops from both sectors. This term comes with a loop factor, which reduces its size somewhat. It also includes a negative sign, which often reduces the tuning, until the logarithmic terms grow too large.

Barbieri and Giudice take $\Delta=10$ as an indication of significant tuning~\cite{Barbieri:1987fn}. This is equivalent to saying that a parameter is tuned at level of 10\%. The formula above indicates that for the twin Higgs to be tuned at this level $m_{tB}$ is about 1 TeV. This is equivalent to $\cos\vartheta\approx0.985$, meaning that the Higgs couplings are very close to the SM value. As these couplings are often the leading collider signal of the twin Higgs framework we see that the mirror twin Higgs model provides a natural explanation of the Higgs mass. This tuning can even be reduced in various ways, for instance, by inducing electroweak symmetry breaking through the tadpole of an auxiliary sector~\cite{Harnik:2016koz}.

The logarithmic sensitivity in the Higgs potential parameters is also modified by the $\mathbb{Z}_2$ breaking. This can affect the twin Higgs protection at the two-loop level if, for instance, the SM and twin top Yukawa couplings run differently. As mentioned above, this implies that constructions like the twin Higgs typically only address the ``little'' hierarchy problem. They provide a significant improvement in fine-tuning, but must be completed by some other structure above the few TeV scale. 

These gains in naturalness withstand significant variations from the mirror twin set-up. The so-called fraternal model~\cite{Craig:2015pha} makes $\mathbb{Z}_2$ breaking a main feature, only twinning the SM's third generation and allowing for modest deviations from the twin equality in some couplings. In effect, the model keeps only the minimal ingredients required for a twin-Higgs-like protection of the Higgs mass. In a similar vein, if the twin top quark is taken to be vector-like then even the third generation leptons can be removed without generating anomalies~\cite{Craig:2016kue}. Rather than removing entire generations from the mirror twin model, simply allowing significant $\mathbb{Z}_2$ breaking between the fermion Yukawa couplings (other than the top quark) is enough to eliminate some cosmological concerns~\cite{Barbieri:2016zxn}. In weakly-coupled UV completions, a hard breaking of the $\mathbb{Z}_2$ in the quartic terms of the scalar potential can further reduce the need for fine-tuning~\cite{Katz:2016wtw}.

Other efforts seek to completely preserve the discrete symmetry or explain the origin of it breaking. In~\cite{Csaki:2019qgb} the twin symmetry is preserved by much of the model (additional top-partner multiplets do play a role) leading to electroweak symmetry breaking with the correct hierarchy of VEVs, but without $\mathbb{Z}_2$ breaking in the gauge and top quark sectors of the model. Vector-like leptons can also be introduced in each sector to produce a radiative Higgs potential that spontaneously breaks the $\mathbb{Z}_2$~\cite{Jung:2019fsp}. Spontaneous breaking can also be induced by simply including a new scalar field in each sector whose potential is similar to Eq.~\eqref{e.Hpot} but with $\delta<0$. This leads to a completely $\mathbb{Z}_2$ breaking vacuum with the new scalar field's VEV being only in one sector. Such scalars can have other nonzero quantum numbers like hypercharge~\cite{Batell:2019ptb,Liu:2019ixm} or color~\cite{Liu:2019ixm,Batell:2020qad,Batell:2025iwm} which lead to a wide variety of twin sectors by breaking twin gauge symmetries or giving additional masses to twin fermions. In particular, the low energy confining group can be larger than SU(3), if the spontaneous breaking occurs in the visible sector, or smaller if the breaking occurs in the twin sector. Concrete models have produced twin color groups of SU(4)~\cite{Batell:2025iwm}, SU(2)~\cite{Batell:2020qad}, and SO(3)~\cite{Batell:2020qad}. Scalars that are triplets of SU(2$)_L$ in the two sectors can be used to spontaneously break the $\mathbb{Z}_2$ and produce masses for the visible sector neutrinos~\cite{Bittar:2024ryj}.

Two Higgs doublet extensions of the SM have also been combined with the twin Higgs idea~\cite{Chacko:2005vw,Chacko:2019jgi}. This larger scalar sector can be used to explain the origin of $\mathbb{Z}_2$ breaking. This includes spontaneously breaking the twin $\mathbb{Z}_2$ symmetry~\cite{Beauchesne:2015lva}, radiative breaking~\cite{Yu:2016bku}, and tadpole or quartic induced breaking~\cite{Yu:2016swa}. Other models address neutrinos masses in conjunction with a spontaneous breaking of the discrete symmetry~\cite{Liu:2019ixm,Feng:2020urb}. The additional Higgs doublet can be combined with a flavorful Yukawa structure to address the SM flavor puzzle~\cite{Altmannshofer:2020mfp} or be used to make mirror neutrons an asymmetric dark matter candidate~\cite{Beauchesne:2020mih}.

\subsubsection{UV Completions of the Twin Higgs\label{ss.UVcomp}}
As mentioned above, the twin Higgs mechanism only addresses the ``little'' hierarchy problem.
It is therefore important to construct realistic UV completions of the twin Higgs that address the big hierarchy problem. Additionally, one may then consider higher energy properties such as those related to the SM flavor structure or other puzzles. Weakly-coupled UV completions are typically supersymmetric and often feature a doubling of the minimal supersymmetric standard model (MSSM) structure~\cite{Chang:2006ra,Craig:2013fga}. Older completions focused on eliminating the fine-tuning of supersymmetric quartic couplings~\cite{Falkowski:2006qq,Chang:2006ra}. More recent, simpler constructions~\cite{Craig:2013fga} have been shown to retain small fine-tuning and preserve other virtues of SUSY models such as gauge coupling unification. 

Within these supersymmetric UV completions other aspects of twin Higgs models can be explored unambiguously. In particular, various ways to further reduce fine-tuning have been found. One way is to include a hard breaking of the twin symmetry~\cite{Katz:2016wtw} through a quartic term in the Higgs potential, rather than the soft breaking through a dimension two mass term shown above. The $D$-term contributions from a new U(1) gauge symmetry can be used to generate the SU(4) symmetric quartic of the Higgs potential~\cite{Badziak:2017syq}. This improves the tuning prospects, but the new gauge group suffers from a Landau pole at the $10^2$\textendash$10^3$ TeV scale. Using an SU(2)~\cite{Badziak:2017kjk} gauge symmetry allows the Landau pole to be pushed to the $10^6$\textendash$10^{10}$ TeV scale. In both these cases the new gauge symmetry and its gauge bosons are singletons~\cite{Bishara:2018sgl}, not doubled by the twin symmetry. If the new SU(2) is doubled, then the consequent reduction in fields charged under each symmetry implies that they are asymptotically free~\cite{Badziak:2017wxn}, with no Landau pole at all.

A ``turtle" construction is a UV extension of the twin Higgs, which can also precede a SUSY completion~\cite{Asadi:2018abu}. In such a scenario the scale at which one global symmetry breaks (for example the scale $f$ at which the approximate global symmetry of the twin Higgs scalar potential is broken) is stabilized by breaking of yet another global symmetry, again reducing fine-tuning. Supersymmetric completions of the twin Higgs can also be extended to solve the strong CP problem~\cite{Albaid:2015axa} by including an axion field in each sector. Of course, all supersymmetric theories need some explanation of SUSY breaking. The simple scenario of gravity mediated SUSY breaking suffers from a so-called Polonyi problem~\cite{Coughlan:1983ci}, which is that the charge neutral Polonyi field $S$ has no symmetry enhanced point in field space. As a part of a SUSY twin Higgs model the Polonyi field can provide an origin of $\mathbb{Z}_2$ breaking in the scalar potential~\cite{Choi:2023eus}. At the same time, by assuming the Polonyi field to be a singleton field that is odd under the twin symmetry
\beq
S\xrightarrow{\mathbb{Z}_2}-S~,
\eeq
the field space origin becomes a symmetry enhanced point, solving the Polonyi problem.

There are also several strongly-coupled UV completions of the twin Higgs framework. In~\cite{Burdman:2006jj} the twin Higgs framework was extended with a theory of universal extra dimensions. The cut-off for the extra dimensional theory was shown to be significantly beyond the usual twin Higgs limit. In contrast to this flat extra dimension, the authors of~\cite{Geller:2014kta} considered the warped extra dimension of the Randall-Sundrum scenario~\cite{Randall:1999ee}. This holographic construction takes the UV brane to have the full SM doubled gauge structure of the standard twin Higgs. The bulk symmetry is SU(7)$\times$SO(8) and is broken down to SU(7)$\times$SO(7) on the IR brane, with the Higgs a pNGB of the SO(8)$\to$SO(7) breaking. The SU(7) contains the visible and twin SU(3)$\times$U(1) gauge group factors which are exchanged by the discrete $\mathbb{Z}_2$ symmetry. The two factors of SO(4) contained within the bulk SO(8) are also exchanged by this symmetry, allowing the twin exchange to be a symmetry of the UV brane and the bulk. By extending the bulk symmetry, the $\mathbb{Z}_2$ is broken holographically to agree with LHC phenomenology. The theory protects the Higgs mass first by the twin Higgs structure, which allows the Kaluza-Klein modes to be at the few TeV scale while retaining a natural Higgs sector. 

Strongly coupled composite completions, without reference to extra dimensions, have also been developed~\cite{Low:2015nqa,Barbieri:2015lqa}. Both constructions rely on the SO(8)$\to$SO(7) symmetry breaking pattern to ensure that the low energy Higgs physics exhibits custodial symmetry. Despite this similarity, a variety of other characteristics\textemdash such as how the $\mathbb{Z}_2$ symmetry is broken\textemdash distinguish the phenomena of different specific models. In addition, the EFT form factors of composite twin Higgs models are determined by the discrete symmetry, in contrast to other scenarios. This framework  can also accommodate anarchic flavor with significantly less fine-tuning~\cite{Csaki:2015gfd} than traditional composite Higgs models. Precision electroweak constraints primarily apply to the Higgs quartic and are satisfied with only moderate tuning~\cite{Contino:2017moj}. 

Clearly, the twin Higgs scenario can be incorporated within various high scale designs. In some cases a particular completion need not be selected, while for others a specific high-energy construction is essential. Interestingly, a renormalization-group-improved analysis of the twin Higgs effective potential shows that the Higgs mass can be a largely UV independent prediction~\cite{Greco:2016zaz}.

\subsection{Neutral Naturalness and SUSY\label{ss.NNsusy}}
In contrast to the supersymmetric UV completions of the twin Higgs discussed above, SUSY can also be the fundamental symmetry that protects the Higgs mass, although it manifests in less familiar ways. These models share many characteristics with folded SUSY~\cite{Burdman:2006tz}, so we outline its broad aspects in some detail.

\begin{figure}[ht]
\centering
\begin{tabular}{ccc}
    \begin{tabular}{c}
\begin{fmffile}{FHiggsloopG}
\begin{fmfgraph*}(120,80)
\fmfpen{1.0}
\fmfstraight
\fmfleft{p1,i1,p2}\fmfright{p3,o1,p4}
\fmfv{l= $H_u^\dag$}{i1}\fmfv{l=$H_u$}{o1}
\fmf{dashes,tension=1}{i1,v1}
\fmf{fermion,left=1,tension=0.4,label.side=left,label=$t$,label.dist=5,foreground=(0.01,,0.75 ,,0.24)}{v1,v2,v1}
\fmf{dashes,tension=1}{v2,o1}
\fmfv{decor.shape=circle,decor.filled=full,decor.size=1.5thick,l=$\lambda_t$,l.a=120,l.d=7}{v1}
\fmfv{decor.shape=circle,decor.filled=full,decor.size=1.5thick,l=$\lambda_t$,l.a=60,l.d=7}{v2}
\end{fmfgraph*}
\end{fmffile}
\end{tabular}
\begin{tabular}{c}
\hspace{5mm}$+$ \hspace{5mm}
\end{tabular} 
\begin{tabular}{c}
\begin{fmffile}{FsusystopG}
\begin{fmfgraph*}(110,80)
\fmfpen{1.0}
\fmfleft{p1,i1,i2,p3} 
\fmfright{p2,o1,o2,p4}
\fmf{phantom}{p3,v2,p4}
\fmf{dashes}{i1,v1,o1}
\fmfv{l=$H_u^\dag$}{i1}\fmfv{l=$H_u$}{o1}
\fmffreeze
\fmf{double,right,tension=0.1,l.side=right,foreground=(0.01,,0.75 ,,0.24)}{v2,v1}
\fmf{double,right,tension=0.1,l.side=right,foreground=(0.01,,0.75 ,,0.24)}{v1,v2}
\fmfv{decor.shape=circle,decor.filled=full,decor.size=1.5thick, l=$\lambda_t^2$,l.a=-90,l.d=5}{v1}
\fmfv{l=$\widetilde{t}$,l.d=15,l.a=0}{i2} 
\end{fmfgraph*}
\end{fmffile}
\end{tabular}
\\
    \begin{tabular}{c}
\begin{fmffile}{FHiggsloopB}
\begin{fmfgraph*}(120,80)
\fmfpen{1.0}
\fmfstraight
\fmfleft{p1,i1,p2}\fmfright{p3,o1,p4}
\fmfv{l= $H_u^\dag$}{i1}\fmfv{l=$H_u$}{o1}
\fmf{dashes,tension=1}{i1,v1}
\fmf{fermion,left=1,tension=0.4,label.side=left,label=$t$,label.dist=5,foreground=blue}{v1,v2,v1}
\fmf{dashes,tension=1}{v2,o1}
\fmfv{decor.shape=circle,decor.filled=full,decor.size=1.5thick,l=$\lambda_t$,l.a=120,l.d=7}{v1}
\fmfv{decor.shape=circle,decor.filled=full,decor.size=1.5thick,l=$\lambda_t$,l.a=60,l.d=7}{v2}
\end{fmfgraph*}
\end{fmffile}
\end{tabular}
\begin{tabular}{c}
\hspace{5mm}$+$ \hspace{5mm}
\end{tabular} 
\begin{tabular}{c}
\begin{fmffile}{FsusystopB}
\begin{fmfgraph*}(110,80)
\fmfpen{1.0}
\fmfleft{p1,i1,i2,p3} 
\fmfright{p2,o1,o2,p4}
\fmf{phantom}{p3,v2,p4}
\fmf{dashes}{i1,v1,o1}
\fmfv{l=$H_u^\dag$}{i1}\fmfv{l=$H_u$}{o1}
\fmffreeze
\fmf{double,right,tension=0.1,l.side=right,foreground=blue}{v2,v1}
\fmf{double,right,tension=0.1,l.side=right,foreground=blue}{v1,v2}
\fmfv{decor.shape=circle,decor.filled=full,decor.size=1.5thick, l=$\lambda_t^2$,l.a=-90,l.d=5}{v1}
\fmfv{l=$\widetilde{t}$,l.d=15,l.a=0}{i2} 
\end{fmfgraph*}
\end{fmffile}
\end{tabular}
\end{tabular}
\vspace{-0.5cm}
\caption{\label{fig:FsusyAllLoops} Two identical copies of the supersymmetric cancellation of quadratic corrections to the Higgs mass parameter (through $H_u$ of the two Higgs fields in SUSY). The two colors indicate that each copy is charged under a distinct SU(3) gauge group. 
}
\end{figure}

The essential idea behind folded SUSY is quite simple. Very schematically, SUSY connects bosonic and fermionic degrees of freedom. Roughly, we can consider that the top quark is extended to a SUSY multiplet
\beq
t\Rightarrow \left.\left(\begin{array}{c}
t \\
\widetilde{t}
\end{array}\right)\right\downarrow\hspace{-1.0mm}\text{SUSY}~,
\eeq
where $\widetilde{t}$ denotes the scalar stop. This scalar partner to the top quark is a manifestation of the SUSY structure in a given model. This symmetry can also explain why the Higgs mass is insensitive to higher mass scales. Having two copies of these particles would also preserve Higgs naturalness; the cancellation between fermionic and bosonic contributions to the Higgs mass renormalization would simply happen twice in exactly the same way, as shown in Fig.~\ref{fig:FsusyAllLoops}.  

The authors of folded SUSY observed a generalization of this doubling idea, where each copy is charged under a different color group. If we label SM color as SU(3$)_A$ then we can label the copy (or folded color) as SU(3$)_B$ and write the top sector particles as
\beq
\overset{\text{SU}(3)_A\times \text{SU}(3)_B\hspace{5mm}}{\left.\overrightarrow{\hspace{-3.5mm}\left(\begin{array}{cc}
t_A & t_B\\
\widetilde{t}_A & \widetilde{t}_B
\end{array}\right)\hspace{-3.5mm}}\hspace{3.5mm}\right\downarrow}\hspace{-2.5mm}\text{SUSY}~.
\eeq
This extension keeps the Higgs natural through the SUSY structure. In this construction the $A$ and $B$ sector fields are assumed to be related by a discrete $\mathbb{Z}_2$ exchange symmetry to ensure the coupling structure is the same in each sector.

But what if the SM colored stop $\widetilde{t}_A$ (the green scalar in Fig.~\ref{fig:FsusyAllLoops}) and the folded top quark $t_B$ (the blue fermion in Fig.~\ref{fig:FsusyAllLoops}) are removed from the low-energy theory? Then the Higgs remains natural because the contributions from the SM top (the green fermion in Fig.~\ref{fig:FsusyAllLoops}) and the folded stop (the blue scalar in Fig.~\ref{fig:FsusyAllLoops}) combine in precisely the same way as in standard SUSY. The fact that they are charged under different color groups is irrelevant to the cancellation of the $\Lambda_\text{UV}^2$ terms. 

\begin{figure}[ht]
    \centering
    \includegraphics[width=0.75\linewidth]{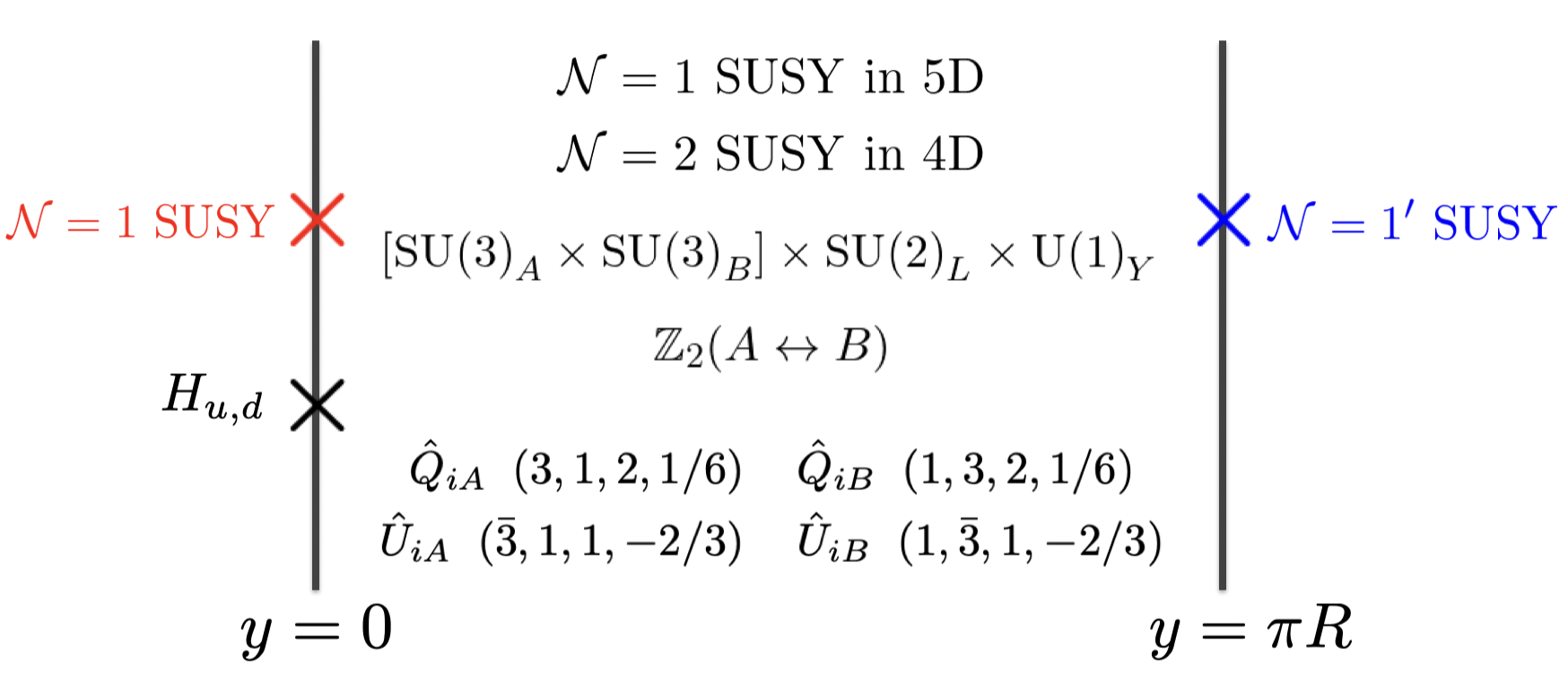}~~
    \caption{Schematic of a realistic folded SUSY construction~\cite{Burdman:2006tz}. The UV theory features an extra spatial dimension of radius $R$ compactified on a $S^1/\mathbb{Z}_2$ orbifold. Gauge and matter fields propagate in the bulk and fill out 5D ${\cal N} = 1$ SUSY multiplets, which, from the 4D perspective, correspond to ${\cal N} = 2$ SUSY multiplets. SUSY is broken by the Scherk-Schwarz mechanism~\cite{Scherk:1978ta,Scherk:1979zr}, with each brane preserving a distinct 4D ${\cal N} = 1$ SUSY, such that  SUSY is completely broken below the compactification scale. The bulk gauge symmetry is [SU(3)$_{A}\times$SU(3)$_{B}]\times$SU(2)$_{L}\times$U(1)$_{Y}$ and there is a discrete $\mathbb{Z}_2$ symmetry that interchanges SM fields (labeled $A$) and their folded partners (labeled $B$). The Higgs fields are localized to the brane at $y = 0$.  
    }
    \label{fig:folded_susy_schematic}
\end{figure}

In this scenario the folded stops are much more difficult to produce at hadron colliders because they do not carry SM color. Note that in this case no true SUSY is manifest in the low-energy theory as the fermions and bosons belong to different SUSY multiplets. The protection of the Higgs mass results from an accidental SUSY realized by the low-energy theory.

The folded SUSY structure finds a natural realization in a realistic UV completion involving an additional spatial dimension, as schematically depicted in Fig.~\ref{fig:folded_susy_schematic}.
The extra dimension is taken to be a finite, flat interval between two 4D branes. It is also compactified on $S^1/\mathbb{Z}_2$, making it an orbifold, from which folded SUSY takes its name. The 5D superfields with $\mathcal{N}=1$ SUSY can be naturally viewed as $\mathcal{N}=2$ SUSY in 4D. The $\mathcal{N}=2$ SUSY is broken to $\mathcal{N}=1$ on each brane by the boundary conditions each bulk field has there. The SM gauge and matter fields propagate in the bulk, while the Higgs multiplets are localized on one of the branes. 

\begin{figure}[ht]
    \centering
    \includegraphics[width=0.6\linewidth]{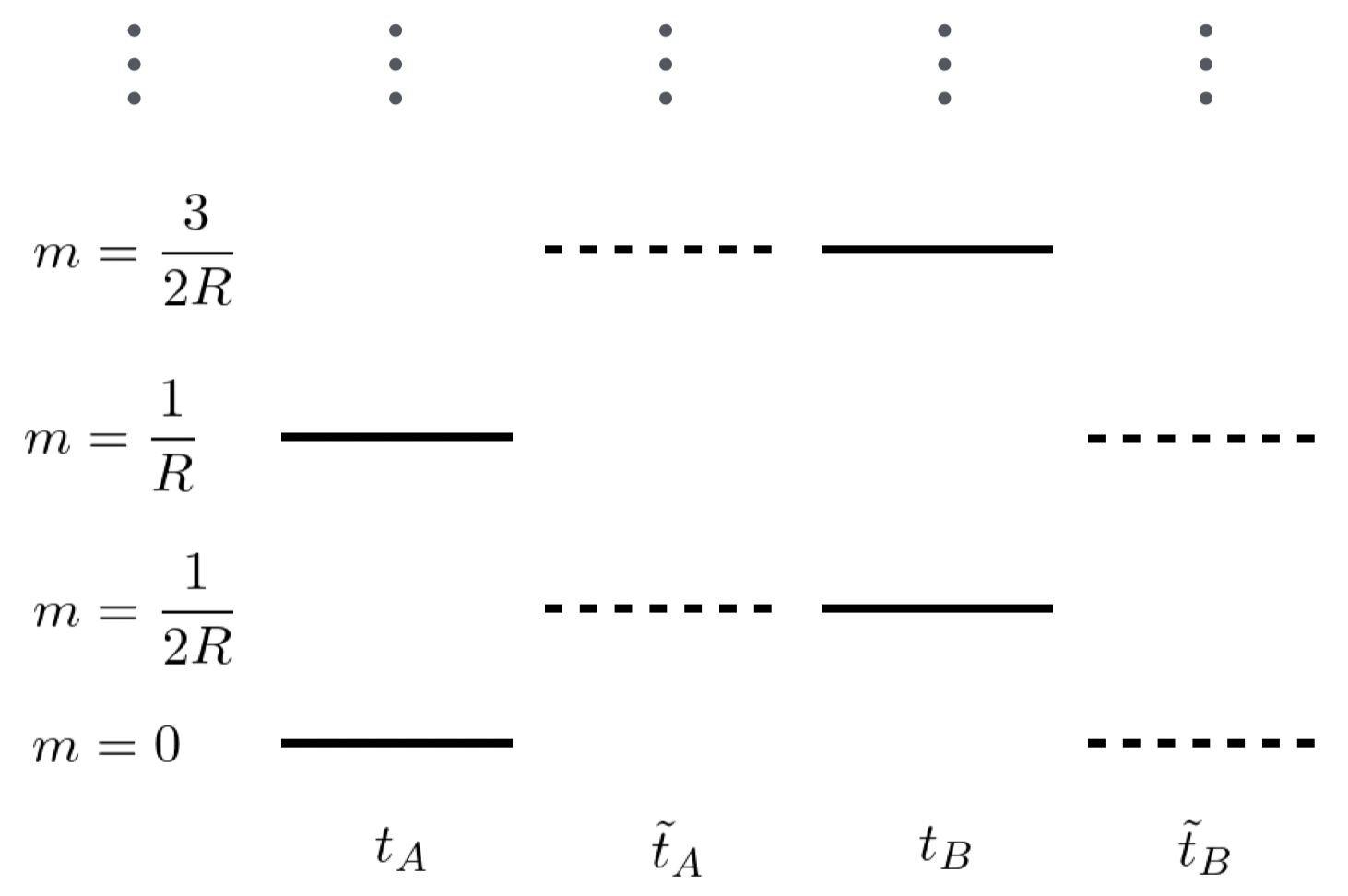}
    \caption{The spectrum of the top sector in folded SUSY.  SUSY is broken via the Scherk–Schwarz mechanism, leading to non-degenerate fermions and bosons within each Kaluza-Klein tower. However, unlike in conventional Scherk–Schwarz scenarios, folded SUSY ensures an overall boson–fermion degeneracy by pairing each such tower with a complementary one from a folded sector.
    }
    \label{fig:folded_susy_spectrum}
\end{figure}

In the original folded-SUSY model the boundary conditions of the $A$ sector colored scalars and $B$ sector colored fermions are chosen such that their Kaluza-Klein tower has no zero-mode, as illustrated in Fig.~\ref{fig:folded_susy_spectrum}. This effectively eliminates them from the low-energy theory below the Kaluza-Klein scale. The top Yukawa coupling is then given by 
\begin{align}
&\int dy d^2 \theta\delta(y) \lambda_t\left[ Q_{3A} H_u U_{3A} +Q_{3B} H_u U_{3B} +{\rm H.c.} \right] \\
& \quad\quad\supset \lambda_t\left[ q_{3A}^{(0)} H_u u_{3A}^{(0)} + {\rm H.c.}\right]  + \lambda_t^2 \left[|\tilde q_{3B}^{(0)} H_u|^2 + |\tilde u_{3B}^{(0)} H_u|^2 \right], \nonumber\end{align}
where the right-hand side indicates the Higgs couplings to the zero mode top quark and folded stop,  essential elements of the protection mechanism.  
In this model, the gauge sector is not folded, so the gauge boson zero modes remain massless while the lightest gauginos acquire a mass of $1/(2R)$.

The $\mathcal{N}=1$ SUSY preserved by the field boundary conditions on each brane can be different. This means that in the low-energy theory SUSY is completely broken by the Scherk-Schwarz~\cite{Scherk:1978ta,Scherk:1979zr} mechanism.\footnote{This mechanism can also improve the naturalness of SUSY models without color charge neutral top partners~\cite{GarciaGarcia:2015yfz}.} Crucially, the complete breaking of SUSY by the mismatch of the $\mathcal{N}=1$ SUSYs on the two branes is nonlocal. Consequently, corrections to the brane localized Higgs potential are finite and calculable. Thus, the simple concept of a SUSY-like protection of the Higgs mass through stops that are neutral to SM color is concretely realized.

Similar to the twin Higgs framework, extra dimensional constructions come with a UV cutoff that is typically not too far above the TeV scale. As shown in~\cite{Craig:2014fka}, however, the extra dimension can be deconstructed while preserving the natural virtues of folded SUSY.  An estimate of the two-loop calculation of the folded-SUSY Higgs potential indicates that it may not produce the correct vacuum without modification. A simple change is to vary the Scherk-Schwarz twist away from its maximum value~\cite{Cohen:2015gaa}. This can ensure electroweak symmetry breaking occurs at one-loop. The tuning in folded SUSY can also be improved by including gauginos that are bifundamentals of both the $A$ and $B$ color groups~\cite{Gherghetta:2016bcc}.

The squark fields of folded SUSY and its variations are color neutral, but carry SM electroweak charges. Two frameworks have been discovered that have scalar top partners that are complete SM gauge singlets. The tripled-top scenario~\cite{Cheng:2018gvu,Cheng:2019yai} demonstrates that the folded-SUSY-like cancellation can be accomplished in 4D using two copies of the top sector all related by a discrete $\mathbb{Z}_3$ symmetry. In this construction the right-handed stops of each copy act as top partners, so they are not charged under SU(2$)_L$. The hypercharge of the top-sector copies may be freely specified, including the choice of scalar top partners with no hypercharge, which are SM gauge singlets. The Hyperbolic Higgs model~\cite{Cohen:2018mgv} obtains SM singlet scalar top partners from a 5D set-up. In contrast to the approximate U(4) symmetry of the scalar potential of twin Higgs models, this scalar potential enjoys an approximate, non-compact U(2,2) symmetry. The kinetic terms of the scalar fields are not invariant under U(2,2), but the form of the scalar potential is sufficient for SM gauge singlet scalars to cancel the divergent loop contributions of SM quarks. 

\subsection{Neutral Naturalness and pNGB Higgs\label{ss.pNGB}}
Neutral naturalness began as a pNGB construction. The twin Higgs scenario posits that the observed Higgs boson is the pNGB of an approximate SU(4) global symmetry. However, the neutral natural pNGB possibilities include a rich variety of realizations. There is a class of theories that may be seen as the direct generalization of the twin Higgs, called orbifold Higgs models~\cite{Craig:2014aea,Craig:2014roa}. From this perspective the twin Higgs gauge structure, SU(3$)_A\times \text{SU}(2)_A\times \text{SU}(3)_B\times \text{SU}(2)_B\times \mathbb{Z}_2$, results from the orbifold $\left[\text{SU}(6)\times \text{SU}(4) \right]/\mathbb{Z}_2$. This provides an origin for the discrete symmetry that relates the gauge and Yukawa couplings in the two sectors which is essential to the one-loop protection of the Higgs. What is more, this orbifolding method can be used with other symmetry groups to produce different numbers and varieties of hidden sectors whose SM singlet top partners all cooperate to keep the Higgs natural. 

There are other pNGB Higgs models that resemble folded SUSY in their set-up. The usual pNGB Higgs scenario assumes an approximate global symmetry is spontaneously broken at some scale $f$ and the Higgs is one of the resulting pNGBs. The top quark is typically taken to belong to a representation of the larger global symmetry. The new fields required to fill out the multiplet are the top partners. 

For an SU(3) global symmetry we have the schematic form
\beq
\left(\begin{array}{c}
t\\
b
\end{array}\right)=q_L\Rightarrow \left.\left(\begin{array}{c}
q_L \\
T
\end{array}\right)\right\downarrow\hspace{-1.0mm}\text{SU}(3)~,
\eeq
where $T$ is the top-partner field. The protection of the Higgs mass works just as well when it happens twice, including new fields charged under a new color group
\beq
\overset{\text{SU}(3)_A\times \text{SU}(3)_B\hspace{5mm}}{\left.\overrightarrow{\hspace{-3.5mm}\left(\begin{array}{cc}
q_{L_A} & q_{L_B}\\
T_A & T_B
\end{array}\right)\hspace{-3.5mm}}\hspace{3.5mm}\right\downarrow}\hspace{-2.5mm}~\text{SU}(3)~.
\eeq
The one-loop protection persists even if the $T_A$ and $q_{L_B}$ fields are removed from the low-energy theory. In this case the top partner $T_B$ is not charged under SM color, though it would carry SM hypercharge. This is the essence of the Quirky Little Higgs model~\cite{Cai:2008au}, where the unwanted fields are removed by the brane boundary conditions in an extra dimensional space, similar to folded SUSY. This symmetry breaking pattern also plays a role in some Dark Top constructions~\cite{Poland:2008ev}, though differences allow the top partner to be a stable dark matter candidate. The neutral top partners can also be right-handed neutrinos~\cite{Batell:2015aha} for various symmetry breaking patterns.

While the SU(3)/SU(2) construction is in some sense minimal, it does not provide the Higgs with a custodial SU(2) symmetry that is highly motivated by precision electroweak measurements. Orthogonal groups can ensure the custodial form of Higgs structure. The minimal realization is SO(5)/SO(4)~\cite{Xu:2018ofw,Xu:2019xuo} where boundary conditions in a warped extra dimension are employed to lift unwanted fields from the low-energy spectrum. 

While the SO(5)/SO(4) symmetry breaking pattern preserves the custodial symmetry, it differs from the twin Higgs construction in how the top partners relate to the top quark. In the twin Higgs set-up, which can be thought of as SO(8)/SO(7), the top and its partner can be related by a discrete symmetry because they are embedded within the SO(8) in similar ways
\begin{align}
\mathcal{T}_{LA}=&\frac{1}{\sqrt{2}}\left(it_{LA},\,-t_{LA},\,ib_{LA},\,b_{LA},\,0,\,0,\,0,\,0 \right)^T \\
\mathcal{T}_{LB}=&\frac{1}{\sqrt{2}}\left(0,\,0,\,0,\,0,\, it_{LB},\,-t_{LB},\,ib_{LB},\,b_{LB}\right)^T~.
\end{align}
This cannot be done when the symmetry breaking pattern is SO(5)/SO(4). In this case the embedding takes the form
\begin{align}
    \mathcal{T}_{LA}=&\frac{1}{\sqrt{2}}\left(it_{LA},\,-t_{LA},\,ib_{LA},\,b_{LA},\,0 \right) \\
    \mathcal{T}_{LB}=&\frac{1}{\sqrt{2}}\left(it_{LB},\,-t_{LB},\,ib_{LB},\,b_{LB},\,\sqrt{2}T_{LB} \right)~.
\end{align}
Note the appearance of the $T_{LB}$ state, which has no corresponding state among the SM fields. The smallest orthogonal group structure that allows the top quark and its partner to be related by a twin parity is SO(6)/SO(5)~\cite{Serra:2017poj,Csaki:2017jby}. The quark fields have the form
\begin{align}
    \mathcal{T}_{LA}=&\frac{1}{\sqrt{2}}\left(it_{LA},\,-t_{LA},\,ib_{LA},\,b_{LA},\,0,\,0 \right) \\
    \mathcal{T}_{LB}=&\frac{1}{\sqrt{2}}\left(0,\,0,\,0,\,0,\,it_{LB},\,-t_{LB} \right)\\
    \mathcal{B}_{LB}=&\frac{1}{\sqrt{2}}\left(0,\,0,\,0,\,0,\,ib_{LB},\,b_{LB} \right)~,
\end{align}
which includes only fields that are related to the SM quarks by a discrete symmetry. This construction has also been extended into the UV holographically~\cite{Dillon:2018wye}.

Though it allows for twin parity, the SO(6)/SO(5) symmetry breaking pattern also produces another pNGB which must be removed from the low-energy theory. Another construction which allows twin parity, leads to a custodial symmetry, and an embedding of both SM and twin SU(2$)_L$ is SO(7)$/G_2$~\cite{Serra:2019omd}. When SO(7) is spontaneously broken to $G_2$ seven pNGBs are produced, but six are eaten by the SM and twin SU(2$)_L$ gauge bosons, leaving behind a single physical Higgs, just as in the twin Higgs. The breaking pattern of SO(7)/SO(6) leads to custodial symmetry for the pNGB Higgs while the additional pNGBs become a viable complex scalar dark matter candidate~\cite{Ahmed:2020hiw}. 

Models employing a SO($2N)/\left[\text{SO}(N)\times \text{SO}(N) \right]$ breaking pattern have also been investigated~\cite{Durieux:2022sgm}. These models naturally produce a hierarchy between the Higgs VEV $v$ and the global symmetry breaking scale $f$. In most pNGB models such a hierarchy is required to agree with experimentally measured Higgs coupling data, but is a source of tuning. These so-called twin Gegenbaur constructions agree with current Higgs measurements without additional tuning and have been UV completed supersymmetrically~\cite{McCullough:2025zeg}.

\section{Collider Phenomenology\label{s.collider}}
The primary characteristic of neutral naturalness is that there are no new colored states up to the few TeV scale. Consequently, some of the most powerful hadron collider searches for new states do not apply. However, there are many other signatures of neutral naturalness that may appear at present-day and future colliders. Indeed, it has been argued~\cite{Curtin:2015bka} that the combination of current machines along with future lepton and hadron colliders will be able to thoroughly probe symmetry-based solutions to the hierarchy problem.

\subsection{Higgs Physics\label{ss.HiggsPhys}}
The most robust prediction of neutral naturalness is the existence of new states that couple to the Higgs. Consequently, examining the Higgs for indications of these structures is highly motivated. In pNGB Higgs models the tree-level couplings of the Higgs to SM fields are modified, which already constrains such theories~\cite{Burdman:2014zta,Thrasher:2017rpa}. The deviations from SM couplings can also have significant effects on, for instance, rare top quark decays~\cite{Bie:2020sro}. In supersymmetric theories, or general bottom-up models with scalar top partners, the tree-level Higgs couplings may be unaffected. However, loop-level couplings like those shown in Fig.~\ref{fig:HiggsLoopVV}, including $h\gamma\gamma$~\cite{Fan:2014txa} and $hZ\gamma$~\cite{Archer-Smith:2020gib} (or even contributions to $hZZ$~\cite{Craig:2013xia,Haisch:2023aiz}), can be modified by the contributions from top partners or other BSM particles in the loop. Loop corrections to di-Higgs production can also be used to probe heavy fermions that couple to the Higgs~\cite{Haisch:2023aiz}.

\begin{figure}
    \centering
    \begin{fmffile}{higgsVVLoop}
\begin{fmfgraph*}(130,80)
\fmfpen{1.0}
\fmfstraight
\fmfleft{p1,i1,p2}\fmfright{o1,p2,o2}
\fmfv{l= $h$}{i1}\fmfv{l=$Z,,\gamma$}{o1}
\fmfv{l=$Z,,\gamma$}{o2}
\fmf{dashes,tension=0.9}{i1,v1}
\fmf{plain,left=0.7,tension=0.7,label=BSM,label.side=left}{v1,v2}
\fmf{plain,left=0.33,tension=0.6}{v2,v3}
\fmf{plain,left=0.7,tension=0.7}{v3,v1}
\fmf{photon,tension=0.9}{v2,o2}
\fmf{photon,tension=0.9}{v3,o1}
\fmfv{decor.shape=circle,decor.filled=full,decor.size=1.5thick}{v1}
\fmfv{decor.shape=circle,decor.filled=full,decor.size=1.5thick}{v2}
\fmfv{decor.shape=circle,decor.filled=full,decor.size=1.5thick}{v3}
\end{fmfgraph*}
\end{fmffile}
\vspace{0.5cm}
    \caption{Schematic form of one-loop BSM contributions to Higgs couplings to $Z$ and $\gamma$. Precision measurements of this process can be sensitive to the new states running in the loop.}
    \label{fig:HiggsLoopVV}
\end{figure}

In neutral naturalness the Higgs often connects the SM to a new sector of particles. If some of these particles are lighter than half the Higgs mass, the Higgs can acquire new, sometimes referred to as exotic, decay modes. For example, in the mirror twin Higgs model the Higgs couples to all twin quarks and leptons. The strengths of these couplings and the mass of the twin fermions are both controlled by the ratio of the SM Higgs VEV to the twin Higgs VEV, $v/f$. This quantity is also directly proportional to the tuning of the model, making both Higgs coupling deviations and the invisible Higgs width direct probes of the naturalness of this scenario. Even when these states are too heavy for an on-shell Higgs to decay into them, they can be produced through an off-shell Higgs~\cite{Craig:2014lda,Goncalves:2017iub,Goncalves:2018pkt}. The loop effects of the new heavy particles can also affect well-known processes~\cite{Haisch:2023aiz}, allowing precision measurements to reveal their presence.

There can be other scalars, besides the observed 125 GeV Higgs, in the Higgs sector, such as the twin Higgs of twin Higgs models or the heavy Higgses of SUSY models. These scalars often have couplings to both SM and hidden-sector fields, making them an interesting portal between the sectors. These motivate specific heavy scalar targets at current and future colliders~\cite{Buttazzo:2015bka,Ahmed:2017psb,Buttazzo:2018qqp} and can be used in conjunction with other collider signals to prove such scalars belong to a neutral naturalness scenario~\cite{Chacko:2017xpd}. This may be accomplished through prompt heavy Higgs signals, and often with powerful complementarity through the long-lived particles that can result from heavy Higgs decays~\cite{Kilic:2018sew,Alipour-fard:2018mre}. 

\subsubsection{Hidden Sector Glueballs \label{sss.glueballs}}
Most models require that some of the hidden-sector particles are charged under a hidden SU(3) gauge group, though in some cases other confining groups\textemdash such as SU(4)~\cite{Batell:2025iwm}, SU(2)~\cite{Batell:2020qad}, or SO(3)~\cite{Batell:2020qad}\textemdash play a similar role. In many scenarios the fields charged under hidden color are much heavier than the  hidden confinement scale, implying that the lowest lying hidden hadrons are glueballs. The spectrum of the SU(3) glueball states has been determined by lattice calculations~\cite{Morningstar:1999rf,Chen:2005mg}. Similar calculations have been performed for SU$(N)$~\cite{Athenodorou:2021qvs} and Sp$(2N)$~\cite{Bennett:2020qtj} gauge groups.

\begin{figure}[th]
    \centering
    \includegraphics[trim={0 2cm 0 10cm},width=0.5\linewidth]{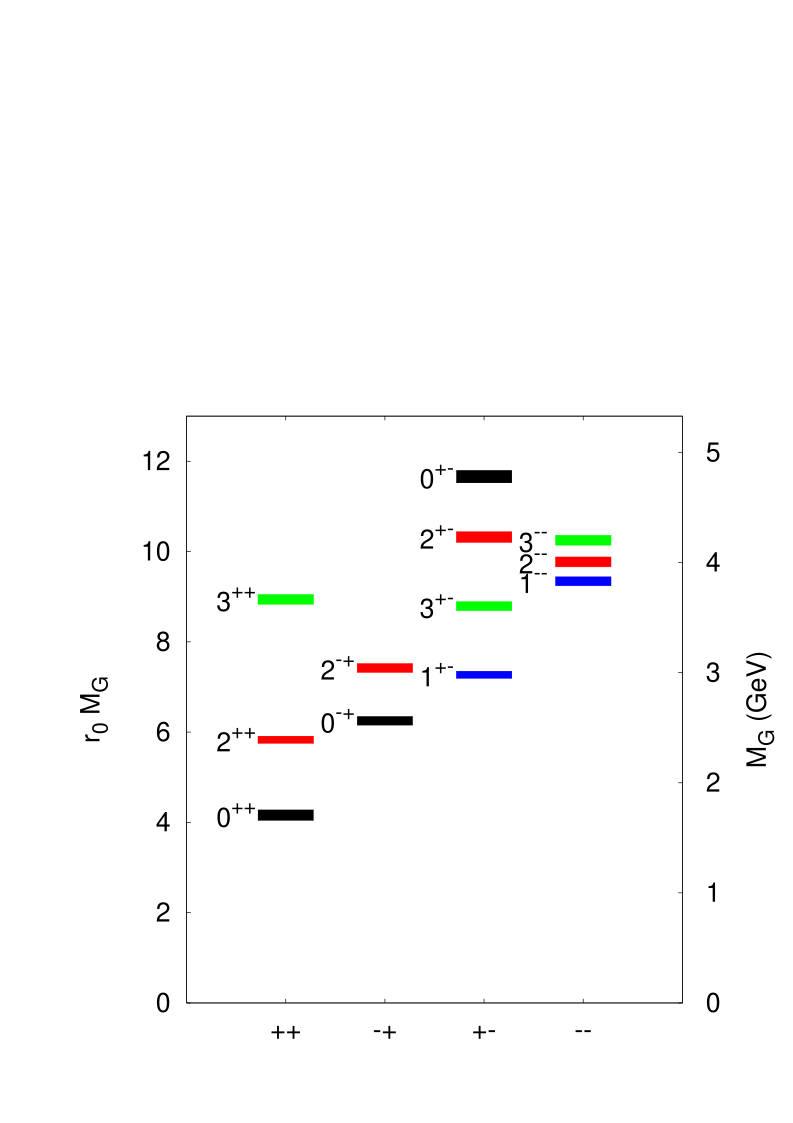}
    \caption{Left: Glueball spectrum of SU(3) gauge theory from~\cite{Chen:2005mg}. Organized by $J^{PC}$ quantum numbers. The width of each box denotes numerical uncertainty.}
    \label{fig:su3GBspectrum}
\end{figure}

The various glueball states are often labeled by angular momentum $J$, parity $P$, and charge conjugation $C$ quantum numbers as $J^{PC}$. The lightest state is the $0^{++}$, and other masses are typically reported in terms of the mass $m_0$ of this glueball. For SU(3) there are 12 glueball states whose masses range from $m_0$ to about $3m_0$, as shown in Fig.~\ref{fig:su3GBspectrum}. These masses are given in terms of the hadronic scale $r_0$, which can be related to the confinement scale~\cite{Ishikawa:2017xam} by $\Lambda_c r_0\approx 0.6$, leading to the result $m_0\approx 7\Lambda_c$.

In realizations of neutral naturalness the hidden ($B$) sector glueballs are often the lightest states of that sector. They are not, however, the lightest states of the theory. Loop effects produce higher dimension operators~\cite{Juknevich:2009ji,Juknevich:2009gg} 
\beq
\mathcal{L}^{(6)}=\frac{\alpha_By^2}{6\pi M^2}H^\dag H G^a_{B\mu\nu}G_B^{a\mu\nu}~,\label{eq:dim6Glue}
\eeq
which connect the gluons of the hidden sector to the Higgs.\footnote{It is argued in~\cite{Davoudiasl:2024epi} that couplings of this type can also lead to significant instanton corrections to the Higgs potential.} The coupling $y$ and mass $M$ follow from the hidden sector fields that couple to the Higgs and the gluons. Additional operators can arise at dimension eight, but the effects from these are typically much smaller than what follows from the dimension six operator in~\eqref{eq:dim6Glue}. They may, however, play an important role in glueball lifetimes on cosmological scales~\cite{Curtin:2015jcv,Curtin:2022oec}.

These operators allow many of the heavier glueballs to decay into lighter ones by radiating a Higgs~\cite{Juknevich:2009gg,Batz:2023zef}. These decay widths have been expressed in terms of nonperturbative gluon to gluon matrix elements, most of which have not yet been calculated. However, there is an important exception. 

The lightest glueball, $0^{++}$, has the right quantum numbers to mix with the Higgs boson. Consequently, the Higgs can develop a small mixing with this glueball state. This allows the lightest glueball to decay through an off-shell Higgs into SM states. The decay width~\cite{Juknevich:2009gg} of the $0^{++}$ glueball into a pair of SM fermions is given by
\beq
    \Gamma\left(0^{++}\to f_\text{SM}f_\text{SM}\right)=\left|\frac{y}{M}\right|^2\left( \frac{\alpha_B}{3\pi}\frac{f_{0^{++}}}{m_h^2-m_0^2}\right)^2\Gamma\left(h(m_0)\to f_\text{SM}f_\text{SM}\right)~, \label{eq:decay-m0}
\eeq
where $m_h$ is the mass of the Higgs boson. The quantity
\beq
f_{0^{++}}=\langle0|\text{Tr}\,G_{B\mu\nu}G^{\mu\nu}_B|0^{++}\rangle~,
\eeq
is a gluon field strength matrix element. This has been calculated in lattice gauge theory~\cite{Chen:2005mg,Meyer:2008tr} to be about $4\pi\alpha_Bf_{0^{++}}\approx 3m_0^3$. Finally, the decay width
\beq
\Gamma\left(h(m_0)\to f_\text{SM}f_\text{SM}\right)~,
\eeq
is the decay width of a SM Higgs boson (of mass $m_0$) into the fermion pair. The total decay width for the glueball can be quite small. In some cases the decay lengths are of the right size to produce displaced vertices at colliders, which makes them a powerful tool for probing neutral naturalness.

\begin{figure}[tbh]
    \centering
    \begin{fmffile}{exoticHiggs}
\begin{fmfgraph*}(300,100)
\fmfpen{1.0}
\fmfstraight
\fmfleft{p1,p2,i1,p3,p4} \fmfright{o1,o2,p5,o3,o4}
\fmfv{l= $\bm{h}$}{i1}
\fmfv{l=$f_\text{SM}$}{o1}
\fmfv{l=$f_\text{SM}$}{o2}
\fmfv{l=$f_\text{SM}$}{o3}
\fmfv{l=$f_\text{SM}$}{o4}
\fmf{dashes,tension=1.6}{i1,v1}
\fmf{fermion,tension=0.7}{v1,v2} 
\fmf{fermion,tension=-0.4}{v2,v3}
\fmf{fermion,tension=0.7}{v3,v1} 
\fmf{gluon,tension=0.9,right=0.2,label.side=right,label=$g_B$,label.dist=5\thick}{v4,v2}
\fmf{gluon,tension=0.9,right=0.2,label.side=right,label=$g_B$,label.dist=5\thick}{v3,v5}
\fmf{plain,tension=0.9}{v4,v6}
\fmf{plain,tension=0.9}{v5,v7}
\fmf{dashes,tension=1.0}{v6,v8}
\fmf{dashes,tension=1.0}{v7,v9}
\fmf{fermion,tension=1.0}{o4,v8,o3}
\fmf{fermion,tension=1.0}{o2,v9,o1}
\fmf{plain,tension=0.4}{v4,v10,v5}
\fmfv{decor.shape=circle,decor.filled=full,decor.size=1.5thick,l=$\lambda_t$,l.a=90,l.d=5}{v1} 
\fmfv{decor.shape=circle,decor.filled=full,decor.size=1.5thick,l=$T$,l.a=188,l.d=25}{v2} 
\fmfv{decor.shape=circle,decor.filled=full,decor.size=1.5thick}{v3} 
\fmfv{decor.shape=circle,decor.filled=full,decor.size=1.5thick,l=$\bm{h}$,l.a=170,l.d=16}{v8} 
\fmfv{decor.shape=circle,decor.filled=full,decor.size=1.5thick,l=$\bm{h}$,l.a=190,l.d=16}{v9} 
\fmfv{decor.shape=cross,l=$0^{++}$,l.a=158,l.d=11}{v6}\fmfv{decor.shape=cross,l=$0^{++}$,l.a=202,l.d=11}{v7}
\fmfv{decor.shape=circle,decor.filled=30,decor.size=26thick}{v10}
\end{fmfgraph*}
\end{fmffile}
    \caption{Decay of the Higgs boson through a  top-partner $T$ loop into two $0^{++}$ glueballs. These decay through mixing with the Higgs into SM fermions.}
    \label{fig:exoticHiggsDecy}
\end{figure}

For instance, despite the uncertainties of glueball hadronization~\cite{Curtin:2022tou,Batz:2023zef} when $2m_0<m_h$ the Higgs can decay into a pair of glueballs. Each of these glueballs then decays, often with long decay length, back into SM fermions, dominantly to $b$-quarks, see Fig.~\ref{fig:exoticHiggsDecy}. These exotic displaced decays of the Higgs~\cite{Curtin:2015fna,Csaki:2015fba,Wang:2024ieo} can be a significant collider probe of neutral natural frameworks.

These exotic Higgs decays are the most studied example of hidden sector glueballs in neutral naturalness, and the most obviously tied to Higgs physics. However, at some level all of the useful signals have the $0^{++}$ mixing with the Higgs at heart. Other effects of glueballs include how they play an important role in understanding rich hidden sector spectra and dynamics~\cite{Craig:2015pha,Cheng:2015buv,Cheng:2021kjg} including dark matter~\cite{Curtin:2022oec}. When hidden sector states are produced at colliders final state radiation can form glueballs that decay after a macroscopic distance. These displaced vertices can also be used to enhance the discovery power of collider searches~\cite{Forsyth:2025wks}. 

\subsection{Dark Hadrons}
The hadrons of the dark sector can also lead to interesting collider signals. In some fraternal twin Higgs models~\cite{Craig:2015pha} bound states of twin bottom quarks are lighter than the twin glueballs and the some of these bound states can mix with Higgs, leading to signals similar to those of the glueballs discussed in the previous section. However, the specifics of these decays depend on other aspects of the twin sector mass spectrum, like the twin neutrino masses and mass difference between the twin bottomonia states, leading to greater variation in the potential outcomes.

Considering the states associated with UV completions of the fraternal twin Higgs increase the number of twin bound states that can decay into SM states within the LHC detectors~\cite{Cheng:2015buv,Cheng:2016uqk}. The vector-like twin Higgs scenario~\cite{Craig:2016kue} resembles the fraternal in many respects, including the bottomonia phenomenology. In addition it has mixed flavor bound states in the twin sector that can also produce displaced vertices from exotic Higgs and heavy Higgs decays.

More mirror-like twin Higgs constructions can lead to visible collider signals of twin hadrons, such as exotic Higgs decays to twin sector mesons~\cite{Hochberg:2018vdo}. In the standard twin Higgs the decay of twin pions to SM states is suppressed by parity and twin isospin. If, however, a new portal between the sectors is introduced the twin pions can decay on cosmologically interesting time scales~\cite{Freytsis:2016dgf}.

By looking beyond the twin Higgs structure a rich variety of dark pion phenomenology is possible, including both prompt and displaced signatures~\cite{Cheng:2021kjg,Cheng:2024hvq}. Similarly, the tripled top framework~\cite{Cheng:2018gvu} can also include an intricate dark sector which includes many dark bound states that can be discovered at colliders. The SM neutral mesons of the dark sector can be produced from the $Z$ and Higgs bosons. Some of the mesons decay promptly back into SM states while others have a long lifetime~\cite{Cheng:2019yai}, providing many experimental opportunities for discovery.

The richness of these dark sectors leads to a wide variety of experimental signatures. Beyond what is outlined in this section, there are also possibilities for dark jets and dark showers. The connections of the dark sector particles to the SM can then lead to semi-visible jets or emerging jets. A further discussion of these phenomena is found in Sec.~\ref{ss.newHad}.

\subsection{Other Portals to SM Neutral Sectors}
While the Higgs is a robust portal to the hidden sector, there are several other interesting ways that the sectors may be linked. The most obvious is by SM gauge bosons, if the hidden sectors include states with SM gauge charges. These scenarios are considered in the following section, \ref{ss.newEW}.

In the case of SM neutral hidden sectors, like in the twin Higgs, the visible and hidden sectors can still be joined through several possible portals. These analyses have, so far, focused on the twin Higgs set-up, so the term twin sector is appropriate. When twin hypercharge is gauged then there may be a kinetic mixing with the SM hypercharge gauge field. Within the low-energy theory such mixing requires at least four-loops~\cite{Chacko:2005pe,Koren:2019iuv}, which can be consistent with unbroken twin hypercharge~\cite{Davidson:2000hf,Vogel:2013raa}. However, for kinetic mixing large enough to be relevant at colliders the twin hypercharge boson must have a mass. In this scenario both the twin photon and twin $Z$ can be discovered at current or future hadron colliders~\cite{Chacko:2019jgi}.

A right-handed neutrino extension of the SM can also act as a neutral fermionic portal to the twin sector~\cite{Bishara:2018sgl}. Such a connection can keep the sectors in thermal contact below the SM's QCD phase transition, which can alleviate cosmological constraints~\cite{Csaki:2017spo}. In some cases these interactions can also produce the $\mathbb{Z}_2$ breaking required to produce a hierarchy between the visible and twin sector Higgs VEVs. 

The Higgs, hypercharge, and neutrino portals rely on fields with discrete symmetry partners, like the twin Higgs field or twin hypercharge boson. However, if there are fields that the discrete symmetries take back to themselves, these can also serve as portals between the sectors. Such singletons\textemdash completely neutral fields with no $\mathbb{Z}_2$ partners\textemdash may be scalars, fermions, or vectors. Singleton models often include additional states which can be discovered at colliders using leptoquark and diquark searches as well as through meson decays~\cite{Liu:2019ixm}. The singletons themselves are often challenging to detect, but heavy vector singletons can be discovered in resonance searches~\cite{Bishara:2018sgl}. One example of a scalar singleton can arise when strongly-coupled UV completions of the twin Higgs include a spontaneously broken scaling symmetry. This produces a light dilaton which directly connects the visible and twin states~\cite{Ahmed:2019kgl} and is unaffected by the discrete symmetry that exchanges the two sectors.

\subsection{New SM Charged States\label{ss.newEW}}
Though the defining characteristic of neutral natural models is colorless top partners, specific realizations often include new states charged under the SM gauge groups. For instance, non-SUSY UV completions of the twin Higgs predict exotic quarks charged under SM color in the few to tens of TeV mass range, which can be thoroughly probed at the LHC and future colliders~\cite{Cheng:2015buv}. Other TeV scale states are SM color neutral but carry electroweak charge. If the discrete twin symmetry is preserved in the UV then these states may be out of reach of the LHC, but can be probed by future machines~\cite{Cheng:2016uqk}. It may also be that the new electroweak charged states are significantly lighter than the colored states. In this case the LHC can produce them at a meaningful rate. The twin color force draws the produced particles into bound states whose resonant decays can be discovered at the LHC~\cite{Cheng:2016uqk,Li:2017xyf}.

Supersymmetric models can also lead to novel electroweak states. Folded-SUSY sleptons~\cite{Burdman:2015oej} carry electroweak charges and can be discovered effectively at the LHC. In tripled-top constructions the $Z$ portal can be the dominant connection to hidden-sector states~\cite{Cheng:2019yai,Cheng:2021kjg} with electroweak charges. These states are often bound by the hidden color force, which in some cases leads to ``quirky'' dynamics, which we focus on in the following section.

\subsection{New Hadronic Signals\label{ss.newHad}}

\begin{figure}
    \centering
    \includegraphics[width=0.6\linewidth]{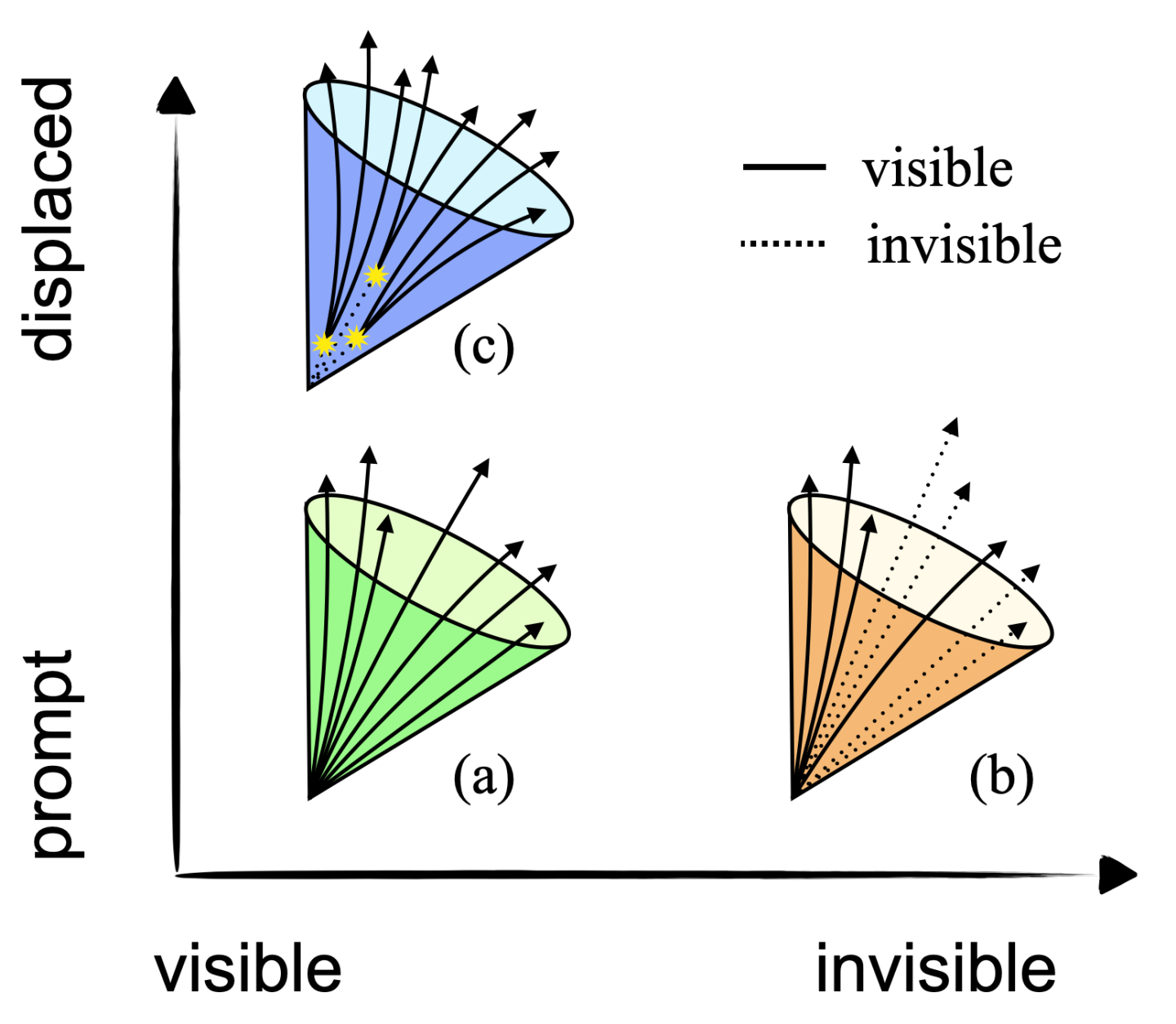}
    \caption{Schematic diagram differentiating the parameter space of dark showers into (a) dark jets, (b) semivisible jets, and (c) emerging jets.  Figure from Ref.~\cite{Collaboration:2931950}.}
    \label{fig:hadronic_signals}
\end{figure}

In the SM the dynamics of QCD that leads to the formation of jets is well established and routinely observed.  In a new, so far unobserved, strongly-interacting hidden sector a wide variety of behaviors are imaginable.  The most common approach is to assume that an analog of the parton shower occurs in the hidden sector which then results in different signals depending on how the hidden parton shower eventually produces visible particles.  This phenomenon is generically called dark showers and it is often connected to neutral naturalness~\cite{Barbieri:2005ri,Burdman:2006tz,Han:2007ae,Strassler:2008bv,Strassler:2008fv,Harnik:2008ax,Juknevich:2009ji,Juknevich:2009gg,Carloni:2010tw,Craig:2015pha,Barbieri:2015lqa,Cohen:2015toa,Garcia:2015loa,Kribs:2016cew,Forestell:2016qhc,Knapen:2016hky,Beauchesne:2017yhh,Cohen:2018mgv,Cheng:2018gvu,Renner:2018fhh,Beauchesne:2018myj,Alimena:2019zri,Cheng:2019yai,Beauchesne:2019ato,Batell:2020qad,Knapen:2021eip,Borsato:2021aum,Barron:2021btf,Kilic:2021zqu,Cheng:2021kjg,Beauchesne:2021qrw,Curtin:2022tou,Albouy:2022cin,Cheng:2024hvq}.

After the parton shower occurs in the hidden sector, at the hidden confinement scale, fragmentation is expected to form hidden sector mesons.  These hidden sector mesons could be stable, could decay within the hidden sector, or could decay to SM particles, either promptly or with a displacement.  It is convenient to subdivide the parameter space according to how the particles in the hidden sector decay back to the SM.  Figure~\ref{fig:hadronic_signals} schematically illustrates this classification by contrasting prompt decays with displaced decays and contrasting invisible decays with visible decays.  The distinction is also useful because it guides how the experimental search is approached and the expected backgrounds.

\begin{figure}
    \centering
    \vspace{-2.5cm}
    \includegraphics[width=0.6\linewidth]{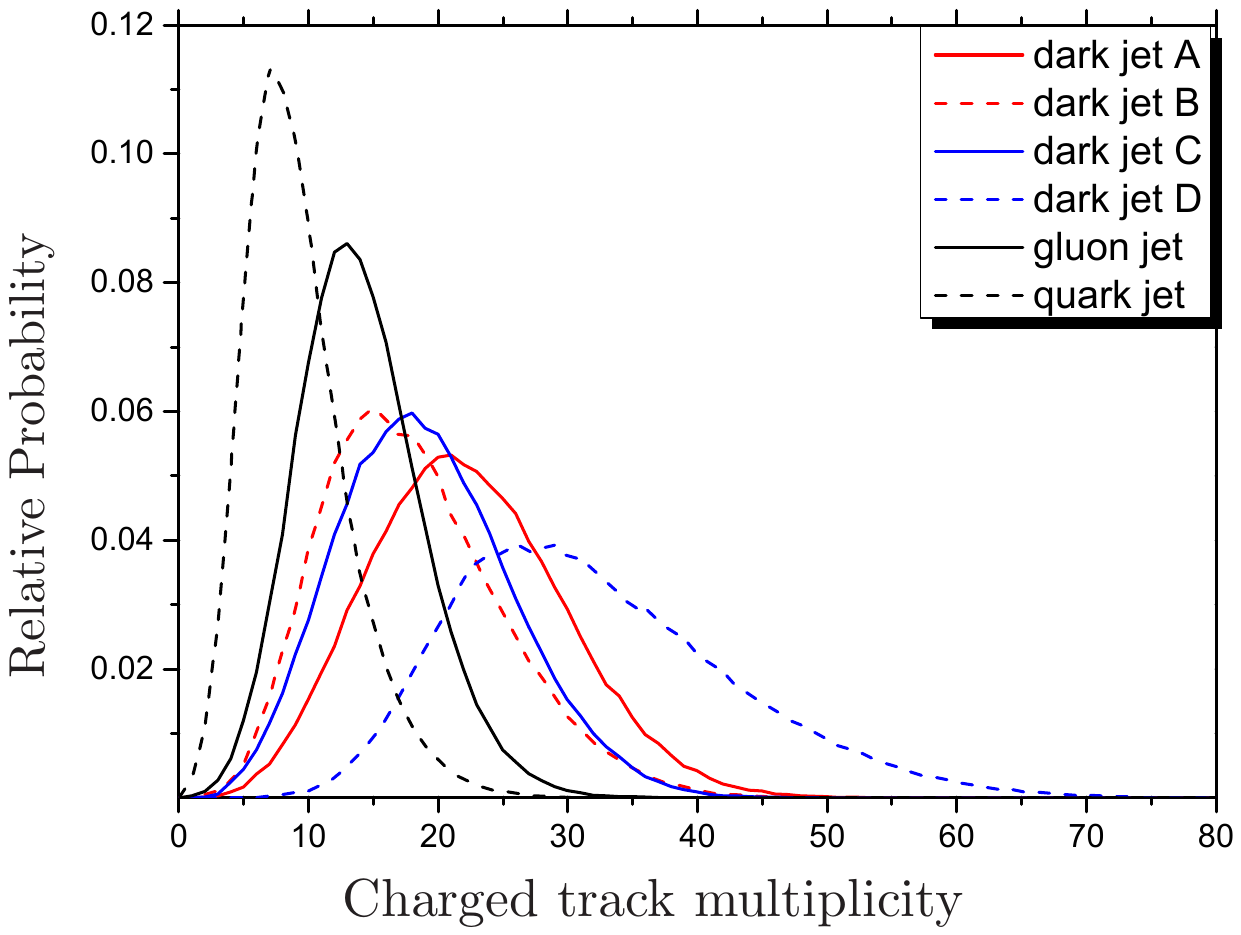}
    \vspace{-2.5cm}
    \caption{The charged track multiplicity of gluon jets (solid back), quark jets (dashed black), and for various models of dark jets (colored lines).  Models B and D are scenarios that could occur in neutral naturalness models.  Figure from Ref.~\cite{Park:2017rfb}.}
    \label{fig:charged_track_multiplicity}
\end{figure}

The first case of dark jets, corresponding to Fig.~\ref{fig:hadronic_signals}(a), occurs when all hidden sector particles decay promptly to the SM~\cite{Park:2017rfb}.  Such a dark jet would appear similar to a standard QCD jet with possible differences present in observables that probe the detailed structure of the jet.  Examples include the jet mass, the multiplicity of charged particles\textemdash called the charged track multiplicity\textemdash within the jet, and energy correlations of the jet~\cite{Park:2017rfb}.  The charged track multiplicity has been demonstrated to be useful in LHC searches~\cite{ATLAS:2023kao} and is shown in Fig.~\ref{fig:charged_track_multiplicity}.

The second possibility, shown in Fig.~\ref{fig:hadronic_signals}(b), is that some of the particles in the hidden sector shower result in stable hidden sector particles while others decay promptly back into SM particles.  The collider objects in this scenario are called semivisible jets~\cite{Cohen:2015toa,Cazzaniga:2022hxl} as only a subset of the total energy of the jet is visible.  Generally, in this situation the direction of the event's missing momentum is aligned with the direction of the hard jets.  One of the standard analysis cuts in collider studies is to veto events where the missing energy is aligned with hard jets because this signal can be imitated by mismeasuring a jet.  Analysis cuts should be carefully selected so as not to increase the mismeasurement background.

One of the key phenomenological quantities that characterizes a semivisible jets signal is $r_{\rm inv}$, which is the fraction of invisible particles in a jet
\begin{equation}
r_{\rm inv} = \left\langle \frac{\text{number of stable hadrons}}{\text{number of hadrons}} \right\rangle.
\end{equation}
Stable hadrons refers to dark sector hadrons that do not eventually decay into SM states.  The larger the value of $r_{\rm inv}$, the more missing energy is present in a collider signal. 

\begin{figure}
    \centering
    \includegraphics[width=0.8\linewidth]{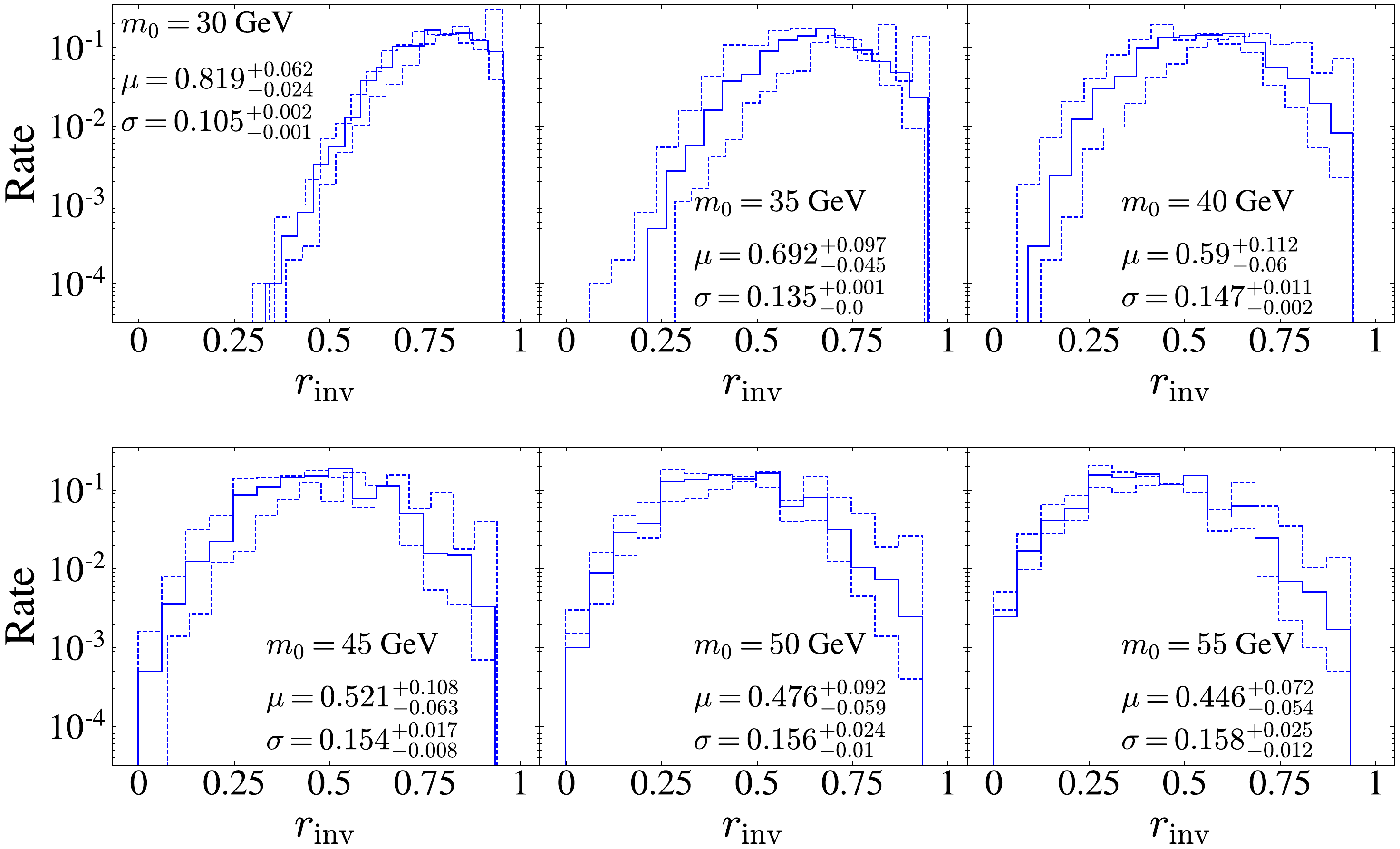}
    \caption{The distribution of $r_{\rm inv}$, the fraction of hidden sector hadrons that are not visible in a collider.  The mass $m_0$ of the lightest hidden sector glueball varies in each panel.  Figure from Ref.~\cite{Batz:2023zef}.}
    \label{fig:rinv_distributions}
\end{figure}

Such a scenario is possible in the fraternal twin Higgs model when the twin QCD sector does not contain any light quarks.  With only twin gluons below the twin QCD scale the hadron spectrum consists of twin glueballs and has been estimated in Ref.~\cite{Batz:2023zef}.  A subset of the twin glueballs are stable while another subset either decays via mixing with the SM Higgs or decays via the emission of an offshell SM Higgs, leading to the semivisible signal.  Figure~\ref{fig:rinv_distributions} shows the distribution of $r_{\rm inv}$ from a simulation of this hidden sector glueball model~\cite{Batz:2023zef}.  The various panels vary the mass $m_0$ of the lightest hidden sector glueball state.  A wide range of $r_{\rm inv}$ values can be generated in models of neutral naturalness.

Searches for semivisible jets have been performed both by ATLAS and by CMS.  The ATLAS collaboration has looked for semivisible jets originating from a $Z'$~\cite{ATLAS:2025kuz} and non-resonantly from a $t$-channel mediator~\cite{ATLAS:2023swa}.  The searches from the CMS collaboration include semivisible jets originating from a $Z'$ both via semivisible jets~\cite{CMS:2021dzg} and lepton-enriched semivisible jets~\cite{CMS:2025caz}.  They have published a comprehensive summary of all of their dark sector searches~\cite{CMS:2024zqs}.

The third scenario, corresponding to Fig.~\ref{fig:hadronic_signals}(c), is when all of the hidden sectors decay to SM, but with a displacement.  Since the parton shower in the hidden sector is initiated at the interaction point, the particles within the resulting jet point back to the interaction point, even though they only become visible at the point where they decay into visible SM particles. Therefore, the jet emerges within the detector. This is distinct from displaced decays where final particle trajectories do not point back to the interaction point.

\begin{figure}
    \centering
    \includegraphics[width=0.45\linewidth]{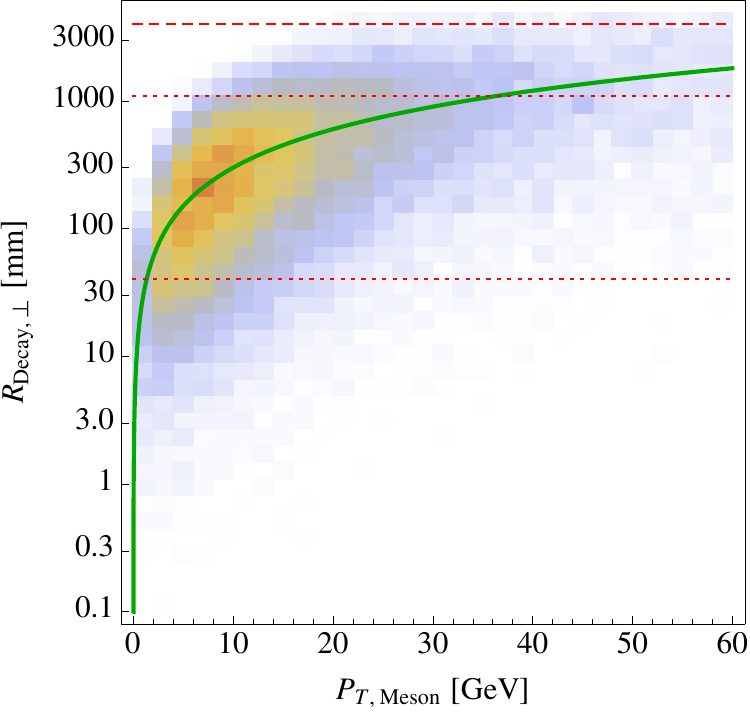}
    \caption{The distribution of transverse decay distances versus the transverse momentum of individual hidden sector hadrons in a representative model.  The green line shows average decay length in the laboratory frame.  Figure from Ref.~\cite{Schwaller:2015gea}.}
    \label{fig:decay_distance_emerging}
\end{figure}

Reference~\cite{Schwaller:2015gea} identified the case when most of the hidden sector particles decay into SM particles within the tracker as interesting and labeled this collider object as an emerging jet.  They suggested defining this collider object similarly to a jet with the added restriction that there should be not too many high momentum particles whose tracks originate close to the interaction point.  This would exclude standard QCD jets from this class but retain certain hadronic signals from a hidden sector.  An example of the distribution of transverse decay distances is shown in Fig.~\ref{fig:decay_distance_emerging}.

The emerging jet object typically contains hidden sector hadrons whose decay lengths are within the tracker. When the decay lengths are longer, such that the decay vertices that produce the SM particles are outside of the tracker, this signal is typically called a trackless jet~\cite{Bai:2011wy}. It has been suggested to look for energy deposits in the calorimeter without any associated tracks and with a relatively small electromagnetic energy fraction compared to typical QCD jets. Both ATLAS~\cite{ATLAS:2025bsz} and CMS~\cite{CMS:2024gxp,CMS:2024zqs} have performed searches for emerging jets.

Another possible signal that can arise in models where there is an additional confining sector is a soft unclustered energy pattern (SUEP).  This signal was first proposed in models where the new confining sector exhibits conformal dynamics~\cite{Strassler:2008bv}.  In these scenarios, it is expected that rather than a jet-like shape, SM particles that originate from dynamics within the hidden sector produce a more diffuse pattern. Without the traditional grouping of the signal into high-momentum objects, triggering on such scenarios is challenging~\cite{Knapen:2016hky}.  One possibility is to allow the system producing the SUEP to recoil off of an initially-radiated jet~\cite{DAgnolo:2019cio}.  Searches, using global event variables, have been performed by CMS~\cite{CMS:2024nca,CMS:2025azd,CMS:2025cwf}.  Machine learning is also expected to be especially helpful in these searches~\cite{Barron:2021btf}.

In the context of neutral naturalness, the SUEP signal can arise within the quirk parameter space (see Sec.~\ref{ss.quirks}).  In addition to the direct annihilation products of the quirk and antiquirk, there is often substantial soft radiation emitted as the quirk and the antiquirk accelerate and decelerate.  This soft radiation could be separated from the hadronic background at colliders using supervised machine learning~\cite{Curtin:2025ksm}.

\section{Neutral Natural Quirks\label{ss.quirks}}
Neutral natural models typically include a confining gauge group in the hidden sector. While the mirror twin Higgs model includes states at or below the hidden confining scale, this appears to be more the exception than the rule. In all frameworks with electroweak charged partner particles the bounds on new states are near the 100 GeV scale or above~\cite{Egana-Ugrinovic:2018roi}. However, the hidden strong coupling is related by a discrete symmetry to the SM strong coupling in the TeV range. This typically leads to a hidden confinement scale in the neighborhood of a few GeV. Models like these, with no particles charged under the confining gauge group at or below the confinement scale~\cite{Okun:1979tgr,Okun:1980mu}, are said to exhibit quirky~\cite{Carlson:1991zn,Kang:2008ea} dynamics. 

This situation is at once foreign and familiar. The SM includes a confining non-Abelian gauge theory in QCD. However, much of our intuition regarding this theory, including all we have gleaned from experiment, is built on the assumption of charged states below the confinement scale. As the intuition for quirky states is less general, we review many of their general properties in this section and emphasize the typical properties that arise in models of neutral naturalness. 

\subsection{Tubes of Color Flux}
It is well known that non-Abelian gauge theories often have a scale $\Lambda_c$ at which the perturbative coupling becomes large. This is sometimes called the confinement scale. At lengths larger than this scale, perturbation theory is not a good guide to the dynamics. However, we can make analytical predictions using nonperturbative insights. 

When a particle and an antiparticle experiencing a confining gauge theory are further apart than $\Lambda_c^{-1}$ they can be modeled as experiencing a confining, linear potential\cite{Wilson:1974sk,Eichten:1974af,Creutz:1980wj,Creutz:1980zw} 
\beq
V(r)\approx\sigma \,r~, 
\eeq
where the string tension $\sigma$ is related to the lightest glueball mass $m_0$ by $\sigma\approx m_0^2/(3.5)^2$~\cite{Lucini:2004my,Teper:2009uf}, corresponding to roughly $\sqrt{\sigma} \approx (0.5\ {\rm fm})^{-1}$ in SU(3) pure-gauge theory \cite{Teper:2009uf}. Then, the string tension can also be related to the confinement scale by taking $m_0\approx 7\Lambda_c$ and finding $\sigma\approx 4\Lambda_c^2$.  The term string tension is used because the gauge field configuration in this limit resembles a tube or string stretching from the particle to the antiparticle. 

This string produces real, physical effects. The potential leads to the force
\beq
F=-\frac{dV}{dr}=-\sigma~,
\eeq
that draws the particles together. This result also justifies labeling $\sigma$ as a tension, seeing it has dimensions of force.

This form for the potential and the characteristics of the field have been studied using lattice gauge theory. These efforts fill in additional specifics of this picture. For instance, the radius of the flux tube is roughly given by~\cite{Bicudo:2017uyy}
\beq
R\sim\frac12 \frac{1}{\sqrt{\sigma}}~.\label{eq:tubeR}
\eeq

The length of the string depends on the energy of the state. Suppose two particles, one in the fundamental representation of the gauge group and the other in the antifundamental, are produced at a given energy 
\beq
E=m_1+m_2+K~,
\eeq
where $K$ is the kinetic energy contained in the system. In the center of momentum frame, the particles move away from each other, but are subject to the constant force from the string. This eventually brings them to a stop, when the string has achieved a length $L$ defined by
\beq
K=\sigma L~.
\eeq
The particles then begin to accelerate toward each other. Their kinetic energy grows as the energy stored in the string decreases along with its length.

\begin{figure}
    \centering
    \includegraphics[width=0.9\linewidth]{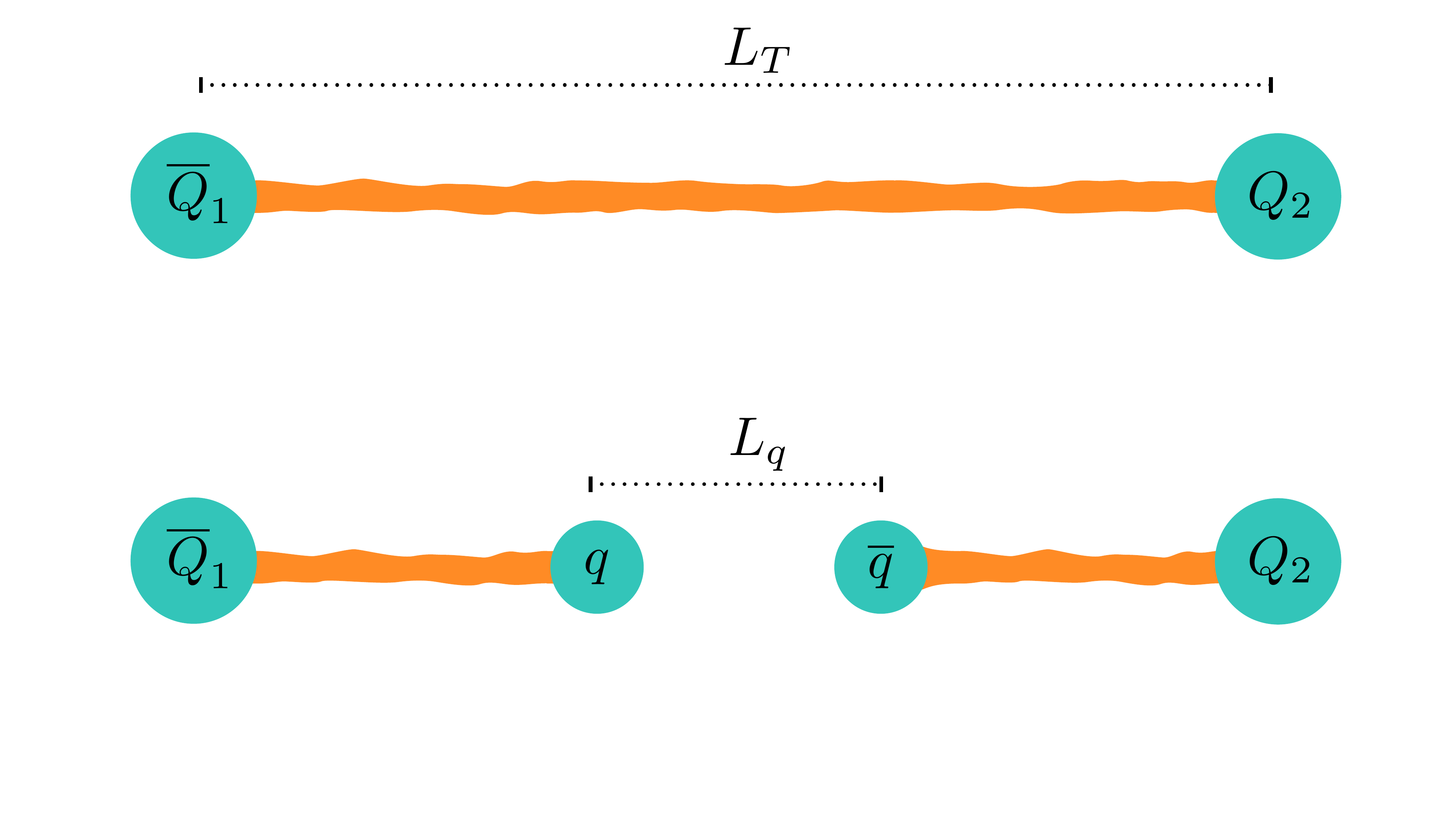}
    \vspace{-2cm}
    \caption{Schematic illustration of particles connected by tubes of color flux. In the lower illustration a pair of particles has been pulled from the vacuum to break the string.}
    \label{fig:quirks}
\end{figure}

This simple picture of gauge theories leads to a schematic understanding of jets. Consider the two field configurations shown in Fig.~\ref{fig:quirks}. The case shown at the top of the figure is simply the two particles $Q_1$ and $Q_2$ separated by some distance $L_T$ and connected by a tube of color flux. The lower part of the figure describes a similar situation, except a length $L_q$ of flux has been removed by the production of another particle-antiparticle pair. These particles (which need not be the same type as either $Q_1$ or $Q_2$) have been pulled from the vacuum in order to break the string.

What sets $L_q$? The production of the $q$ particles is related to their vacuum fluctuations. From the worldline formulation of quantum field theory (see for instance~\cite{Gelis:2019yfm}) the diameter of particle-antiparticle vacuum fluctuations scales like the inverse of their mass. Therefore, we expect
\beq
L_q\sim\frac{1}{m_q}~,
\eeq
where $m_q$ is the mass of the $q$ particle.

Let us compare the energy of the two configurations shown in Fig.~\ref{fig:quirks}. For the unbroken string we have
\beq
E_\text{unbroken}=m_1+m_2+L_T\sigma~.
\eeq
In the broken string case we have
\beq
E_\text{broken}=m_1+m_2+2m_q+(L_T-L_q)\sigma~.
\eeq
The difference in these energies is
\beq
E_\text{unbroken}-E_\text{broken}=L_q\sigma-2m_q\sim2\left(\frac{\Lambda_c^2}{m_q}-m_q \right)~.
\eeq

Roughly speaking, if there exist particles charged under the confining gauge group with masses well below the confinement scale, $m_q\ll\Lambda_c$ then the broken string configuration has lower energy than the unbroken state. In this scenario, after the $Q$ particles are produced we expect the string to fragment and the (perhaps heavy) $Q$ particles to end up bound to lighter $q$ particles. This is qualitatively what is observed in SM QCD. The up, down, and strange quarks have masses well below $\Lambda_\text{QCD}\approx$ 300 MeV~\cite{Karbstein:2018mzo}. Consequently, all color flux strings break and quarks pair produced at high energies lead to jets of quark bound states.

Nearly all known realizations of NN include new particles charged under a confining gauge group that is similar to QCD. These particles are thought to have masses in the hundreds of GeV, while the confinement scale is typically on the order of a few GeV. If all particles charged under a confining gauge group have $m_q\gg\Lambda_c$ then the unbroken string configuration is the lower energy state. This radically changes the dynamics of these states compared to our SM QCD intuition.

The discussion above is somewhat schematic. A more precise argument (following~\cite{Kang:2008ea}) can be made using Schwinger pair production~\cite{Schwinger:1951nm}. The nonperturbative rate, per volume, for the pair production of particles with mass $m_q$ in an electric field $\mathcal{E}$ is given by
\beq
\frac{\Gamma}{V}=\frac{\mathcal{E}^2}{4\pi^3}e^{-\pi m_q^2/\mathcal{E}}~.
\eeq
To take this result over to a non-Abelian gauge theory we use the fact that the field flux is confined to a tube. This allows us to relate the energy density per length ($\sigma$) to the energy density per volume ($\frac12\mathcal{E}^2$) by  $\sigma \approx \mathcal{E}^2 A$ to find production rate per length of flux tube
\beq
\frac{\Gamma}{L}\sim\frac{\sigma}{4\pi^3}e^{-\pi m_q^2/\mathcal{E}}~.
\eeq

By using the lattice result for the radius of the flux tube~\eqref{eq:tubeR} we find
\beq
\mathcal{E}\sim\frac{\sqrt{\pi}}{2}\sigma~.
\eeq
This leads to
\beq
\frac{\Gamma}{L}\sim\frac{\sigma}{4\pi^3}e^{-\frac{2m_q^2}{\sigma\sqrt{\pi}}}\approx \frac{\Lambda_c^2}{\pi^3}e^{-\frac{m_q^2}{2\sqrt{\pi}\Lambda_c^2}}~.
\eeq
The important point of this result is that for $m_q\gg\Lambda_c$ the pair production rate is exponentially suppressed. This agrees with our analysis above. If there are no states with gauge charge at or below the confinement scale then produced particles are confined by physical tubes, or strings, of color flux. The particles connected to these tubes have been dubbed `quirks.'

What are the dynamics of quirky bound states? A quirk produced in a higher energy collision begins to move with three-momentum $p_0$. As it moves it experiences a force
\beq
\frac{dp}{dt}=-\sigma~.
\eeq
This means that a quirk with mass $m_q$ has a velocity given by
\beq
v(t)=\frac{p_0-\sigma t}{\sqrt{m_q^2+\left( p_0-\sigma t\right)^2}}~,
\eeq
until the classical turning point ($v=0$) is reached at
\beq
t_\text{turn}=\frac{p_0}{\sigma}~.
\eeq
Assuming (for the moment) that no energy is lost, the total period of an oscillation\textemdash from production to tuning point, back through the production point to the turning point on the other side and back to the initial point\textemdash is $T=4t_\text{turn}$. The angular frequency of these oscillations is then
\beq
\omega=\frac{\sigma\pi}{2p_0}~.\label{eq:Qfreq}
\eeq

Of course, this classical analysis is only a rough guide to the quantum mechanical quirky bound state dynamics. For instance, there is a nonzero probability that the two quirks annihilate when passing by each other during one period of oscillation. This probability is decreased, however, by the radiation that we have so far neglected to include. 

\subsection{De-excitation and Decay}
Again taking classical physics as a guide, it is well known that an accelerating charge radiates according to the Larmor equation. For each quirk, the above trajectory leads to a radiated power of
\beq
\mathcal{P}_q= \frac{8\pi Q_q^2\alpha\sigma^2}{3m_q^2}~,\label{eq:RadPower}
\eeq
where $Q_q$ is the gauge charge (or representation factor) and $\alpha=g^2/(4\pi)$ is defined in terms of coupling $g$ associated to the gauge field that is carrying the radiation. As shown in~\cite{Forsyth:2025wks} this result also applies when relativistic dynamics are included. 

The quirks must, of course, couple to the gauge field that binds them. This seems to indicate a possible way to radiate energy. However, soft radiation into the confining gauge field does not escape to large distances. In order for energy to leave the system the energy in the gauge field must be able to form a glueball. The lightest glueball mass $m_0$ has been shown to satisfy $m_0\gtrsim \Lambda_c$~\cite{Lucini:2008vi,Lucini:2010nv}. For instance, for SU$(N)$ gauge theories $m_0\gtrsim6\Lambda_c$~\cite{Athenodorou:2021qvs} with comparable results for Sp$(2N)$~\cite{Bennett:2020qtj}. 

It seems that in order for the confining field to remove energy from the system the energy in the radiation needs to exceed $\Lambda_c$ by a factor of a few. However, the authors of~\cite{Harnik:2008ax} argue that the distribution of the radiation is sharply peaked at the classical frequency given in Eq.~\eqref{eq:Qfreq}, which we can write as
\beq
\omega\approx \Lambda_c2\pi\frac{\Lambda_c}{p_0}~.
\eeq
For quirks produced well above threshold $p_0\gg\Lambda_c$. This seems to indicate that it is difficult for very excited bound states to radiate glueballs, though it may be possible for quirks produced close to threshold.

One might suppose that a sufficient accumulation of softer radiation can combine into one glueball. The characteristic timescale to form a glueball is $\Lambda_c^{-1}$. So, we need the radiated power to satisfy
\beq
\mathcal{P}_q>\frac{m_0}{\Lambda_c^{-1}}~.
\eeq
This leads to the requirement
\beq
\frac{\sigma}{m_q^2}\alpha_s>\frac{3m_0\Lambda_c}{8\pi\sigma}~,
\eeq
which is generally far from satisfied, even if we take $\alpha_s=1$. For SU(3), for instance this relation becomes approximately
\beq
\frac{\Lambda_c}{m_q}\sqrt{\alpha_s}>\frac14~.
\eeq
It is not clear that states that satisfy this requirement are quirky, $m_q\gg\Lambda_c$.

Another possibility is that glueballs can be produced nonperturbatively when the quirks pass within a radius of $\Lambda_c^{-1}$ of each other. Within this region the radiated power might be larger as there is effectively a background of gluons\textemdash sometimes called brown muck\textemdash surrounding each quirk. When the brown muck around one quirk collides with the other quirk and its muck, one can imagine glueballs being produced. On the other hand, in heavy-quark effective theory the momentum carried by this gluon background is taken to be of order $\Lambda_c$~\cite{Manohar:2000dt}. Because even the lightest glueball is a factor of a few times this scale, it is not clear how often, if ever, this can produce an on-shell glueball.

In summary, glueball radiation may play a role in quirk bound state de-excitation, but for certain classes of quirks the size of that role is not clear. Some arguments have been made~\cite{Kang:2008ea,Cheng:2018gvu} in an attempt to quantify possible radiation rates using the form of hard gluon emission in Feynman diagrams. It is not obvious, however, that such estimates are a good guide to nonperturbative processes. The accurate modeling of this process in order to determine its effects on quirky bound states remains an interesting open question.

While not required by the minimal quirk scenario, often quirks also carry other charges, similar to the quarks of the SM which have both color and electroweak charges. If the constituent quirks have electric charge then they will radiate photons as they oscillate. Each quirk radiates power according to Eq.~\eqref{eq:RadPower}. The time it takes to de-excite from a state with kinetic energy $K$ to the ground state can be estimated as
\beq
t_\text{de-excite}\sim\frac{K}{\mathcal{P}_{q_1}+\mathcal{P}_{q_2}}~.\label{eq:deTime}
\eeq
The distribution of radiated photon energies is sharply peaked at the oscillation frequency given in Eq.~\eqref{eq:Qfreq}. If the quirk constituents do not have electric charge but couple to the $Z$ boson they can radiate light fermions through an off-shell vector~\cite{Cheng:2018gvu}. This process is much slower due to the suppression coming from the off-shell effects. 

Radiating vector particles does more than decrease the energy of the bound state. Each radiated vector changes the orbital angular momentum by one unit, either up or down. After radiating many particles the bound state will have randomly walked to large angular momentum. This nonzero angular momentum effectively prevents the constituents from becoming close enough to annihilate until they shed enough energy to reach their ground state. 

This can be understood schematically as discussed in~\cite{Kang:2008ea,Evans:2018jmd}. In the center of momentum frame, the four-momentum produced by the quirk annihilation is essentially the sum of the masses $m_{q_1}+m_{q_2}=m_T$. This implies that the quirks need to be within a distance of $m_T^{-1}$ of each other in order to annihilate. Therefore, the partial waves with $\ell\gtrsim k/m_T$ do not contribute to the annihilation cross section, where $k$ is the momentum transverse to the oscillation. In practice, this implies that the quirky bound state must completely de-excite to the $\ell=0$ ground state before it has an appreciable chance of annihilating. 

\subsection{Coulombic Regime}
Lower energy bound states are not modeled well by the linear potential. A Coulombic interaction~\cite{Eichten:1974af,Kribs:2009fy,Fok:2011yc,Harnik:2011mv} is better suited
\beq
V_c=-\frac{\text{C}_2\alpha_s(r^{-1})}{r}~,
\eeq
where $\text{C}_2$ is the quadratic Casimir of the gauge group. For SU$(N)$ theories, for instance, 
\beq
\text{C}_2=\frac{N^2-1}{2N}~.
\eeq
The ground state energy is given in terms of the reduced mass $m_\text{red}$ by the usual formula
\beq
E_\text{gs}=-\frac{m_\text{red}}{2}\text{C}_2^2\left[\alpha_s(a_0^{-1})\right]^2~,
\eeq
where the Bohr radius is given by
\beq
a_0^{-1}=\text{C}_2\alpha_s(a_0^{-1})m_\text{red}~.
\eeq

The value of the coupling at the Bohr radius can be obtained from the two-loop formula given in~\cite{Ellis:1996mzs}
\beq
\Lambda_c=\text{C}_2\alpha_s(a_0^{-1})m_\text{red}e^{-1/[2b_0\alpha_s(a_0^{-1})]}\left[\frac{b_1}{b_0}+\frac{1}{b_0\alpha_s(a_0^{-1})} \right]^{\frac{b_1}{2b_0}}~,
\eeq
where for SU$(N)$ gauge theories below the scale of any charged scalars or fermions
\beq
b_0=\frac{11 N}{12\pi}~, \ \ \ \ b_1=\frac{17 N}{22\pi}~.
\eeq
For a given hierarchy between the quirk masses and the confinement scale $\Lambda_c/m_\text{red}$, this determines the value of the coupling at the Bohr radius. The value of the coupling at other length scales can be found  by simply using the two-loop running of $\alpha_s$. In Table~\ref{tab:alphaInfo} we provide values of $\alpha_s\left(a_0^{-1}\right)$ for an SU(3) gauge group at several values of $m_\text{red}/\Lambda_c$. We see that the value of the coupling goes down as the hierarchy between the bound state particle masses and the confinement scale increases. Even if the ratio of scales is only an order of magnitude the coupling at the Bohr radius is still perturbative.

\begin{table}[th]
    \centering
    \begin{tabular}{|p{0.8in}||p{0.8in}|p{0.8in}|p{0.8in}|p{0.8in}|p{0.8in}|}\hline
        \begin{center}
            $\displaystyle m_\text{red}/\Lambda_c$
        \end{center} 
          & \begin{center}
              10
          \end{center} & \begin{center}
              $10^2$
          \end{center} & \begin{center}
              $10^3$
          \end{center} & \begin{center}
              $10^6$
          \end{center} & \begin{center}
              $10^9$
          \end{center}  \\ \hline
    \begin{center}
    $\displaystyle\alpha_s\left(a_0^{-1}\right)$
              \end{center} &\begin{center}
                  0.285
              \end{center}  &\begin{center}
                  0.147
              \end{center}  &\begin{center}
                  0.096 
              \end{center} &\begin{center}
                  0.046
              \end{center}  &\begin{center}
                  0.030
              \end{center}  \\ \hline
      \begin{center}
    $\displaystyle\Lambda_c^{-1}/a_0$
              \end{center} &\begin{center}
                  3.8
              \end{center}  &\begin{center}
                  20
              \end{center}  &\begin{center}
                  130
              \end{center}  &\begin{center}
                  6.1$\times10^5$
              \end{center}  &\begin{center}
                  4.0$\times10^7$
              \end{center}  \\ \hline
    \end{tabular}
    \caption{Numerical values for the value of the coupling at the Bohr radius and the ratio of the confinement length to the Bohr radius for various values of the ratio of the quirk masses to the confinement scale. We assume the confining gauge group is SU(3).}
    \label{tab:alphaInfo}
\end{table}

With these results we can also compare the Bohr radius with scale $\Lambda_c^{-1}$ at which the linear potential is expected to begin to dominate the dynamics. This is 
\beq
\frac{\Lambda_c^{-1}}{a_0}=\frac{m_\text{red}}{\Lambda_c}\text{C}_2\alpha_s\left(a_0^{-1}\right)~.
\eeq
Values of this ratio are also shown in Table~\ref{tab:alphaInfo} for an SU(3) gauge group. We see that for even a small hierarchy between the quirk masses and the confinement scale that the confinement length is somewhat larger than the Bohr radius. This indicates that the ground state is dominated by the Coulombic dynamics. For much larger hierarchies we see that the Coulombic regime can dominate up to very high bound state energies.

Because the ground state is always Coulombic, we can use the standard, nonrelativistic wavefunctions for the Coulomb potential to approximate it. Decay widths related to decay of these states depend on the squared magnitude of the bound state wavefunction evaluated at the origin. This is approximately given by
\beq
|\psi(0)|^2=\frac{1}{\pi a_0^3}~.
\eeq

\subsection{Quirks in Neutral Naturalness}
In most models of neutral naturalness the top partners transform under the fundamental representation of an SU(3) gauge group that is related to the SU(3$)_c$ of the SM by a discrete symmetry. If the partners of the SM quarks carry SM hypercharge and SU(2$)_L$ couplings then LEP searches imply their masses are greater than about 100 GeV~\cite{Egana-Ugrinovic:2018roi}. This causes the strong gauge coupling of the hidden sector to confine at higher energies, typically on the order of a few GeV~\cite{Curtin:2015fna}. Therefore, the particles in these new sectors have $\Lambda_c/m_q\lesssim10^{-2}$, putting them squarely in the quirky regime.

While variations from this simple expectation are certainly possible, we here discuss the aspects of quirky particles of this type. For instance, the radiated power~\eqref{eq:RadPower} into photons for this kind of quirk is 
\beq
\mathcal{P}_q\sim 7\times10^{-4}\left( \frac{300\text{ GeV}}{m_q}\right)^2\left( \frac{\Lambda_c}{3\text{ GeV}}\right)^4\text{ GeV}^2~.
\eeq
The de-excitation time that follows from this radiation is roughly
\beq
t_\text{de-excite}\sim 5\times 10^{-19}\left( \frac{K}{1\text{ TeV}}\right)\left( \frac{m_q}{300\text{ GeV}}\right)^2\left( \frac{3\text{ GeV}}{\Lambda_c}\right)^4 \text{s}~.
\eeq
This shows that de-excitation is typically quite prompt compared to collider timescales. In addition, the majority of photons produced are quite soft, with energies well below one GeV.

In the case of folded SUSY the folded squark bound states with net electric charge cannot decay into a pair of dark gluons. Consequently, these states decay into SM particles, making them the most attractive discovery channel in many circumstances. If the constituents have equal masses they dominantly decay to a $W\gamma$ final state~\cite{Burdman:2008ek}. Resonance searches at the LHC have been compared to these predictions for the both 8 TeV~\cite{Burdman:2014zta} and 13 TeV~\cite{Forsyth:2025wks} runs. The SM exhibits a cancellation in $W\gamma$ production which can be used to strengthen these searches~\cite{Capdevilla:2019zbx}. Displaced vertices from hidden sector glueballs can also be used to enhance experimental sensitivity in scenarios where the $W\gamma$ rate is suppressed~\cite{Forsyth:2025wks}.

These and other SM vector decay modes can be used to discover the quirks associated to other pNGB Higgs models as well, such as SO(7)/SO(6)~\cite{Ahmed:2020hiw}. While the final decay product is the most conspicuous signal of these quirky states, the pattern of radiation they produce as they de-excite can also lead to signals in the underlying event~\cite{Harnik:2008ax}. However, the softness of the photons, as shown above, may make the discovery of this pattern quite challenging.

\begin{figure}
    \centering
    \begin{fmffile}{quirkProd}
\begin{fmfgraph*}(250,150)
\fmfpen{1.0}
\fmfstraight
\fmfleft{p1,p2,i1,p3,p4} \fmfright{o1,o2,o3,o4,o5,o6,p5,o7,o8,o9,o10,o11,o12}
\fmfv{l= $Z,,\gamma$}{i1}
\fmf{photon,tension=1.6}{i1,v1}
\fmf{fermion,tension=0.8}{v1,v2} 
\fmf{fermion,tension=-0.3,right=0.3}{v2,v3}
\fmf{fermion,tension=0.8}{v3,v1} 
\fmf{gluon,tension=0.9,label.side=right,label=$g_B$,label.dist=5\thick}{v2,v4}
\fmf{gluon,tension=0.9,label.side=right,label=$g_B$,label.dist=5\thick}{v5,v3}
\fmf{plain,tension=0.7}{v4,o3}
\fmf{plain,tension=0.7}{v5,o10}
\fmfv{decor.shape=circle,decor.filled=full,decor.size=1.5thick}{v1}
\fmfv{decor.shape=circle,decor.filled=full,decor.size=1.5thick,l=$\overline{T}$,l.a=186,l.d=25}{v3} 
\fmfv{decor.shape=circle,decor.filled=full,decor.size=1.5thick,l=$T$,l.a=170,l.d=25}{v2} 
\fmfv{decor.shape=circle,decor.filled=30,decor.size=15thick}{v4} 
\fmfv{decor.shape=circle,decor.filled=30,decor.size=15thick}{v5}
\fmfv{l=$G_B$}{o3}\fmfv{l=$G_B$}{o10}
\fmffreeze
\fmf{plain,tension=0.7}{v4,o1}\fmf{plain,tension=0.7}{v5,o12}
\fmf{plain,tension=0.7}{v4,o2}\fmf{plain,tension=0.7}{v5,o11}
\fmf{plain,tension=0.7}{v4,o4}\fmf{plain,tension=0.7}{v5,o9}
\fmf{plain,tension=0.7}{v4,o5}\fmf{plain,tension=0.7}{v5,o8}
\end{fmfgraph*}
\end{fmffile}
    \caption{Illustration of the production of a heavy quirk-antiquirk pair that annihilates into a pair of hard hidden sector gluons. These gluons shower and hadronize into hidden sector glueballs.}
    \label{fig:glueShower}
\end{figure}

In addition to decays into SM vectors, the quirks or squirks have a significant branching fraction into hidden gluons. Thus, heavy quirks can produce showers of glueballs~\cite{Lichtenstein:2018kno,Curtin:2022tou,Batz:2023zef}, as shown in Fig.~\ref{fig:glueShower}. Some fraction of these glueballs, primarily the $0^{++}$ states, have displaced decays back into the SM, leading to striking signatures at current and upcoming colliders~\cite{Chacko:2015fbc,Forsyth:2025wks}. As the simulation of the showering and hadronization of these pure glue showers improves~\cite{Curtin:2022tou,Batz:2023zef} increasingly robust collider searches can be made. We also see that neutral naturalness provides a concrete motivation for exploring dark showers~\cite{Albouy:2022cin,Born:2023vll}. An important exception to this analysis was explored in~\cite{Barela:2023exp}. The authors showed that when there are large $A$-terms in folded SUSY that squirk-antisquirk bound states can dominantly decay to a di-Higgs final state, which allows prompt visible signals to be used for discovery.

The precise nature of the quirks can also play a significant role in determining bound state lifetimes. When the top partners are SM singlets or only carry SU(2$)_L$ charge the slow de-excitation can lead to displaced signals, as demonstrated in~\cite{Cheng:2018gvu} for some tripled-top models. While the quirk scenario is more general than neutral naturalness, these models point to a particularly motivated region of quirky models. As such, they serve as motivation for developing quirk collider searches~\cite{Farina:2017cts,Knapen:2017kly,Li:2019wce,Li:2020aoq,Curtin:2025ksm}. This includes recent work that has considered quirks with lower confinement scales which can lead to interesting signals at LHC far detectors~\cite{Li:2023jrt}, FASER~\cite{Li:2021tsy,Feng:2024zgp}, or the Forward Physics Facility~\cite{MammenAbraham:2024gun,FPFWorkingGroups:2025rsc}.

\section{Connections to Lattice Gauge Theory\label{s.lattice}}

A recurring theme in neutral naturalness models is the appearance of a ``QCD-like" SU(3) gauge group in a twin sector.  As with ordinary QCD, the hidden SU(3) is a confining gauge group, with many of its important properties determined at strong gauge coupling so that perturbation theory is often not a reliable tool.  If the twin SU(3) were a perfect copy of ordinary QCD, then lattice QCD calculations \cite{Kronfeld:2012uk, Davoudi:2022bnl} could be used to directly determine masses, matrix elements, and other important inputs to the rich phenomenology described elsewhere in this review.  Unfortunately, the real world cannot be $\mathbb{Z}_2$-symmetric, so further work is required to obtain lattice results relevant for a hidden SU(3) sector with rather different properties than ordinary QCD. Fortunately, there are a number of useful lattice results that can still be applied for neutral naturalness models, with room for further improvements with additional lattice calculations in the future.

\subsection{Repurposing Lattice QCD Calculations for Neutral Naturalness}

Here, we discuss how lattice QCD calculations, particularly calculations done at unphysically heavy quark masses, can be reinterpreted to give insights relevant for neutral naturalness models.  For a similar and more detailed discussion of reusing lattice QCD results for BSM phenomenology more broadly, see \cite{DeGrand:2019vbx}.

Numerical lattice gauge theory calculations work directly with the path integral; schematically, to find the expectation value of some operator $\mathcal{O}$,
\begin{equation}
\ev{\mathcal{O}} = \frac{1}{\mathcal{Z}} \int [DA] [D\psi_i]  \mathcal{O} e^{-S[A, \psi_i] (g_0, m_1, m_2, ...)}
\end{equation}
Here $\mathcal{Z} = \ev{1}$ is the partition function, and the expectation value is computed in Euclidean space-time signature.  In addition to the gauge field $A$ and quark fields $\psi_i$, the action depends on a number of free parameters, namely the (bare) gauge coupling $g_0$ and the quark masses $m_i$.

To evaluate $\ev{\mathcal{O}}$ numerically, the theory is discretized on a finite grid, introducing a lattice spacing $a$.  Then, Markov chain Monte Carlo methods are used to generate a set of lattice field configurations weighted by the action $e^{-S}$.  In practice, the quark fields are formally integrated over, so that only configurations of the gauge field $A$ are stored.  A simple average of $\mathcal{O}[A]$ evaluated on the gauge field configurations then gives a direct estimate of the desired $\ev{\mathcal{O}}$:
\begin{equation}
\ev{\mathcal{O}} \approx \frac{1}{N_A} \sum_{\{A\}} \mathcal{O}[A(g_0, am_1, am_2...)]
\end{equation}
where $N_A$ is the number of gauge-field configurations.  This outline of the standard procedure for numerical lattice calculations allows us to emphasize two important points that are relevant for applications to neutral naturalness.  

First, by way of the action, the lattice gauge configurations obtained in the Monte Carlo procedure depend on various free parameters.\footnote{Note that there may be additional parameters in the observable $\mathcal{O}$ itself that could be set separately; for example, $\mathcal{O}$ may contain ``valence'' quark-mass values that can differ from the ``sea'' quark-mass values appearing in the action, a technique known as ``partial quenching'' \cite{Bernard:1993sv,Sharpe:2006pu}.  We do not consider this possibility further here.} The set of parameters $\{g_0, am_1, am_2...\}$ are bare parameters, but through an appropriate choice of scale setting they can be traded for a set of physical parameters.  Specifically, we may think of $g_0$ as setting the overall scale $\Lambda_c$ of the QCD-like gauge theory through dimensional transmutation.  The quark masses are then determined as dimensionless ratios to this scale, $\{m_1/\Lambda_c, m_2/\Lambda_c, ...\}$.  The creation of these field configurations is computationally intensive, making it highly non-trivial to adjust the quark masses after the fact.  Thus, a typical lattice calculation presents results based on one or a handful of specific values of the quark masses; interpolation or extrapolation may be necessary to match the values corresponding to a particular twin SU(3) sector.

On the other hand, adjusting the value of the overall scale $\Lambda_c$ is trivial.  One way to think of this is that numerical lattice calculations can only predict dimensionless ratios, so that a single overall scale-setting input is always required.  For example, a twin SU(3) sector with a twin proton mass of $m_{p,B} \sim 5$ GeV would (up to differences in the quark-mass ratios $m_{q,B}/\Lambda_{c,B}$) be described by lattice QCD results scaled up by a ratio of $m_{p,B} / m_p \sim 5$.

Second, due to technical considerations, it is significantly less computationally expensive to generate gauge field configurations with heavier quark masses.  This means that particularly in earlier lattice QCD calculations when the available computing power was lesser and algorithmic developments were less sophisticated, many calculations were carried out with light quark masses that were significantly heavier than the physical up and down quark masses, and then extrapolated to the physical point.  As discussed above, phenomenological and model-building considerations often result in a hidden QCD sector in which the hidden quark masses are significantly larger.\footnote{This also tends to raise the hidden confinement scale $\Lambda_c$, since the quarks have an anti-screening effect, slowing the running of the strong gauge coupling; assuming that the QCD strong coupling and the twin strong coupling are set at a common high-energy scale, the hidden quarks being integrated out at higher energies leads to a higher $\Lambda_c$.}  Thus, older lattice calculations at heavier light-quark masses can provide useful and direct inputs to twin-sector physics.  A variety of such results are collected and tabulated in \cite{DeGrand:2019vbx}.

A related parameter to the set of $\{m_q / \Lambda_c\}$ is the number of dynamical light quarks, $N_f$; in modern lattice QCD calculations this is often written as ``$N_f = 2+1+1$", with $N_f = 2$ denoting the nearly massless up and down quarks and the two additional $+1$ denoting the strange and charm quark, which have increasingly small effects on the calculation results due to their larger masses.  (Schematically, only quarks with $m_q \ll \Lambda_c$ present in the action modify the gauge-field configurations significantly, so that heavier quarks are to a good approximation ``not dynamical".)

In the limit that the hidden-quark masses are very large, $m_q \gg \Lambda_c$, the strong-coupling dynamics of the hidden SU(3) essentially occurs in the ``pure-gauge" limit, $N_f = 0$.  Lattice calculations without dynamical fermions benefit from certain algorithms that can allow much higher computational efficiency, so that even older pure-gauge calculations have provided definitive results, for example on the SU(3) glueball mass spectrum \cite{Morningstar:1999rf}.  Pure-gauge results may be the most applicable for twin SU(3) sectors that are in this heavy-quark limit.  For a review of pure-gauge lattice calculations in the context of large-N expansion, see \cite{Lucini:2012gg}.  See also Ref.~\cite{Hatton:2021dvg} for studies of bottomonium which can be directly repurposed for use in neutral naturalness models with very heavy hidden-sector quark masses.

\subsection{Glueballs on the Lattice\label{ss.lattice.glueball}}

As highlighted in section \ref{sss.glueballs}, glueball states can be an important part of the phenomenology of neutral-natural sectors.  Some references to specific lattice calculations are given as part of that section; here, we reiterate and expand on lattice results for glueball properties.  Glueball lattice calculations can be extremely challenging in terms of signal-to-noise, especially in QCD where the glueballs are relatively heavy and unstable; as a result, the most precise glueball results are from pure-gauge ($N_f = 0$) calculations, which benefit from no light fermionic states and highly efficient algorithms for improved statistics.

The mass spectrum of glueball states in the pure-gauge limit has been very well-studied on the lattice, both for SU(3)~\cite{Morningstar:1999rf,Chen:2005mg} and for more general SU$(N)$~\cite{Athenodorou:2021qvs}, including extrapolation to the large-$N$ limit~\cite{Lucini:2004my,Lucini:2010nv}.  There have also been calculations for several Sp$(2N)$ groups~\cite{Bennett:2020qtj}, although this is of less direct relevance for existing neutral naturalness realizations (see the discussion in section~\ref{sss.glueballs}.)

\begin{figure}[th]
    \centering
    \includegraphics[width=0.8\linewidth]{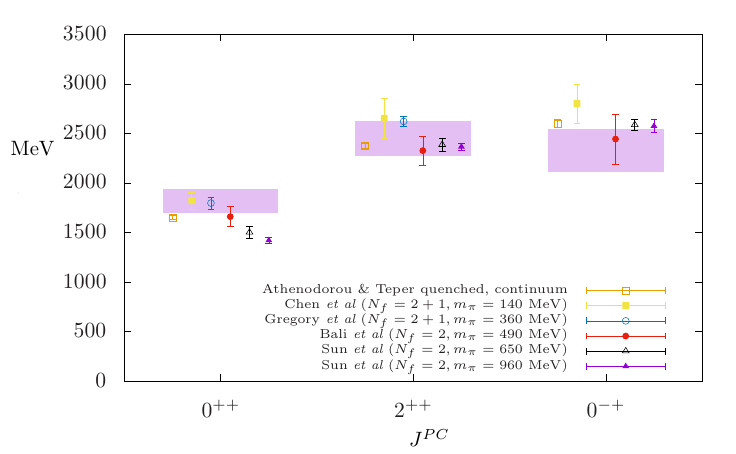}
    \caption{From~\cite{Athenodorou:2023ntf}, glueball spectrum of SU(3) gauge theory with $N_f = 4$ light quarks (purple boxes) compared to pure-gauge theory (``quenched", orange left-most points) and various other lattice calculations \cite{Bali:2000vr, Chen:2005mg, Gregory:2012hu, Sun:2017ipk, Athenodorou:2020ani, Chen:2021dvn} (other data points).  Results are shown for the ground-state glueballs with select indicated $J^{PC}$ quantum numbers.  Weak dependence on the number of active light quarks and on their masses (represented in the form of the pion mass) is evident from the comparison.}
    \label{fig:Nf4GBspectrum}
\end{figure}

While many of these pure-gauge results date back several years, there has been a renewal of recent activity in studying glueballs with dynamical light flavors for application to QCD, reviewed in \cite{Vadacchino:2023vnc,Morningstar:2024vjk}.  One broad conclusion from recent work on glueballs with light dynamical fermions, including for example \cite{Athenodorou:2023ntf} (see Fig.~\ref{fig:Nf4GBspectrum}), is that the presence of light quarks does not qualitatively impact the glueball spectrum compared to the pure-gauge case.  Although more precise answers could be obtained for specific theories of interest, this may indicate that the pure-gauge lattice results are a useful guide for the expected glueball spectrum in neutral naturalness models.

Lattice calculations can also provide matrix elements for glueball states, which are crucial inputs for determining their interactions and decays.  These are generically more difficult to study than the glueball masses, but some results are available.  Glueball-to-vacuum matrix elements, which determine glueball decay rates in processes with no glueballs in the final state, have been studied in \cite{Chen:2005mg,Meyer:2008tr}.  Recently, calculations of the radiative decay of a scalar glueball to a photon plus a strange-vector meson $G \rightarrow \gamma \phi$ have been presented in \cite{Zou:2024ksc}.  Gravitational form factors of glueball states (i.e.~matrix elements of the stress-energy tensor) have been studied in \cite{Abbott:2024bre}.  Finally, glueball scattering is sensitive to glueball-glueball interactions, but is extremely challenging to calculate. Recently, however, a first such calculation including the estimation of a trilinear glueball coupling was presented at a conference, as described in the review  \cite{Morningstar:2024vjk}.

\subsection{Baryon Properties and Interactions\label{ss.lattice.nucleon}}

The properties of mirror equivalents of baryons can be important inputs to phenomenology, particularly in the context of dark matter where dark atoms or dark nucleons can provide composite dark matter candidates (see Sec.~\ref{s.DM} below.)  Baryon lattice calculations can be challenging due to signal-to-noise problems at large time separation \cite{Lepage:1989hd,Beane:2010em}, and even more so for light nuclei, but as a consequence there are many calculations available at heavier-than-physical light-quark masses.  A number of such results for baryon masses are compiled in \cite{DeGrand:2019vbx}, including approximate formulas to allow for interpolation in the quark mass.

One particularly interesting feature of the baryon mass spectrum is the neutron-proton mass difference, which is highly sensitive to isospin-breaking effects, i.e.~the mass difference $m_d - m_u$ and electromagnetic effects sensitive to the different charges of the up and down quarks.  Small changes to these quantities in a mirror sector can result in interesting effects on the dark matter cosmology e.g.~\cite{Barbieri:2017opf,Chacko:2018vss}.  Lattice calculations of the neutron-proton mass difference \cite{BMW:2014pzb,Romiti:2022uyl} isolate the electromagnetic and quark-mass difference contributions and can be readily adapted to estimate the corresponding splitting for mirror protons and neutrons.

\begin{figure}[th]
    \centering
    \includegraphics[width=0.8\linewidth]{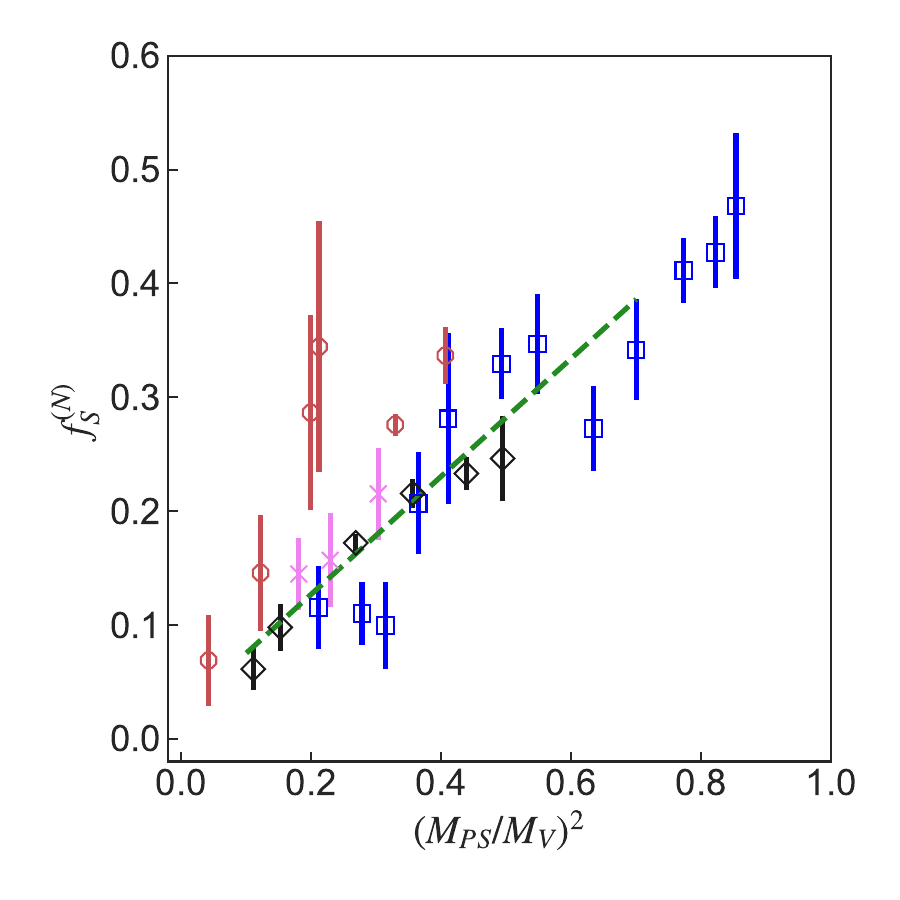}
    \caption{From~\cite{DeGrand:2019vbx}, dependence of the $N_f = 2$ nucleon sigma term $f_S^{(N)}$ as defined in the text on the light-quark mass, parameterized by the ratio $M_{PS}^2 / M_V^2$ (which would be $M_\pi^2 / M_\rho^2$ in QCD and is roughly linear in $m_q$, see the reference.)  Lattice data shown are from the reference and from Refs.~\cite{Walker-Loud:2008rui, PACS-CS:2008bkb, EuropeanTwistedMass:2008pab, Jansen:2009hr, ETM:2009ztk}.  A clear linear trend versus light-quark mass is evident.}
    \label{fig:latticeSigmaTerm}
\end{figure}

Other baryon matrix elements can also be computed for interactions and decays.  For example, the nucleon ``sigma term'' describes the coupling of nucleons to the scalar quark current; for quark species $q$ and nucleon $N$, a conventional definition which is renormalization-group invariant is
\begin{equation}
f_q^{(N)} \equiv \frac{1}{M_N} \langle N | m_q \bar{q} q | N \rangle.
\end{equation}
For the QCD neutron and proton, these matrix elements for the light $u,d,s$ quarks determine the Higgs coupling of the baryons, and is a crucially important input for Higgs-portal dark matter direct detection.  In the context of neutral naturalness, the dark-sector analogue of the sigma term appears on the Higgs coupling for mirror baryons.  Figure~\ref{fig:latticeSigmaTerm} shows a compilation of lattice results from \cite{DeGrand:2019vbx} for the nucleon sigma term with $N_f = 2$ light quarks as a function of the quark mass, showing a clear and regular trend from which results suitable for phenomenology can be taken.

Finally, the binding energy of light nuclei can be an important input for cosmology and dark matter in some neutral naturalness models in which mirror atoms or mirror nuclei are significant.  Direct lattice calculations beyond the very smallest nuclei are extremely challenging, but some early lattice results for the two-nucleon system were obtained at heavier than physical light-quark masses in \cite{NPLQCD:2012mex,Yamazaki:2012hi, NPLQCD:2013bqy, Yamazaki:2015asa, Berkowitz:2015eaa, NPLQCD:2020lxg}, indicating the existence of bound di-nucleon states at these masses.  However, these calculations are difficult and the dependence on choice of interpolating operators and other technical details can be significant; more recent calculations in \cite{Horz:2020zvv,Amarasinghe:2021lqa} have cast doubt on whether or not bound two-nucleon states actually exist at heavy quark masses.  Further references and a review of the present status of these studies can be found in \cite{Detmold:2024iwz,BaSc:2025yhy}.  For the purposes of phenomenology, the question of whether two-nucleon states are bound in SU(3) at heavy quark masses should be viewed as unsettled.

\subsection{String Tension and Confinement\label{ss.lattice.confinement}}

Lattice calculations can give helpful insights into the physics of confinement in QCD and beyond.  One such quantity which can be highly important for early-universe physics in a sector with a mirror version of QCD is the critical temperature $T_c$ at which confinement occurs.  Finite-temperature lattice calculations can also probe the order of the confinement transition; a strong first-order confinement transition in the early universe from a mirror sector could give rise to primordial gravitational waves (see Sec.~\ref{ss.PT} below.)

In the pure-gauge limit ($N_f = 0$), the confinement transition is well known to be first-order \cite{Iwasaki:1991pc,Boyd:1996bx}.  On the other hand, for physical QCD the transition is known to be a crossover (see \cite{Schmidt:2025ppy} for a recent review.)  More recent lattice work studying the order of the confinement transition as a function of the number of massless flavors \cite{Cuteri:2021ikv,Klinger:2025xxb} has shown evidence that the transition is second-order or crossover in the massless limit up to $N_f = 7$.  Lattice results also show a clear trend for the critical temperature to decrease somewhat relative to other hadronic scales as $N_f$ increases; for example, recent lattice calculations indicate a confinement temperature for $N_f = 2$ in the massless-quark limit of around $T_c \sim 130$ MeV \cite{HotQCD:2019xnw,Kotov:2021rah}, while for $N_f = 3$ the temperature reduces to $T_c \sim 100$ MeV \cite{Dini:2021hug}.  For larger values of $N_f$, results shown and/or compiled in \cite{Lombardo:2014fea,Klinger:2025xxb} indicate this trend continuing to larger values of $N_f$.  Although the dependence is relatively mild, it may be helpful to incorporate lattice results as a quantitative input for $T_c$ in the context of neutral naturalness models.

In the limit that their masses are very large, the properties of the heavy ``quirk'' states can also be studied in pure-gauge theory as static quarks.  The string tension $\sigma$, which characterizes the potential between static quarks, has been studied in pure gauge theory~\cite{Athenodorou:2021qvs} and in the large-$N_c$ context by~\cite{Lucini:2005vg,Lucini:2012wq}.  Away from the pure-gauge limit, the presence of light quarks can reduce the string tension significantly, as seen for example in \cite{Athenodorou:2023ntf} with $N_f = 4$.  In addition, light quarks allow ``string breaking'' to occur (as described in Sec.~\ref{ss.quirks}) due to pair production; lattice studies of this phenomenon include \cite{Bali:2005fu,Bulava:2019iut} for $N_f = 2$ with somewhat heavier than physical light-quark masses.

\subsection{Other Quantities and Other Gauge Theories\label{ss.lattice.other}}

One of the advantages of lattice calculations is that many observables can be investigated using a single Markov chain of gauge-field configurations.  For specific quantities of interest not listed here, ongoing communication between the lattice and pheno fields will ensure that lattice calculations can give nonperturbative inputs where they are needed.

Lattice calculations of parton distributions are maturing rapidly, and may be of interest in the context of modeling hadronization effects in a hidden sector.  While we are not aware of direct application of these results to hidden-sector hadronization thus far, see \cite{Constantinou:2022yye,Lin:2025hka} for recent reviews of lattice parton distribution calculations.

Lattice calculations can also be used to study the ultraviolet completions of composite twin Higgs or other pNGB Higgs neutral naturalness models.  There is already a substantial literature on lattice calculations targeting composite Higgs models, see~\cite{Svetitsky:2017xqk,Witzel:2019jbe,Drach:2020qpj,Lee:2024xhb} for recent reviews.  This connection is most readily made for purely gauge-fermion UV completions, such as those discussed in~\cite{Batra:2008jy,Serra:2017poj}; extensions of work on fermionic UV completions for general composite Higgs theories~\cite{Barnard:2013zea,Ferretti:2013kya} to the case of composite neutral naturalness could be particularly interesting.

Lattice calculations for other gauge groups aside from SU$(3)$ could also be of interest for application to neutral naturalness models.  As discussed in Sec.~\ref{sss.glueballs}, models involving gauge-symmetry breaking can realize different  hidden-sector gauge groups, with concrete examples of SU(4), SU(2), or SO(3) gauge groups already being proposed \cite{Batell:2025iwm, Batell:2020qad}.  Lattice calculations for composite Higgs, composite dark matter, or other BSM targets have studied each of these theories before; lattice calculations which may be of interest include the following for SU(4) gauge \cite{LSD:2014obp,DeGrand:2015lna,Appelquist:2015zfa,Ayyar:2017qdf,Ayyar:2018ppa,LatticeStrongDynamicsLSD:2023vtk}, SU(2) gauge \cite{Bursa:2011ru,Hietanen:2013fya}, SO(3) gauge \cite{deForcrand:2002vs,Barresi:2003yb,Barresi:2004qa} and SO(4) gauge \cite{Hietanen:2012sz} in four spacetime dimensions.  Various other references for lattice calculations in different theories (including some other choices of gauge groups) may be found in the review articles noted above.

\section{Connections to Dark Matter\label{s.DM}}

The bulk of neutral naturalness dark matter studies stem from the twin Higgs framework. We divide these into those more closely aligned to the mirror set-up and those that assume a fraternal starting point. After discussing twin Higgs dark matter we describe some of the other possibilities that have begun to be explored. 

\subsection{Mirror Twin Higgs Dark Matter}

As described in Sec.~\ref{s.models} there are a variety of possible particle spectra in the twin sector and consequently many possibilities for stable twin particles. With the correct interactions and abundance, a stable twin particle could comprise part or all of the dark matter. In the fully $\mathbb{Z}_2$ symmetric mirror twin Higgs model the list of stable twin particles includes the twin photon, the twin electron, twin neutrinos, and twin atoms. The twin photon, twin electron, and twin neutrinos are not sufficiently cold, while self-interactions of twin atoms are too large to make up the dark matter.  Thus, in addition to the model building reasons for breaking the $\mathbb{Z}_2$ symmetry, a viable dark matter candidate also requires $\mathbb{Z}_2$ breaking.

In the mirror twin set-up, if the twin neutron were lighter than the twin proton, it would be stabilized by twin baryon number, yielding a viable dark matter prospect. Its abundance can be set by an asymmetry in the twin baryons making the twin neutron a natural asymmetric dark matter candidate~\cite{Farina:2015uea,Barbieri:2016zxn,Barbieri:2017opf}.  The asymmetry in baryon number and twin baryon number could have a common origin~\cite{Farina:2016ndq,Kilic:2021zqu}.  In the simplest case, the twin neutron $n_B$ would have a mass~\cite{Farina:2015uea}
\begin{equation}
\label{eq:aDM}
m_{n_B} \approx 
\bigg(\frac{\Omega_{\rm DM}}
{ \Omega_{\text{baryon}}} \bigg)
m_n \approx 5~{\rm GeV}.
\end{equation}
If one allows the masses of the twin up quark, twin down quark, and twin electron to freely vary, then the twin neutron, twin Hydrogen atom, and twin Helium atom are all possible dark matter candidates, depending on the hierarchy among masses~\cite{Barbieri:2016zxn,Barbieri:2017opf}.  When the twin quark masses are taken above the twin QCD scale the preferred mass range for the dark matter is 1\textendash30~GeV~\cite{Barbieri:2016zxn}.  Obtaining a twin neutron that is lighter than the twin proton can be accomplished by expanding the Higgs sector to a mirror twin two Higgs doublet model~\cite{Beauchesne:2020mih} or adding a pair of vector-like quarks with up-type quantum numbers~\cite{Beauchesne:2021opx}.

An alternative possibility for asymmetric dark matter is to compensate for the factor of $\Omega_{\rm DM} / \Omega_{\text{baryon}} \approx 5$ with a factor of $|\sin\tilde{\phi} / \sin\phi| \approx 5$ which can be generated by a relative phase between the SM sector couplings and the twin sector couplings~\cite{Bittar:2023kdl}, where $\phi$ and $\tilde{\phi}$ are the effective phases of the SM sector Higgs couplings and the twin sector Higgs couplings respectively.  In such a model, the mass of the dark matter is predicted to be
\begin{equation}
\label{eq:aDM-CP}
m_{n_B} \approx 
\bigg(\frac{\Omega_{\rm DM}}
{ \Omega_{\text{baryon}}} \bigg)
\bigg| \frac{\sin\phi}{\sin\tilde{\phi}} \bigg|
m_n \approx 
m_n.
\end{equation}
This allows the twin baryon mass to be comparable to the SM baryon mass which avoids the requirement of substantial $\mathbb{Z}_2$ breaking between the QCD sectors.  The relative phase between the SM and twin couplings does not affect any naturalness considerations~\cite{Bittar:2023kdl}.  Typically, in asymmetric dark matter models, signals of indirect detection are very small since the symmetric component of dark matter is largely absent.  Sizable indirect detection signals, however, may be generated in the presence of a population of primordial black holes~\cite{Agashe:2024dkq}.  These black holes act as a source for all particles in the spectrum, one of which may induce interactions among the dark matter that leads to the indirect detection signal.

Whether the dark matter is composed of twin neutrons, twin Hydrogen, or twin Helium, it would interact with visible nucleons via the Higgs portal.  Above 10 GeV the cross section is detectable with $\sigma_{\rm DD} \gtrsim 10^{-47}~{\rm cm}^2$~\cite{Barbieri:2016zxn,Barbieri:2017opf}, while below a mass of 10 GeV this cross section falls below the neutrino floor~\cite{Farina:2015uea}.  For these low masses, introducing kinetic mixing between twin hypercharge and visible hypercharge allows for additional interactions between the dark matter and visible nucleons.  For a kinetic mixing of $\epsilon \sim 10^{-11}$ twin Hydrogen or twin Helium dark matter could be detected through nuclear recoils and twin electrons could be detected through electron recoils~\cite{Chacko:2021vin}.  Kinetic mixing also allows for the possibility that the twin Hydrogen accumulates in the center of stars.  Even in small amounts, the twin Hydrogen could modify the luminosity function of white dwarfs at observable levels~\cite{Curtin:2020tkm}.

Other than the twin neutron, there are a few more exotic possibilities that exist within the mirror twin Higgs model.  If the masses of the twin leptons and twin photon are raised, the lightest twin particle becomes the twin pion.  When the twin pion has a mass of a few hundred MeV the correct relic abundance can be achieved via the strongly interacting massive particle (SIMP) mechanism~\cite{Hochberg:2018vdo}.  Alternatively, the mirror electron, at a few MeV, could also be the dark matter with an abundance set by freeze-in if a kinetic mixing of $\epsilon \sim 10^{-12}$ between the visible and twin hypercharges is present~\cite{Koren:2019iuv}.

As discussed in Sec.~\ref{ss.UVcomp} at energies above the $\sim 10~{\rm TeV}$ scale the twin Higgs model needs to be UV completed.  The most common UV completions employ supersymmetry or a composite Higgs.  In a supersymmetric UV completion, in addition to the visible electroweakinos, the twin electroweakinos are also possible dark matter candidates.  One example is the twin bino with its freeze-out abundance determined by annihilation to twin fermions~\cite{Badziak:2019zys}.  The twin stau is another candidate with analogous annihilation channels to the visible stau~\cite{Badziak:2022eag}.  Twin supersymmetric particles as dark matter typically retain many of the appealing features of the corresponding supersymmetric visible sector particles as dark matter candidates, but with weaker constraints from direct detection~\cite{Badziak:2023gpz}.

\begin{figure}
    \centering
    \begin{fmffile}{stauDecay}
\begin{fmfgraph*}(175,100)
\fmfpen{1.0}
\fmfstraight
\fmfleft{p1,p2,i1}\fmfright{o1,o2,o3}
\fmfv{l= $\widetilde{\tau}_{RA}$}{i1}
\fmfv{l=$\widetilde{\tau}_{1B}^\ast$}{o1}
\fmfv{l=$\tau_{RB}$}{o2}
\fmfv{l=$\tau_{RA}$}{o3}
\fmf{dashes,tension=1.2}{i1,v1}
\fmf{fermion,tension=0.33}{v1,o3}
\fmf{fermion,tension=0.33,label.side=left,label=$\widetilde{H}_A$,,\;$\widetilde{B}_A$}{v2,v1}
\fmf{fermion,tension=0.33,label.side=right,label=$\widetilde{H}_B$,,\;$\widetilde{B}_B$}{v2,v3}
\fmf{fermion,tension=0.3}{v3,o2}
\fmf{dashes,tension=0.4}{v3,o1}
\fmfv{decor.shape=circle,decor.filled=full,decor.size=1.5thick}{v1}
\fmfv{decor.shape=cross,decoration.angle=-30,decor.size=5thick}{v2}
\fmfv{decor.shape=circle,decor.filled=full,decor.size=1.5thick}{v3}
\end{fmfgraph*}
\vspace{0.5cm}
\end{fmffile}
    \caption{Visible ($A$) sector right-handed stau decay into visible and twin ($B$) sector taus and twin stau. The decay is mediated by either the Higgsinos or binos of the two sectors. }
    \label{fig:stauDecay}
\end{figure}

\subsection{Fraternal Twin Higgs Dark Matter\label{ss.fraternalDM}}
The fraternal twin Higgs model automatically includes several dark matter candidates. In this model there is only a single generation of twin leptons, the twin $\tau$ and twin $\nu$, at least one of which is stabilized by twin lepton number.  When twin hypercharge is not gauged, the twin $\tau$ freezes out via interactions mediated by the twin $W$ and is a viable weakly interacting massive particle (WIMP) for masses between roughly 60 GeV and 150 GeV~\cite{Garcia:2015loa,Craig:2015xla}. If one allows for the twin $\nu$ mass to be comparable to the twin $\tau$, then the dark matter becomes multicomponent, consisting of the twin $\tau$, twin $\nu$, and possibly also the twin $W$~\cite{Garcia:2015loa}. Weakly gauging twin hypercharge opens up another annihilation channel for the twin $\tau$ which lowers the preferred mass range to about 1\textendash20~GeV~\cite{Craig:2015xla}.  In the fraternal twin model the twin $\tau$ unavoidably interacts with the Higgs leading to tension with current direct detection bounds~\cite{LUX:2016ggv,PandaX-II:2017hlx,XENON:2018voc}.  These bounds can be avoided in models where twin hypercharge is spontaneously broken by additional hypercharged scalars. This allows for new Yukawa couplings involving the hypercharge scalar that dynamically generate contributions to the twin $\tau$ mass, thereby decoupling it from the direct-detection cross section and allowing rates below the neutrino floor~\cite{Curtin:2021spx}.

Small variations on the fraternal twin Higgs set-up can lead to interesting experimental signals and model building. If the first and second generation of twin quarks are re-introduced, the lightest twin particles can be long-lived pion-like mesons. The annihilation of the twin $\tau$s in indirect detection leads to a QCD-like shower within the twin QCD sector ending either with stable twin pions or visible photons~\cite{Freytsis:2016dgf}.  Alternatively, if twin electromagnetism is broken, the twin photon obtains a mass and the twin $\tau$ and twin $\nu$ can mix.  After this mixing, the dominantly twin $\nu$ state can be the dark matter, between 0.1 and 10 GeV, where the relic density can be set by either co-annihilation or co-scattering, depending on the mass of the twin $\tau$ and twin photon~\cite{Cheng:2018vaj}.

With only a single generation of twin quarks, the lightest twin baryon is the spin$-3/2$ baryon $\Delta_B$ with the quantum numbers of three twin $b$ quarks $(b_B b_B b_B)$.  When this state is lighter than the twin $\tau$, it is a viable asymmetric dark matter candidate both when $m_{b_B}$ is parametrically above the twin QCD scale~\cite{Terning:2019hgj} or below the twin QCD scale~\cite{Garcia:2015toa}.  When twin hypercharge is gauged, the lightest neutral state is a twin atom which consists of a $\Delta_B$ bound with a twin $\tau$. This $\Delta_B$ asymmetry and the visible baryon asymmetry can be generated simultaneously, provided there is a sufficiently large kinetic mixing~\cite{Feng:2020urb}.

With a little additional structure, the fraternal twin Higgs model contains the correct moving pieces to implement either partially acoustic dark matter (PAcDM) or quasi-acoustic dark matter (QuAcDM)~\cite{Prilepina:2016rlq}. These scenarios are motivated by their potential to alleviate the longstanding $\sigma_8$ and $H_0$ tensions.  For PAcDM, the dark matter consists of a component that acts like traditional cold dark matter and a second sub-dominant component that interacts efficient with dark radiation.  This interaction suppresses the matter power spectrum at small scales.  The twin baryon $\Delta_B$ acts as the cold dark matter while the twin $\tau$ can comprise the sub-dominant dark matter component provided that the twin U(1$)_L$ is gauged to mediate efficient interactions between the twin $\tau$ and twin $\nu$.  QuAcDM, on the other hand, allows all species of DM to couple, slightly inefficiently, to dark radiation, also suppressing the matter power spectrum at small scales.  Gauging U(1$)_{B-L}$ realizes this set-up where the $\Delta_B$ and twin $\tau$ couple to the twin $\nu$ via the weakly gauged U(1$)_{B-L}$.

\subsection{Exotic Dark Matter\label{ss.exoticDM}}

Additional particles introduced into twin models can also be the dark matter.  Adding a singlet scalar $S$ that couples to both Higgs doublets via $S^2 (|H_A|^2 + |H_B|^2)$ adds a scalar portal on top of the twin Higgs model.  The dark matter in this case is the scalar $S$ and the twin mechanism, in addition to its usual advantages, now additionally lowers the direct detection cross section, re-opening viable masses for the $S$ roughly above 100 GeV~\cite{Curtin:2021alk}.

A twin sterile neutrino is a possible dark matter candidate.  Ref.~\cite{Holst:2023hff} considered one species of sterile neutrino with an abundance generated by the Dodelson-Widrow mechanism~\cite{Dodelson:1993je},  wherein sterile neutrinos are generated through oscillations with active neutrinos in the primordial plasma.  Similar to Ref.~\cite{Chacko:2016hvu}, the sterile neutrino of the SM sector is used to asymmetrically reheat the SM and twin sectors, however, in the model of Ref.~\cite{Holst:2023hff} there is substantial $\mathbb{Z}_2$ breaking among the sterile neutrinos that lets the twin sterile neutrino mass be $\sim 100~{\rm keV}$.

Twin models have also been proposed where the twin color SU(3) is broken to SU(2).  This breaking causes the twin quarks to split into an SU(2) doublet and SU(2) singlet.  The SU(2) singlet that results from the twin $b$ quark is a viable asymmetric dark matter candidate~\cite{Kilic:2021zqu}.  Another variant of color groups is that both the visible and twin sectors start with a color group of SU(4)~\cite{Batell:2025iwm}.  In the visible sector color is spontaneously broken to SU(3) while in the twin sector it remains as SU(4).  If there are stable twin baryons that result from the twin SU(4) these could form part of the dark matter.

Dark matter has been explored in neutral naturalness models other than the twin Higgs.  The dark top model cancels the leading top quark divergences with 3 species of a fermionic top partner~\cite{Poland:2008ev}.  In the twin model there are 3 species of top partner due to the twin SU(3) group while in dark top models other group structures are possible.  One example is the coset SU(6$)_c \times \text{SU}(3)_L / \text{SU}(6)_c \times \text{SU}(2)_L$ where the left-handed top quark can be embedded in a $({\bf 6}, {\bf \bar{3}})$ of the global group which contains a $({\bf 3}, {\bf 1}, {\bf 2})$, the SM top, and a $({\bf 1}, {\bf 3}, {\bf 1})$, the dark top, where the first two labels are the SU$(3) \times$SU(3) subgroup of SU$(6)_c$ and the third label is SU$(2)_L$.  Without the additional structure of a twin Higgs model, the dark top is stable and thus a dark matter candidate.  When the dark top is an electroweak singlet, its primary interactions with the SM are through the Higgs which mediates both freeze-out and direct detection. For dark tops that are partly electroweak doublets, there is a possible $Z$ interaction which greatly increases both the annihilation cross section and direct detection rate.

In expanded cosets, the additional pNGBs that accompany the Higgs multiplet could be the dark matter.  The coset SO(7)/SO(6) is an example where the SO(6) contains the electroweak group and an additional unbroken U(1)~\cite{Ahmed:2020hiw}.  This breaking leads to six pNGBs including the Higgs multiplet and a complex scalar $\chi$ that is stabilized by the U(1). The scalar $\chi$ interacts with the Higgs both through higher-order Goldstone interactions and through interactions with heavy fermions. The momentum-dependence of the Goldstone interactions suppresses the direct detection rate relative to the typical Higgs-mediated rate.

\section{Connections to Astrophysics and Cosmology\label{s.AstCosmo}}

In addition to a variety of potential dark matter candidates, the presence of a rich hidden sector in many neutral naturalness models has myriad implications for astrophysics and cosmology.

\subsection{Cosmological Signals\label{ss.cosmosignals}}

One of the most notable challenges for the mirror twin Higgs model is the contribution of the light twin particles to the relativistic energy density, as measured by the parameter $\Delta N_{\rm eff}$.  This is defined via
\begin{equation}
\rho_{\rm rad} = \rho_\gamma + \frac{7}{8}\left(\frac{4}{11}\right)^{4/3} N_{\rm eff} \,\rho_\gamma,
\end{equation}
where $\rho_{\rm rad}$ is the total energy density in radiation, $\rho_\gamma$ is the energy density of visible photons, and $N_{\rm eff} = 3.044 + \Delta N_{\rm eff}$.  For free streaming radiation $\Delta N_{\rm eff} \leq 0.30$~\cite{Planck:2018vyg}.  

The contribution of a single neutrino species with a temperature of $T_\nu$ is
\begin{equation} \label{eq:dneff_neutrino}
\Delta N_{\rm eff}(1\nu) = \left(\frac{T_{\nu}}{T_{\nu}^{\rm SM}}\right)^4,
\end{equation}
where $T_{\nu}^{\rm SM}$ is the temperature of the SM neutrinos.  The contribution for the twin photon with a temperature $T_\gamma$ is
\begin{equation} \label{eq:dneff_photon}
\Delta N_{\rm eff}(\gamma) = 4.4 \left(\frac{T_{\gamma}}{T_{\gamma}^{\rm SM}}\right)^4,
\end{equation}
where $T_{\gamma}^{\rm SM}$ is the temperature of the SM photon bath. Without $\mathbb{Z}_2$ breaking, three twin neutrinos would contribute $\approx 3$ and the twin photon would contribute $\approx 4.4$ for a total contribution of $\Delta N_{\rm eff} = 7.4$ which is strongly excluded. 

When $\mathbb{Z}_2$ breaking is included the energy density of the twin sector decreases slightly.  The two sectors are kept in thermal equilibrium by the Higgs portal interaction which is shown in Fig.~\ref{fig:diagram-higgsportal} which couples a pair of twin fermions with a pair of SM fermions.  The rate for this process scales as
\begin{equation}
n \langle \sigma v \rangle \sim (y^{\rm twin} y^{\rm SM})^2 \frac{v^2}{f^2} \frac{T^5}{m_h^4},
\end{equation}
where $y^{\rm twin}$ is the Yukawa coupling of the twin fermion in the diagram of Fig.~\ref{fig:diagram-higgsportal} and $y^{\rm SM}$ is the Yukawa coupling of the SM fermion.  When this rate becomes comparable to the Hubble expansion the two sectors decouple.  This occurs around 3 GeV~\cite{Chacko:2016hvu,Craig:2016lyx}.

\begin{figure}
  \centering
\begin{fmffile}{higgsCoupling}
\begin{fmfgraph*}(130,80)
\fmfpen{1.0}
\fmfstraight
\fmfleft{i1,p1,i2}\fmfright{o1,p2,o2}
\fmfv{l= $f_A$}{i1}\fmfv{l=$f_B$}{o1}
\fmf{fermion,tension=0.8}{i1,v1,i2}
\fmf{fermion,tension=0.8}{o1,v2,o2}
\fmf{dashes,tension=0.6,label.side=left,label=$H_A$,label.dist=5}{v1,v3}
\fmf{dashes,tension=0.6,label.side=left,label=$H_B$,label.dist=5}{v3,v2}
\fmfv{decor.shape=circle,decor.filled=full,decor.size=1.5thick,l=$y^\text{SM}$,l.a=180,l.d=5}{v1}
\fmfv{decor.shape=circle,decor.filled=full,decor.size=1.5thick,l=$y^\text{twin}$,l.a=0,l.d=5}{v2}
\fmfv{decor.shape=cross}{v3}
\end{fmfgraph*}
\end{fmffile}
\vspace{0.5cm}
  \caption{Feynman diagram of the Higgs portal interaction that maintains thermal equilibrium between the twin and SM sectors.}
  \label{fig:diagram-higgsportal}
\end{figure}

At a temperature of 3 GeV the SM bath contains photons, neutrinos, electrons, muons, $\tau$ leptons, $u$ quarks, $d$ quarks, $s$ quarks, and $c$ quarks.  As long as $f \gtrsim 3 v$ the twin sector bath only contains photons, neutrinos, electrons, muons, $u$ quarks, $d$ quarks, and $s$ quarks.  After the sectors decouple, when the SM particles leave the bath, the additional degrees of freedom heat the SM sector more than the twin sector, namely that $T_\nu^{\rm twin} = 0.93 \; T_\nu^{\rm SM}$ and $T_\gamma^{\rm twin} = 0.93 \; T_\gamma^{\rm SM}$.  By Eqns.~\eqref{eq:dneff_neutrino} and~\eqref{eq:dneff_photon} we have $\Delta N_{\rm eff} = 5.6$ which is still strongly excluded.

Within the mirror twin Higgs framework, there are a few approaches to reducing the contribution to $\Delta N_{\rm eff}$.  The first is to asymmetrically reheat the visible and twin sectors such that the temperature of the twin sector is lower.  Since $\rho_{\rm rad} \propto T^4$, a slight reduction in the twin sector temperature is enough to satisfy observational constraints.  Adding a right-handed neutrino $N_A$ (and its twin $N_B$) can successfully heat the visible sector preferentially~\cite{Chacko:2016hvu}.  The heavy neutrinos dominate the energy density, then decay after the visible and twin sectors decouple, but before well-probed processes, like Big Bang nucleosynthesis, occur. The decays are dominantly to the visible sector because of the mass difference between the gauge bosons of each sector. A scalar singlet $X$ could also be used instead of a right-handed neutrino~\cite{Craig:2016lyx}. With heavy right-handed neutrinos and a new colored scalar (and its twin), baryogenesis can be achieved along with asymmetric reheating~\cite{Beauchesne:2021opx}.

Additional portals between the visible and twin sectors can allow entropy in the twin sector to return to the visible sector.  For example, the neutrino portal allows the twin neutrino, when sufficiently heavy, to undergo a three-body decay into three visible neutrinos~\cite{Liu:2019ixm}.

Another approach to reducing $\Delta N_{\rm eff}$ is to adjust the decoupling temperature between the visible and twin sectors.  While the decoupling temperature cannot be increased without giving up the twin mechanism, it can be lowered if additional interactions are included to keep the sectors in thermal equilibrium longer.  If the decoupling temperature is below the twin QCD phase transition (and above the visible QCD phase transition) then a larger number of degrees of freedom in the visible thermal bath heat the visible sector radiation substantially more~\cite{Csaki:2017spo}.  A small mixing between the visible and twin neutrinos is sufficient to reduce $\Delta N_{\rm eff}$ to $\approx 1$ and additional model building can bring it into agreement with the bound $\Delta N_{\rm eff} \leq 0.30$.  Similarly, if the  scale of twin confinement differs substantially from the visible scale of confinement, like in the model of Ref.~\cite{Batell:2025iwm}, the difference in degrees of freedom reduces $\Delta N_{\rm eff}$.

Yet a different approach to reducing $\Delta N_{\rm eff}$ leverages the fact that the Higgs interacts weakly with the neutrinos.  Therefore, if the twin fermions are taken to be very heavy then the twin neutrinos decouple from the thermal bath before the visible and twin sectors decouple~\cite{Harigaya:2019shz}. Since the twin neutrinos leave the thermal bath earlier, their temperature is lower, reducing their contribution to $\Delta N_{\rm eff}$.  The contribution from the twin photons can be reduced by including a mass and kinetic mixing with the visible photons so that their entropy is transferred to the visible photons.

In the mirror twin Higgs model, the presence of light twin electrons, twin protons, and twin photons results in twin baryon acoustic oscillations, analogous to the baryon acoustic oscillations in the visible sector~\cite{Chacko:2018vss}.  These twin baryon acoustic oscillations occur prior to twin recombination and both suppress structure formation and leave an oscillatory pattern in the matter power spectrum.  The oscillations in the matter power spectrum only occur at small scales (corresponding to modes that are inside the horizon prior to twin recombination) and the inclusion of a twin sector, with a sufficiently low temperature and relative energy density, is fully consistent with existing Cosmic Microwave Background and large-scale structure data~\cite{Bansal:2021dfh}.  Future measurements of the matter power spectrum have the potential to observe twin baryon acoustic oscillations.

One of the practical challenges in extracting constraints from cosmological sources at small scales is that non-linear corrections can substantially modify predictions of the matter power spectrum~\cite{Bansal:2021dfh}.  This can be addressed by performing $N$-body simulations and in dedicated studies it has been shown that some simplified parameterizations capture the overall behavior~\cite{Zu:2023rmc}.

Without the twin electrons, twin protons, or twin photons, the fraternal twin Higgs model alone does not suppress the matter power spectrum.  A suppression can be achieved in the fraternal twin Higgs model by gauging either twin U(1$)_{B-L}$ or twin U(1$)_L$~\cite{Prilepina:2016rlq}. This additional force mediates interactions between some, or all, of the twin matter and the radiation-like twin neutrinos.  This coupling of the dark matter to radiation suppresses the growth of density perturbations.

\subsection{Cosmological Tensions\label{ss.cosmotensions}}

At present, there are two notable cosmological discrepancies. The first is a tension between the inferred value of the Hubble constant from Planck~\cite{Planck:2018vyg} and the measured late-time value~\cite{Riess:2021jrx}. The late-time value of the Hubble constant is larger by 4\textendash5$\sigma$.  The second is a tension between the inferred value of $\sigma_8$ (the amplitude of matter fluctuations at scales of roughly $8h^{-1}~{\rm Mpc}$) from Planck~\cite{Planck:2018vyg} and the value measured from weak lensing and galactic cluster surveys~\cite{Hildebrandt:2018yau}.  The value from Planck is larger by 2\textendash3$\sigma$.

One attractive feature of the mirror twin Higgs model is that it naturally addresses these tensions simultaneously.  Reference~\cite{Bansal:2021dfh} showed that the twin photon and twin neutrinos provide enough additional radiation to increase the value of the Hubble constant while the twin baryons, comprising a subcomponent of the dark matter, suppress the matter power spectrum at small scales.  It is estimated that the fraternal twin Higgs model (with either twin U(1$)_{B-L}$ or twin U(1$)_L$ gauged) might similarly address these tensions~\cite{Prilepina:2016rlq}.  It has also been claimed that the hadrosymmetric twin Higgs model, where only quarks have twin counterparts and the leptons do not, alleviates the Hubble and $\sigma_8$ tensions~\cite{Sotudeh:2025rpr}.

This feature of twin Higgs where a fraction of the dark matter interacts, for a time, with dark radiation is one example of a general category of dark matter called atomic dark matter.  A similar study to Ref.~\cite{Bansal:2021dfh} was performed in Ref.~\cite{Bansal:2022qbi} in the atomic dark matter framework.  In addition to the cosmic microwave background and large-scale structure, the structure of subhalos have been studied~\cite{Gemmell:2023trd} and the inner densities of isolated dwarf galaxies~\cite{Roy:2024bcu}.  These studies were accomplished through the use of hydrodynamical simulations~\cite{Roy:2023zar} and show that these subhalos have cuspier density profiles as do the inner parts of the dwarf galaxies.

\subsection{Phase Transitions, Gravitational Waves, and Baryogenesis\label{ss.PT}}

It is very interesting to consider the potential consequences of the various early Universe phase transitions expected in twin Higgs models. 
Reference~\cite{Schwaller:2015tja} presents the basic conditions required for a strong first-order phase transition in confining SU$(N)$ hidden sectors and studies the associated gravitational wave signals and observational prospects. 
Lattice studies indicate that QCD with physical quark masses undergoes a smooth crossover around $T \sim 160$ MeV~\cite{Schmidt:2025ppy}. However, in a hidden QCD-like sector with $N_f \geq 3$ light quark flavors the chiral transition is argued to be first-order~\cite{Pisarski:1983ms}, and in the heavy-quark (pure-glue) limit the theory is known to undergo a first-order confinement transition~\cite{Iwasaki:1991pc,Boyd:1996bx} (see also discussion in Sec.~\ref{ss.lattice.confinement}). Such first-order cosmological phase transitions proceed via bubble nucleation, growth, and percolation, leading to the production of gravitational waves from bubble wall collisions, sound waves, and magnetohydrodynamic turbulence (see Refs.~\cite{Caprini:2019egz,Athron:2023xlk,Croon:2024mde} for reviews).
In particular, a first-order mirror QCD phase transition and corresponding gravitational wave signal is expected in fraternal twin Higgs models since no light quark flavors are present in the spectrum. This signal could potentially be observed by pulsar timing arrays and satellite-based experiments in the nano-Hz to milli-Hz frequency ranges~\cite{Schwaller:2015tja,Zu:2023olm,Rosenlyst:2023tyj}.  See also Ref.~\cite{Huang:2020crf}, which incorporates pure-gauge lattice studies to make detailed predictions for gravitational wave signals which can be relevant for theories with no light quark flavors.
Complementing these works, detailed studies of the electroweak and global U(4) symmetry breaking phase transitions in mirror twin Higgs models were presented in Refs.~\cite{Fujikura:2018duw,Badziak:2022ltm}. In particular, Ref.~\cite{Badziak:2022ltm} found that strong first-order phase transitions can occur depending on the source of $\mathbb{Z}_2$ symmetry breaking, opening up the possibility for observable gravitational wave signals and new baryogenesis mechanisms in mirror twin Higgs models.  

Other works examine the possibility that electroweak symmetry may not be restored at high temperature in twin Higgs models. Reference~\cite{Kilic:2015joa} demonstrates that in mirror twin Higgs models, while the one-loop quadratic contributions to the Higgs potential at finite temperature from the SM degrees of freedom are canceled by their same spin partners (a consequence of the twin Higgs mechanism), subleading corrections remain which drive the potential to a symmetry-restored phase at high temperatures. On the other hand, Ref.~\cite{Matsedonskyi:2020kuy} points out that if the Yukawa couplings of twin fermions other than those of the twin top are larger than their SM counterparts, the electroweak symmetry may not be restored at scales below the global symmetry breaking scale $f\sim {\rm TeV}$. This could have important implications for electroweak baryogenesis. Furthermore, Ref.~\cite{Badziak:2025fdp} presents a model of symmetry non-restoration in the supersymmetric twin Higgs model.

Besides electroweak baryogenesis, several works propose mechanisms of baryogenesis, leptogenesis, and cogenesis of visible and dark matter asymmetries through the CP-violating out-of-equilibrium decays of heavy states. Reference~\cite{Farina:2016ndq} explores a simple model of twin baryogenesis, in which asymmetries in the twin and visible sectors are generated through the out-of-equilibrium decay of a TeV scale particle charged under both baryon and twin baryon number. The resulting twin baryons can serve as an asymmetric dark matter candidate and explain the relic abundance provided the twin baryon mass is around 5 GeV, which can be realized with the introduction of suitable sources of $\mathbb{Z}_2$ breaking. As another example, Ref.~\cite{Earl:2019wjw} studies a model in which light Dirac neutrino masses are generated through a seesaw mechanism connecting the visible and mirror sectors. Leptogenesis occurs through the decays of the heavy singlet Dirac neutrinos to the SM leptons and mirror leptons. Reference~\cite{Alonso-Alvarez:2023bat} studies a model where baryogenesis is realized through the out-of-equilibrium decays of right-handed neutrinos, but without any hard $\mathbb{Z}_2$ breaking.
 See also Refs.~\cite{Feng:2020urb,Beauchesne:2020mih,Beauchesne:2021opx,Kilic:2021zqu,Bittar:2023kdl} for interesting related mechanisms of baryogenesis or leptogenesis.

\subsection{Mirror Stars} 

In the simplest realizations of mirror twin Higgs models, the $\mathbb{Z}_2$ symmetry implies the presence of both electromagnetism and nuclear physics in the mirror sector. It is therefore natural to speculate about astrophysical compact objects built out of mirror matter, i.e., mirror stars~\cite{Curtin:2019lhm,Curtin:2019ngc,Curtin:2020tkm,Howe:2021neq}. 
References~\cite{Curtin:2019lhm,Curtin:2019ngc} discuss the astrophysical signatures of such mirror stars. If the mirror and visible sector photons interact through a small kinetic mixing, visible matter may be captured in mirror star cores. This visible sector component of the mirror star\textemdash the ``SM nugget''\textemdash is then heated through interactions with mirror matter. This may present discoverable electromagnetic signatures in both optical frequencies, due to bremsstrahlung cooling processes, and in X-rays, due to ``Thomson conversion'' of mirror photons to ordinary photons. 
A detailed search strategy for directly observing mirror stars using Gaia observations is presented in Ref.~\cite{Howe:2021neq}.
Furthermore, Ref.~\cite{Hippert:2021fch} explores the formation and structure of mirror neutron stars using realistic equations of state adapted from those describing ordinary neutron stars. 
The observational prospects for mirror neutron stars using gravitational waves and binary pulsars are also discussed~\cite{Armstrong:2023cis}.  It may be possible, in fact, to distinguish the merger of two standard neutron stars from those containing mirror matter~\cite{Hippert:2022snq}.  

As these studies illustrate, the presence of rich hidden sectors in neutral naturalness theories may lead to highly novel astrophysical systems and signatures. Given the exciting program of current and planned astronomical surveys, further investigations in this direction would be timely.

\section{Connections to Neutrinos\label{s.Neutrinos}}

Neutral naturalness models can impact neutrino physics in a variety of ways, including through novel mechanisms for neutrino masses, new leptogenesis scenarios, connections with the dark radiation problem, and explanations of oscillation anomalies and other experimental discrepancies. First, in many neutral naturalness scenarios such as the twin Higgs, there are a host of new fermions that are neutral under the SM gauge symmetries. With suitable symmetry breaking mechanisms, such fermions can marry with SM neutrinos, effectively serving as right-handed neutrinos and generating neutrino masses through  a seesaw mechanism. Along these lines, Ref.~\cite{Batell:2015aha} proposed that the neutral top partners, responsible for cancelling the dominant corrections to the Higgs potential, may simultaneously function as right-handed neutrinos. In this framework, naturalness considerations robustly predict a TeV-scale seesaw, and large neutrino Yukawa couplings may be achieved through the inverse~\cite{Mohapatra:1986bd} or linear~\cite{Malinsky:2005bi} seesaw mechanisms along with an associated rich phenomenology. It is worth noting that 
the neutrino Yukawa couplings necessarily break the SU(3) symmetry of the right-handed neutrino top partners in this scenario. This implies that, unlike other constructions such as the Twin Higgs, this SU(3) symmetry can be either a spontaneously broken gauge symmetry or an explicitly broken global symmetry. Another interesting possibility for neutrino mass generation arises in models with fermionic 
singletons\textemdash completely neutral fields with no $\mathbb{Z}_2$ partners\textemdash which can connect the visible and mirror lepton sectors~\cite{Bishara:2018sgl}. There is clearly a wide scope for further studies of novel neutrino mass mechanisms within models of neutral naturalness.

The dynamics underlying neutrino masses may also have important consequences for cosmology in the mirror twin Higgs framework, as discussed in Sec.~\ref{s.AstCosmo}. In particular, the challenges associated with a large number of relativistic degrees of freedom in the mirror twin Higgs model can be circumvented in models with a low-scale ($\sim$GeV) seesaw model, as described in Ref.~\cite{Chacko:2016hvu}. 
In this scenario, the right-handed neutrinos freeze-out with a large thermal abundance and later become non-relativistic due to the cosmological expansion, eventually dominating the energy density.  After the SM and mirror sectors decouple, the right-handed neutrinos preferentially decay to SM states, increasing the energy density in the visible sector relative to that of the mirror sector, thereby alleviating the bounds from dark radiation. 
Ref.~\cite{Csaki:2017spo} also investigated the impact of a low-scale seesaw on twin cosmology, demonstrating that mixing between the mirror and SM neutrinos can maintain thermal equilibrium between the two sectors until temperatures below the mirror QCD phase transition, thus lowering the prediction for $\Delta N_{\rm eff}$. Furthermore, Ref.~\cite{Bittar:2024ryj} investigates a mirror twin Higgs model with spontaneous $\mathbb{Z}_2$ and electroweak symmetry breaking, in which an SU(2) triplet scalar generates SM neutrino masses via the type-II seesaw mechanism, while the twin neutrinos acquire large masses, thereby mitigating the dark radiation problem and yielding distinctive predictions for $N_{\rm eff}$ and collider signatures. Interesting scenarios of leptogenesis~\cite{Earl:2019wjw,Feng:2020urb} and twin sterile neutrino dark matter~\cite{Holst:2023hff} within the twin Higgs framework have also been put forth, and it would be of great interest to expand on these investigations; see Sec.~\ref{ss.PT} for further discussion.

In a separate direction, neutral naturalness models can potentially provide explanations for experimental anomalies associated with the neutrino sector. For example, Ref.~\cite{Bai:2015ztj} speculates that three twin neutrinos may explain the LSND and MiniBooNE anomalies while evading the otherwise stringent constraints from disappearance experiments. As another illustration, Ref.~\cite{AristizabalSierra:2018emu} proposes an explanation of the EDGES 21-cm absorption signal that relies on a relic population of mirror sector neutrinos that produce mirror photons through their decays. The mirror photons are then resonantly converted to visible sector photons, enhancing the 21-cm absorption signal. 
Finally, in Refs.~\cite{Liu:2020sim,Liu:2021kqp} a potential resolution of the muon anomalous magnetic moment discrepancy in seesaw extensions of twin Higgs models is explored.

\section{Connections to Flavor\label{s.Flavor}}

There are potentially far-reaching implications for flavor physics in neutral naturalness models, although the existing literature has only scratched the surface of the subject.  An interesting illustration comes from the consideration of flavor in composite twin Higgs models, as in the study of Ref.~\cite{Csaki:2015gfd}. In ordinary composite Higgs models, the attractive hypothesis of quark flavor anarchy generically leads to stringent bounds on the scale of the composite resonances, which in turn requires a rather severe tuning of the composite Higgs potential, typically at the per-mille level. On the other hand, in composite twin Higgs models, the composite resonances can be freely raised to scales of order 10 TeV without incurring significant tuning, since the twin partners (and not the visible sector composite resonances) cancel the dominant quantum corrections to the Higgs potential. Raising the composite resonances then significantly weakens, though does not entirely eliminate, constraints from flavor, allowing a scenario with percent-level tuning. 

Also motivated by the flavor puzzle and the little hierarchy problem, Ref.~\cite{Altmannshofer:2020mfp} studied a twinned flavorful two-Higgs doublet model, in which one Higgs doublet generates the masses of the third generation while a second doublet is responsible for masses of the light generations. This scenario can provide a partial understanding of the hierarchical patterns observed in the fermion masses and also predicts novel signatures associated with the heavy scalar doublet in both flavor physics and at colliders. 

While the examples above primarily concern flavor physics in the visible sector, there are also good motivations to consider the nature of fermion flavor in the mirror sector in twin Higgs models. The fraternal twin Higgs scenario~\cite{Craig:2015pha} represents perhaps the most dramatic breaking of $\mathbb{Z}_2$ symmetry in the fermion flavor sector, with all but the third generation twin fermions removed from the spectrum. This scenario is well motivated from a bottom-up naturalness perspective and has important consequences for phenomenology and cosmology, as discussed above in Secs.~\ref{ss.twinHiggsVar}, \ref{ss.fraternalDM}, and \ref{s.AstCosmo}.  Beyond the fraternal scenario, Ref.~\cite{Barbieri:2016zxn} considers a more modest breaking of the $\mathbb{Z}_2$ exchange symmetry in the Yukawa couplings (except of course in the top Yukawa). This has several interesting implications, including the radiative generation of $\mathbb{Z}_2$ breaking in the Higgs potential, which provides the correct vacuum alignment, the alleviation of bounds from too much dark radiation, and several potential dark matter candidates in the form of mirror baryons and leptons. Following this work, the authors proposed an effective theory that simultaneously explains the fermion mass hierarchies as in, e.g., Froggatt-Nielsen models, and also realizes the $\mathbb{Z}_2$ breaking in the Yukawa couplings through spontaneous $\mathbb{Z}_2$ breaking between the visible and mirror flavor symmetries~\cite{Barbieri:2017opf}. Novel spontaneous twin gauge symmetry breaking patterns can also lead to new flavor structures in the twin fermion sector~\cite{Batell:2019ptb,Liu:2019ixm,Batell:2020qad}, with an array of associated phenomenological signatures.

Quark flavor physics is intertwined with the strong CP problem, as both originate from the structure of the quark mass matrices and their complex phases, which together determine the physical value of the QCD $\theta$ angle. Therefore, it is important to understand how solutions to the Strong CP problem fit within the neutral naturalness paradigm. 
Reference~\cite{Albaid:2015axa} discussed the potential role of the twin Higgs mirror $\mathbb{Z}_2$ symmetry in addressing the Strong CP problem, with particular emphasis on the challenges associated with the $\mathbb{Z}_2$ breaking needed for a viable electroweak vacuum. 
The authors of~\cite{Barbieri:2016zxn} considered the possibility of a heavy QCD axion within the mirror twin Higgs framework, where the $\mathbb{Z}_2$ symmetry is broken by the Yukawa sector. They argued that a consistent solution to the strong CP problem can be achieved if the $\mathbb{Z}_2$ breaking arises dynamically from VEVs of Froggatt–Nielsen symmetry–breaking fields.
Furthermore, 
Ref.~\cite{Badziak:2025fdp} explored the role of the QCD axion within a supersymmetric twin Higgs UV completion in addressing the strong CP problem and generating the baryon asymmetry via the axiogenesis mechanism~\cite{Co:2019wyp}. Beyond these works, the strong CP problem remains largely unexplored in the context of neutral naturalness, making it a compelling direction for further investigation. 

It would be interesting to further explore solutions to the flavor puzzle within the neutral naturalness paradigm, particularly in the setting of concrete UV completions that permit extrapolation between the TeV scale and high scales where flavor is generated. Likewise, a worthy goal is to understand to what extent the structure of flavor in the hidden sectors of neutral naturalness theories can be tied through symmetry to visible sector observables at colliders and in precision measurements. 

\section{Summary\label{s.Sum}}
As experiments have reached ever higher energies without clear evidence of deviations from the SM, understanding the hierarchy between the electroweak scale and higher physics energy scales has become increasingly acute. A particularly interesting explanation of Higgs naturalness is positing a new symmetry which relates the SM quarks to colorless partners, which has become known as neutral naturalness. These are primarily bottom-up constructions that characterize the new partner particles and interactions only up to the few TeV scale and require completion at high energies. Typically, these completions include new colored states with masses of order a few TeV, motivating future high-energy colliders to thoroughly test Higgs naturalness due to new symmetries. 

Realizations of neutral naturalness typically include a hidden sector of particles that are related to at least some of the SM fields by a discrete symmetry. Candidates for the cosmological dark matter, either as a single particle species or from a collection of stable particles, are often obvious consequences of this structure. This discrete symmetry also relates the SM's SU(3$)_c$ color group to a distinct hidden-sector SU(3) gauge group. Consequently, neutral naturalness generically leads to rich dark sectors, with phenomenological features such as dark showers, semivisible jets, quirks, and long-lived particles. Useful nonperturbative inputs for studying this strongly-coupled hidden SU(3) sector can be obtained from lattice gauge theory calculations of the spectrum of bound states, matrix elements, and other observables.  Because at least some of the hidden-sector particles are essential to explaining the mass of the Higgs boson, the Higgs is a robust link between the sectors. Signals of these connections can appear in deviations from the SM Higgs couplings and exotic Higgs decay modes that can be probed at future Higgs factories. 

While the number of novel realizations of the idea of neutral naturalness continues to grow, the majority of past work has focused on the twin Higgs framework. This pertains to higher-energy (UV) completions, collider phenomenology, dark matter candidates, cosmological and astrophysical signals, and intersections with neutrino and flavor physics. That this narrower focus has led to such a rich variety of signals and overlaps with other aspects of beyond the SM physics is impressive, but also points the way to further exploration through other neutral naturalness models. Such explorations may lead to new connections to higher-scale physics that applies to the SM flavor structure, the strong CP puzzle, or gauge unification. New dark matter candidates, including those associated with  UV completions as well as lighter, non-WIMP, candidates may also be discovered. 
 
 Neutral naturalness connections across the range of new phenomena are particularly fascinating due to the discrete symmetries that relate hidden-sector particles to SM counterparts. This assortment of concrete models accommodates a variety of new signals at the intensity frontier, energy frontier, and cosmic frontier which are correlated with each other and, through discrete symmetry, with the known properties of the SM. This makes models of neutral naturalness compelling benchmarks for organizing simplified model signals in addition to suggesting new strategies for discovering new physics.

\section*{Acknowledgments}
We are thankful to the Snowmass theory frontier beyond-the-standard-model model building topical group organizers Patrick Fox, Graham Kribs, and Hitoshi Murayama for the invitation to write this review and to Jonathan Feng for the encouragement to further develop and expand it. The work of B.B. is supported by the U.S. Department of Energy under grant No. DE–SC0007914. The work of M.L. is supported by the US Department of Energy under
grant No. DE-SC0007914, by the National Science Foundation under grant No. PHY-2412696, and by Pitt PACC. The work of E.T.N. is supported by the U.S.~Department of Energy under grant number DE-SC0010005. The work of C.B.V is supported in part by National Science Foundation Grant No. PHY-2210067.

\bibliographystyle{elsarticle-num} 
\bibliography{ReviewNNbib}

\end{document}